\documentstyle[12pt,epsf,epsfig]{article}
\newcount\timecount
\newcount\hours \newcount\minutes  \newcount\temp \newcount\pmhours
\hours = \time
\divide\hours by 60
\temp = \hours
\multiply\temp by 60
\minutes = \time
\advance\minutes by -\temp
\def\hour{\the\hours}
\def\minute{\ifnum\minutes<10 0\the\minutes
            \else\the\minutes\fi}
\def\clock{
\ifnum\hours=0 12:\minute\ AM
\else\ifnum\hours<12 \hour:\minute\ AM
      \else\ifnum\hours=12 12:\minute\ PM
            \else\ifnum\hours>12
                 \pmhours=\hours
                 \advance\pmhours by -12
                 \the\pmhours:\minute\ PM
                 \fi
            \fi
      \fi
\fi
}

\def\monthname{\relax\ifcase\month 0/\or January\or February\or
   March\or April\or May\or June\or July\or August\or September\or
   October\or November\or December\else\number\month/\fi}

\def\bold#1{\setbox0=\hbox{$#1$}%
     \kern-.025em\copy0\kern-\wd0
     \kern.05em\copy0\kern-\wd0
     \kern-.025em\raise.0433em\box0 }

\def\beq{\begin{equation}}
\def\eeq{\end{equation}}

\def\ga{\mathrel{\raise.3ex\hbox{$>$\kern-.75em\lower1ex\hbox{$\sim$}}}}
\def\la{\mathrel{\raise.3ex\hbox{$<$\kern-.75em\lower1ex\hbox{$\sim$}}}}
\def\gev{{\rm \, Ge\kern-0.125em V}}
\def\tev{{\rm \, Te\kern-0.125em V}}
\def\gyr{{\rm \, G\kern-0.125em yr}}
\def\ohsq{\Omega_{\chi} h^2}

\def\gappeq{\mathrel{\rlap {\raise.5ex\hbox{$>$}}
{\lower.5ex\hbox{$\sim$}}}}
\def\lappeq{\mathrel{\rlap{\raise.5ex\hbox{$<$}}
{\lower.5ex\hbox{$\sim$}}}}
\def\Toprel#1\over#2{\mathrel{\mathop{#2}\limits^{#1}}}

\def\sel{{\widetilde e}}

\def\stau{\widetilde \tau}

\def\sbot{\widetilde b}
\def\snu{\widetilde \nu}

\def\m12{m_{1\!/2}}

\begin{document}
\begin{titlepage}
\pagestyle{empty}
\baselineskip=21pt
\rightline{\tt hep-ph/0306219}
\rightline{CERN--TH/2003-138}
\rightline{UMN--TH--2205/03}
\rightline{FTPI--MINN--03/16}
\vskip 0.2in
\begin{center}
{\large {\bf Updated Post-WMAP Benchmarks for Supersymmetry}} \\
\end{center}
\begin{center}
\vskip 0.2in
{\bf M.~Battaglia}$^1$, {\bf A.~De~Roeck}$^1$,
{\bf J.~Ellis}$^1$, {\bf F.~Gianotti}$^1$, {\bf K.~A.~Olive}$^{2}$
and {\bf L.~Pape}$^{1}$
\vskip 0.1in
{\it
$^1${CERN, Geneva, Switzerland}\\
$^2${William I. Fine Theoretical Physics Institute, \\
University of Minnesota, Minneapolis, MN 55455, USA}}\\
\vskip 0.2in
{\bf Abstract}
\end{center}
\baselineskip=18pt \noindent

We update a previously-proposed set of supersymmetric benchmark scenarios,
taking into account the precise constraints on the cold dark matter
density obtained by combining WMAP and other cosmological data, as well as
the LEP and $b \rightarrow s \gamma$ constraints. We assume that $R$
parity is conserved and work within the constrained MSSM (CMSSM)  with
universal soft supersymmetry-breaking scalar and gaugino masses $m_0$ and
$m_{1/2}$. In most cases, the relic density calculated for the previous
benchmarks may be brought within the WMAP range by reducing slightly
$m_0$, but in two cases more substantial changes in $m_0$ and $m_{1/2}$
are made. Since the WMAP constraint reduces the effective dimensionality
of the CMSSM parameter space, one may study phenomenology along `WMAP
lines' in the $(m_{1/2}, m_0)$ plane that have acceptable amounts of dark
matter. We discuss the production, decays and detectability of sparticles
along these lines, at the LHC and at linear $e^+ e^-$ colliders in the
sub- and multi-TeV ranges, stressing the complementarity of hadron and
lepton colliders, and with particular emphasis on the neutralino sector.
Finally, we preview the accuracy with which one might be able to predict
the density of supersymmetric cold dark matter using collider
measurements.

\vfill
\leftline{CERN--TH/2003-138}
\leftline{June 2003}
\end{titlepage}
\baselineskip=18pt

\section{Introduction}

One of the crucial topics in the planning of analyses of data from
experiments at present and future colliders is the search for
supersymmetry~\cite{hierarchy}. Even the minimal supersymmetric extension
of the Standard Model (MSSM), which conserves $R$ parity, has over 100
free parameters once arbitrary soft supersymmetry-breaking parameters are
allowed. For this reason, much attention is focussed on the constrained
MSSM (CMSSM), in which the soft supersymmetry-breaking scalar masses
$m_0$, gaugino masses $m_{1/2}$ and trilinear parameters $A_0$ are each
assumed to be universal at some high input scale, as in minimal
supergravity and other models. These CMSSM parameters are constrained
principally by the absence at LEP of new particles with masses $\lappeq
100$~GeV, by the agreement of $b \to s \gamma$ decay with Standard Model
predictions, by the measurement of the anomalous magnetic moment of the
muon, $g_\mu - 2$, and by the range allowed for relic cold dark matter,
$\Omega_{CDM}$.

Benchmark supersymmetric scenarios have a venerable 
history~\cite{oldBench}. A couple of years ago, a new set of benchmark 
supersymmetric models was proposed in~\cite{Bench},
consistent with all the above experimental constraints, as well as the
cosmological constraint on $\Omega_{CDM}$. Subsequently, there has not
been any significant change in the LEP limits on supersymmetric
particles~\cite{LEPSUSY,LEPHWG}, we are still looking forward to more
sensitive searches at the Fermilab Tevatron collider, the situation of $b
\to s \gamma$ decay has changed little~\cite{bsg}, and the interpretation
of the $g_\mu - 2$ measurements remains unclear~\cite{g-2}. Various other
benchmark scenarios have been proposed, in particular some supplementary
points and lines in the CMSSM and in models with different mechanisms for
supersymmetry breaking~\cite{SPS}, benchmarks for supersymmetric Higgs
physics~\cite{HW}, and scenarios in which the prospects for the Fermilab
Tevatron collider are more favourable~\cite{BenchKane,BKT}.

The road(w)map for supersymmetric phenomenology has, however, been altered
significantly by the recent improved determination of the allowable range
of the cold dark matter density obtained by combining WMAP and other
cosmological data:  $0.094 < \Omega_{CDM} < 0.129$ at the 2-$\sigma$
level~\cite{WMAP}. This range is consistent with earlier indications, but
more precise. Within the MSSM with conserved $R$ parity, if one assumes
that most of the cold dark matter consists of the lightest
supersymmetric particle (LSP) and identifies this with the lightest
neutralino $\chi$~\cite{EHNOS}, this WMAP constraint reduces the 
dimensionality of the
parameter space. In the CMSSM, in particular, whereas generic regions of
the $(m_{1/2}, m_0)$ planes would previously have been allowed for fixed
values of $\tan \beta$ and $A_0$~\cite{CMSSM}, now only thin strips are
permitted~\cite{EOSS,others}~\footnote{Some authors had also considered an
analogous narrow range for $\ohsq$ before this was mandated by the WMAP
data~\cite{earlier}.}.

It is of course possible to relax the restrictions to these narrow `WMAP
lines' in various ways. One could consider models that violate $R$ parity,
though these would also be subject to (different) astrophysical and
cosmological constraints. Or one could consider alternatives
to gravity-mediated supersymmetry breaking, such as anomaly- or gauge-
mediated supersymmetry breaking, in which the sparticle spectra could be
radically different, and/or the LSP might be a gravitino~\cite{SPS}.  
However, we do not discuss such models in this paper.

Here, we consider first the minor modifications of the previous CMSSM
benchmark scenarios that bring them on to the WMAP lines. In most cases,
this requires only a small change in $m_0$, keeping $m_{1/2}$ the same as
before, as seen in Table~\ref{tab:Keith}. An exception is the previous
benchmark H, which had been chosen at the end of one of the coannihilation
`tails', at the largest possible value of $m_{1/2}$. The WMAP constraint
not only reduces the possible spread in $m_0$ for fixed $m_{1/2}$, but
also reduces the allowable range of $m_{1/2}$. Therefore, our modification
of benchmark H has reduced values of both $m_{1/2}$ and $m_0$. Another
example is benchmark M, which has also been shifted significantly because
the rapid-annihilation `funnel' for $\tan \beta = 50$ and $\mu > 0$ is
affected substantially by the WMAP restriction on $\ohsq$. More detailed
discussions of these updated benchmark points are given in Section 2,
where we also discuss the (generally small)  extent to which the decay
branching ratios differ from the previous versions of the benchmark
points~\footnote{Where it seems necessary to avoid confusion with the
previous versions of these benchmark scenarios~\cite{Bench}, we denote the
updated versions with primes: A', B', etc.: otherwise we retain the same
notation.}.

\begin{table}[p!]
\renewcommand{\arraystretch}{0.80}
{\bf Supersymmetric spectra in post-WMAP benchmark scenarios}\\
{~}\\
\begin{tabular}{|c|r|r|r|r|r|r|r|r|r|r|r|r|r|}
\hline
Model          & A' &  B'  &  C' &  D' &  E' &  F' &  G' &  H' &  I' &  J' &  K' &  L' &  M' \\ 
\hline
$m_{1/2}$      & 600 & 250 & 400 & 525 &  300& 1000& 375 & 935 & 350 & 750 & 1300& 450 & 1840 \\
               &     &     &     &     &     &     &     &{\tiny (1500)}&   &  &{\tiny (1150)}&  &{\tiny (1900)} \\
$m_0$          & 120 &  60 &  85 & 110 & 1530& 3450& 115 & 245 & 175 & 285 & 1000& 300 & 1100 \\
               &{\tiny (140)}&{\tiny (100)}& {\tiny (90)}&{\tiny (125)}&{\tiny (1500)}&   &
{\tiny (120)}&{\tiny (419)}& {\tiny (180)}& {\tiny (300)}&     &{\tiny (350)}& {\tiny (1500)} \\
$\tan{\beta}$  & 5   &  10 &  10 & 10  & 10  & 10  & 20  & 20  & 35  & 35  & 35  & 50  & 50   \\
sign($\mu$)    & $+$ & $+$ & $+$ & $-$ & $+$ & $+$ & $+$ & $+$ & $+$ & $+$ & $-$ & $+$ & $+$  \\ 
$\alpha_s(m_Z)$& 121 & 125 & 123 & 121 & 124 & 120 & 124 & 120 & 123 & 120 & 118 & 122 & 117 \\
$m_t$          & 175 & 175 & 175 & 175 & 171 & 171 & 175 & 175 & 175 & 175 & 175 & 175 & 175  \\ \hline
Masses         &     &     &     &     &     &     &     &     &     &     &     &     &      \\ \hline
$|\mu(m_Z) |$  & 741 & 333 & 503 & 634 & 205 & 496 & 471 &1026 & 439 & 843 &1317 & 540 & 1764 \\ \hline
h              & 115 & 113 & 116 & 117 & 114 & 118 & 117 & 122 & 116 & 121 & 118 & 118 &  124 \\
H              & 884 & 375 & 578 & 736 &1532 &3491 & 523 &1185 & 452 & 883 &1176 & 489 & 1652 \\
A              & 882 & 375 & 578 & 735 &1533 &3491 & 523 &1185 & 451 & 883 &1176 & 489 & 1663 \\
H$^{\pm}$      & 886 & 383 & 584 & 740 &1535 &3492 & 529 &1188 & 459 & 887 &1180 & 496 & 1654 \\ \hline
$\chi_1$       & 252 &  98 & 163 & 220 & 115 & 430 & 153 & 402 & 143 & 320 & 573 & 187 &  821 \\
$\chi_2$       & 480 & 181 & 310 & 424 & 182 & 522 & 289 & 774 & 270 & 615 &1105 & 358 & 1583 \\
$\chi_3$       & 761 & 346 & 519 & 655 & 221 & 523 & 489 &1068 & 464 & 897 &1413 & 588 & 1994 \\
$\chi_4$       & 775 & 365 & 535 & 662 & 304 & 885 & 504 &1078 & 478 & 906 &1421 & 599 & 1999 \\
$\chi^{\pm}_1$ & 480 & 180 & 309 & 424 & 174 & 511 & 290 & 774 & 270 & 615 &1105 & 358 & 1583 \\
$\chi^{\pm}_2$ & 775 & 367 & 535 & 664 & 304 & 886 & 505 &1079 & 479 & 907 &1422 & 600 & 1999 \\ \hline
$\tilde{g}$    &1715 & 715 &1145 &1495 & 869 &2914 &1075 &2681 & 999 &1593 &3716 & 994 & 5262 \\ \hline
$e_L$, $\mu_L$ & 425 & 188 & 289 & 375 &1544 &3512 & 285 & 673 & 300 & 581 &1319 & 430 & 1635 \\
$e_R$, $\mu_R$ & 261 & 121 & 180 & 232 &1535 &3471 & 189 & 433 & 224 & 405 &1114 & 348 & 1300 \\
$\nu_e$, $\nu_{\mu}$ 
               & 418 & 171 & 278 & 367 &1542 &3511 & 273 & 669 & 289 & 575 &1317 & 422 & 1633 \\ 
$\tau_1$       & 258 & 112 & 172 & 225 &1522 &3443 & 162 & 403 & 155 & 323 & 971 & 200 & 920 \\ 
$\tau_2$       & 425 & 192 & 291 & 376 &1538 &3498 & 291 & 670 & 310 & 573 &1268 & 420 & 1511 \\ 
$\nu_{\tau}$   & 418 & 187 & 277 & 366 &1542 &3497 & 270 & 661 & 277 & 555 &1261 & 386 & 1502 \\ \hline 
$u_L$, $c_L$   &1202 & 546 & 834 &1064 &1644 &3908 & 792 &1808 & 755 &1493 &2602 & 965 & 3491 \\ 
$u_R$, $c_R$   &1151 & 527 & 803 &1021 &1635 &3867 & 762 &1730 & 723 &1429 &2494 & 930 & 3332 \\ 
$d_L$,$s_L$    &1205 & 552 & 838 &1067 &1646 &3909 & 797 &1810 & 758 &1429 &2603 & 968 & 3492 \\ 
$d_R$, $s_R$   &1144 & 526 & 799 &1016 &1634 &3861 & 759 &1718 & 723 &1495 &2479 & 925 & 3309 \\ 
$t_1$          & 896 & 393 & 618 & 807 &1050 &2580 & 587 &1380 & 553 &1131 &1935 & 710 & 2630 \\ 
$t_2$          &1143 & 573 & 819 &1013 &1387 &3330 & 777 &1677 & 731 &1372 &2237 & 891 & 3054 \\ 
$b_1$          &1101 & 502 & 765 & 976 &1379 &3323 & 717 &1645 & 659 &1325 &2173 & 815 & 2998 \\ 
$b_2$          &1144 & 528 & 798 &1011 &1622 &3834 & 756 &1695 & 711 &1377 &2242 & 880 & 3062 \\ 
\hline 
\end{tabular} 
\caption[]{\it 

Proposed post-WMAP CMSSM benchmark points and mass spectra (in GeV), as
calculated using {\tt SSARD}~\protect\cite{SSARD} and {\tt
FeynHiggs}~\protect\cite{Heinemeyer:2000yj}, using the
one-loop corrected effective potential computed at the electroweak scale
and one-loop corrections to the chargino and neutralino masses. We 
recall (in parentheses) the values of $m_{1/2}$ and $m_0$ used 
in~\protect\cite{Bench}, in cases where they differ. As
in~\protect\cite{Bench}, exact gauge coupling unification is enforced,
resulting in the predictions for $\alpha_s(m_Z)$ that are shown in units
of 0.001. We use $A_0 = 0$, $m_b(m_b)^{\overline {MS}} = 4.25$~GeV and
$m_t = 175$~GeV for most of the points, with the exceptions of points E'
and F', where $m_t = 171$~GeV is used.}

\label{tab:Keith}
\end{table}

Then, in Section 3, we discuss systematically how the phenomenology of
supersymmetric models varies with $m_{1/2}$ along the WMAP lines. We first
present parametrizations of these lines appropriate for the {\tt
SSARD}~\cite{SSARD} and {\tt ISASUGRA 7.67}~\cite{ISASUGRA} codes. We then
discuss the patterns of decays of various selected sparticles. We show in
particular that $\chi_2 \to {\tilde \tau_1} \tau$ decays are generally
important, though other $\chi_2 \to {\tilde \ell} \ell$ decays (where 
$\ell \equiv e, \mu$) are also
important for low $m_{1/2}$ at low $\tan \beta$. The decays
$\chi_2 \to \chi h, \chi Z$ are never very large on the WMAP lines,
despite being kinematically allowed for most values of $m_{1/2}$. We then
discuss the average numbers of $Z, h$ and $\tau$ particles produced per
sparticle production event at the LHC. This study confirms the importance
of the $\tau$ signature, with trilepton signatures also looking promising.
We discuss the extent to which our various benchmark scenarios sample
generic features along the WMAP lines.

Section 4 contains a discussion of sparticle detectability at various
accelerators, starting with the LHC~\footnote{We have nothing to add here
to the discussion in~\cite{Bench} of the prospects for detecting
supersymmetry at the Fermilab Tevatron collider, referring the reader
to~\cite{BenchKane,BKT} for alternative discussions.}. We find that the
lightest MSSM Higgs boson and all the squarks are in principle observable
along the complete WMAP lines, except along parts of the
rapid-annihilation `funnels', whereas sleptons and other neutralinos and
charginos are detectable only in the lower parts of the WMAP ranges of
$m_{1/2}$. This discussion is extended in Subsection 4.2 to linear $e^+
e^-$ colliders with $E_{CM} = 0.5, 1.0$~\cite{LC} and 3.0 or
5.0~TeV~\cite{CLIC}.  Confirming previous studies~\cite{LC,Bench,SPS}, we
see that a machine with $E_{CM} = 0.5$~TeV would be able to explore
supersymmetry in the lower parts of the WMAP ranges of $m_{1/2}$, whereas
a machine with $E_{CM} = 1.0$~TeV would be able to explore supersymmetry
along (almost) the entire WMAP lines. CLIC with $E_{CM} = 3.0$~TeV would
be able to complete the spectrum of electroweakly-interacting sparticles
along the entire WMAP lines, as well as furnish detailed measurements of
squarks for a large fraction of the allowed range of
$m_{1/2}$~\cite{CLICPhys}. We also discuss in more detail the
observability of neutralinos at the different colliders surveyed, with
particular emphasis on CLIC studies.

Finally, in Section 5 we review some conclusions from our results,
addressing in particular the question whether collider measurements will
have the potential to determine if the LSP constitutes most of the cold
dark matter in the Universe~\cite{EHNOS,CDM}.

\section{Updated CMSSM Benchmark Points}

\subsection{Improved Choices of Supersymmetry-Breaking Parameters}

Since the laboratory constraints on the CMSSM have not yet changed
substantially since the termination of LEP~\cite{LEPSUSY}, the only reason 
for updating
the benchmark points we proposed previously~\cite{Bench} is the refined
estimate of $\Omega_{CDM} h^2$ provided by combining the new WMAP data
with those previously available~\cite{WMAP}. Assuming that most of
the cold dark matter is composed of LSPs $\chi$, previously we allowed
$0.1 < \ohsq < 0.3$ (rather conservatively), whereas the WMAP analysis now
allows only the range
\begin{equation}
0.094 \; < \; \ohsq \; < \; 0.129
\label{WMAPrange}
\end{equation}
at the 2-$\sigma$ level~\cite{WMAP}. We see in Table 2 of~\cite{Bench} 
that {\it all} 
the previous benchmark points yielded relic densities $\ohsq$ {\it above} 
the range (\ref{WMAPrange}). Since the relic density calculations have 
not changed significantly since~\cite{Bench} in the regions of interest, 
the previous benchmark points must be moved. 

Of the 13 benchmark points proposed previously, 5 (B, C, G, I, L) were in
the `bulk' regions at low $m_{1/2}$ and $m_0$, 4 (A, D, H, J) were along
the coannihilation `tails' extending to larger $m_{1/2}$~\cite{ourcoann},
2 (K, M) were along rapid-annihilation `funnels' where both $m_{1/2}$ and
$m_0$ may grow large~\cite{rapidann}, and 2 (E, F) were in the
`focus-point' region at very large $m_0$~\cite{fp}. The WMAP constraint
(\ref{WMAPrange}) has slimmed the `bulk' region down considerably, but
only minor reductions in $m_0$ are needed to reduce the relic density of
points (B, C, G, I, L) into the range (\ref{WMAPrange}), as shown in
Table~\ref{tab:Keith}. The coannihilation `tails' are also much thinner
than they were before, and the required reductions in $m_0$ for points
(A, D, J) are also small. The previous point H is an exception to this
rule, since it was chosen at the extreme tip of a coannihilation `tail'.
In this case, since the upper limit in (\ref{WMAPrange}) is considerably
reduced compared with that assumed in~\cite{Bench}, the allowed upper
limit on $m_{1/2}$ has been reduced substantially~\cite{EOSS}. Hence we
must make reductions in both $m_{1/2}$ and $m_0$ for point H, as also
shown in Table~\ref{tab:Keith}. In the case of point K, the
rapid-annihilation `funnel' has become thinner following WMAP, and a minor
adjustment in $m_{1/2}$ is sufficient, but in the case of point M, the
`funnel' has changed substantially, and both $m_{1/2}$ and $m_0$ had to be
changed.

The two focus-point benchmarks (E, F) may also be adapted to the WMAP
constraint (\ref{WMAPrange}) with small changes, as also shown in
Table~\ref{tab:Keith}, which is based on the {\tt SSARD} code. However, we
take this opportunity to underline the extreme delicacy of model
calculations in this region, as may be seen by comparing the results of
different codes~\cite{SSARD,ISASUGRA,SUSPECT,Sabine} in their successive
releases~\footnote{See also the discussion 
in~\cite{Bench}.}. The various codes do agree that there is a
strip in the `focus-point' region where $\ohsq$ is in the WMAP range
(\ref{WMAPrange}), but the error in its location due to the experimental
uncertainty in the mass of the top quark $m_t$ (for example) is much
larger than the intrinsic width of the WMAP strip in this region.  We have
chosen a pole mass $m_t = 171$~GeV for benchmarks E and F, as opposed to
the choice $m_t = 175$~GeV made for the other benchmarks, which would have
required much larger values of $m_0$ in this focus-point region, namely
$m_0 = 2550, 5030$~GeV for the {\tt SSARD} code, respectively. The new D0
value of $m_t \simeq 179$~GeV~\cite{D0mt}, if confirmed, would push the
`focus-point' region up to still larger $m_0$, whatever code is used:
specifically to $m_0 = 5800, 9070$~GeV for benchmarks E and F,
respectively, if the {\tt SSARD} code is used~\footnote{This would put 
all sfermions beyond the reach of all the colliders discussed here.}

There are also significant theoretical uncertainties associated with
higher-order effects~\cite{HO}, that are reflected in differences between 
different
codes in their various versions. For example, for $m_t = 175$~GeV at
benchmarks E and F (with $m_{1/2} = 300, 1000$~GeV, respectively), the 
{\tt
ISASUGRA 7.51}, [{\tt ISASUGRA 7.67}], [{\tt SUSPECT 2.10}], [{\tt SUSPECT
2.11}] codes require $m_0 = 1530, 3450$~GeV, [$m_0 = 3590, 6260$~GeV],
[$m_0 = 2350, 4110$~GeV], [$m_0 = 2590, 3850$~GeV], respectively. We see
that the predictions of successive versions of the {\tt SUSPECT}
code~\cite{SUSPECT} do not vary much, and agree quite well with {\tt
SSARD}~\cite{SSARD}, whereas there has been significant evolution in the
predictions of {\tt ISASUGRA}~\cite{ISASUGRA}. 

\begin{figure}
\begin{center}
\begin{tabular}{c c}
\mbox{\epsfig{file=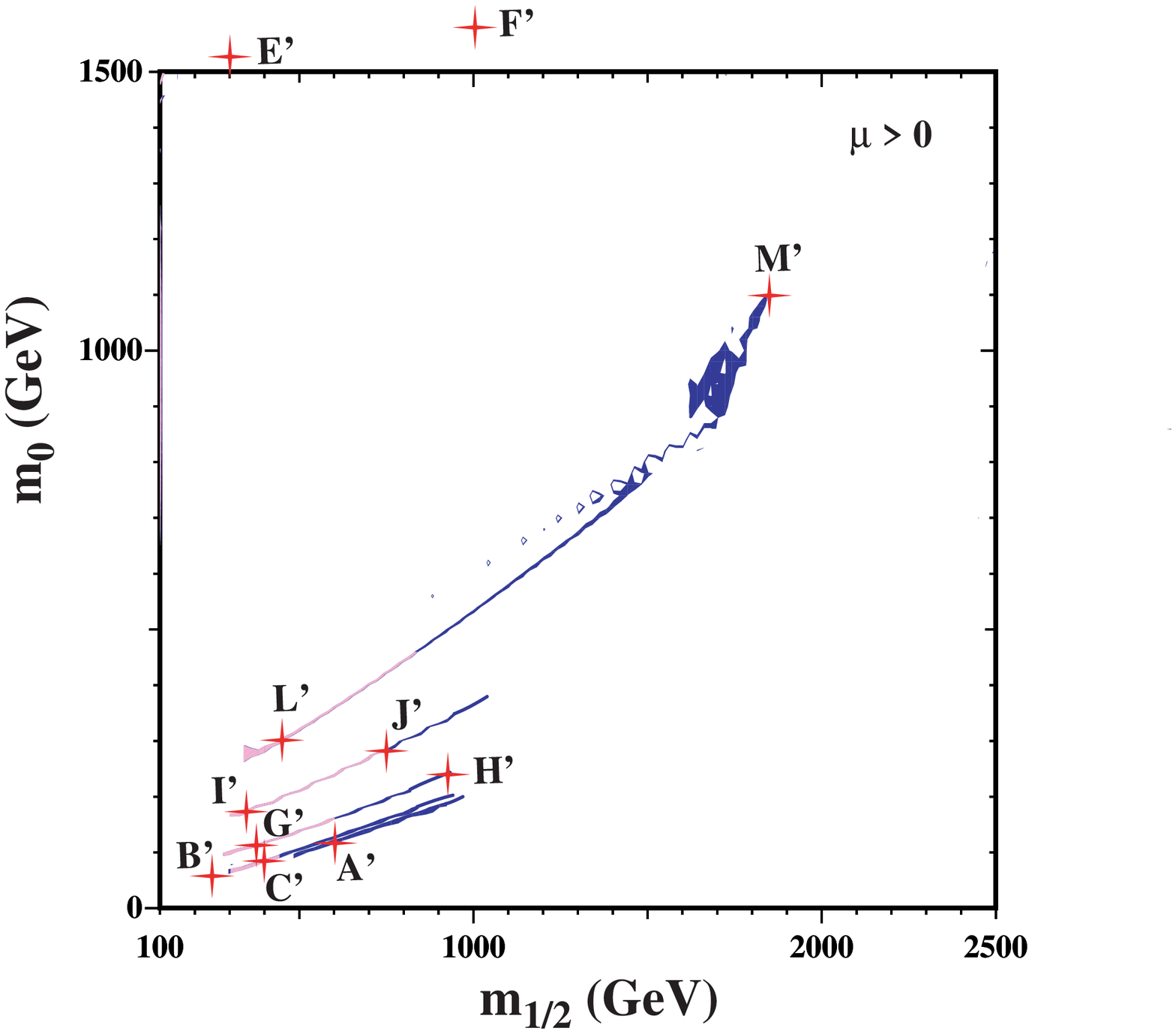,height=8cm}} &
\mbox{\epsfig{file=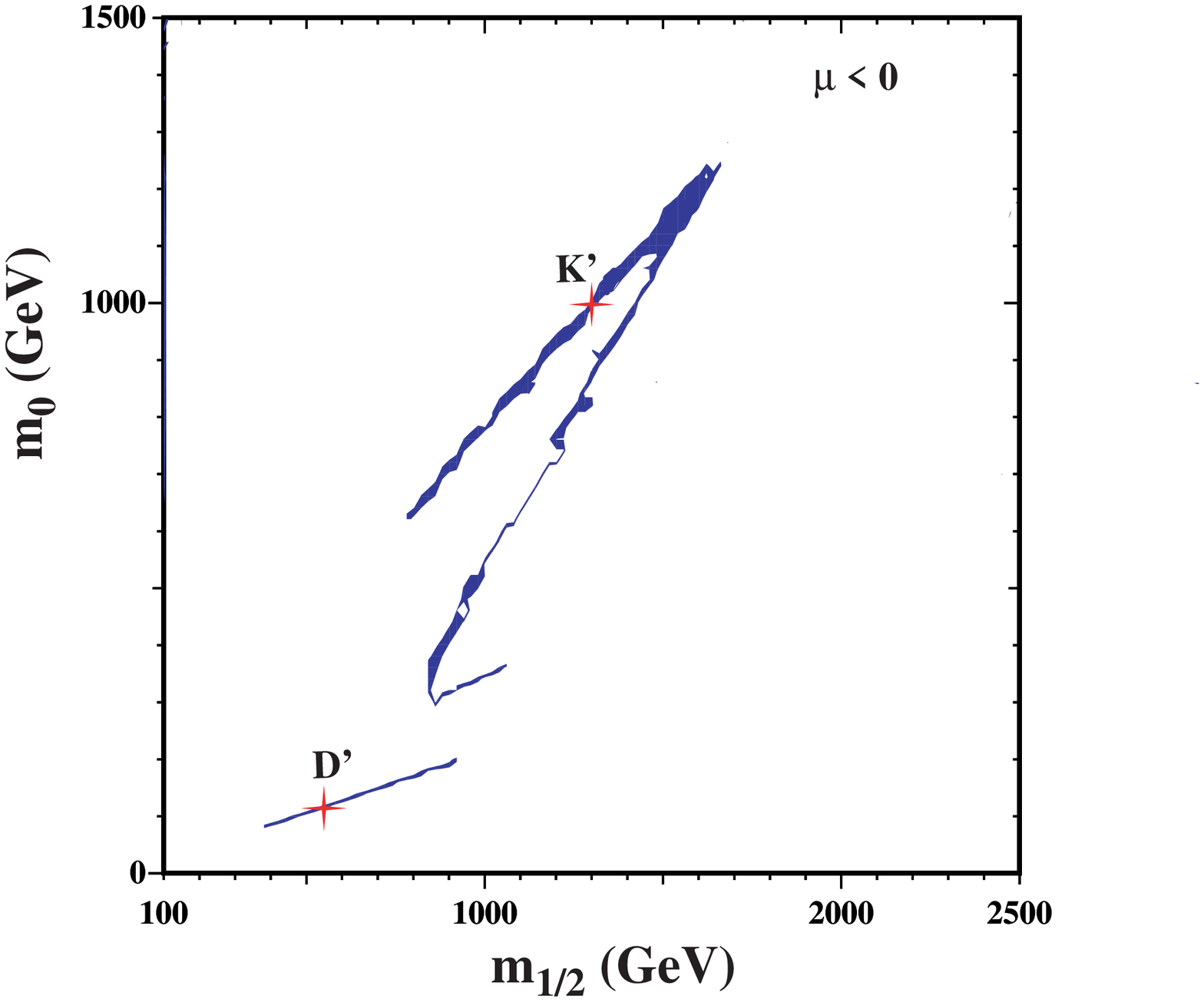,height=8cm}} \\
\end{tabular}
\end{center}   
\caption{\label{fig:strips}\it
The shaded strips display the regions of the $(m_{1/2}, m_0)$ plane that 
are compatible~\cite{EOSS} with $0.094 < \Omega_\chi h^2 < 0.129$ in the 
`bulk', coannihilation `tail', and rapid-annihilation `funnel' regions, as well 
as the laboratory constraints, for (a) $\mu > 0$ and $\tan \beta = 5, 10, 20,
35$ and 50, and (b) for $\mu < 0$ and $\tan \beta = 10$ and 35. The parts
of these `WMAP lines' for $\mu > 0$ compatible with $g_\mu - 2$ at the 
2-$\sigma$ level have lighter (pink) shading~\cite{g-2}. The updated post-WMAP 
benchmark scenarios are marked in red. Points (E',F') in the focus-point 
region have larger values of $m_0$.
}
\end{figure}

In view of these uncertainties in the focus-point region, we concentrate
in the following on the remaining updated benchmark points proposed in
Table~\ref{tab:Keith}. These are located on the `WMAP lines' shown in
Fig.~\ref{fig:strips}(a) for $\mu > 0$ and $\tan \beta = 5, 10, 20, 35$
and 50, and (b) for $\mu < 0$ and $\tan \beta = 10$ and 35~\footnote{There
would be corresponding WMAP lines in the focus-point regions, whose
location is uncertain, but which would generally lie at larger values of
$m_0$.}. We recall that, for given values of $\tan \beta$, $m_{1/2}$ and 
the sign of $\mu$, lower values of $m_0$ generally have values of
$\ohsq$ below the WMAP range~\footnote{Therefore, in this region the LSP
might constitute only a part of the cold dark matter.}, whereas higher
values of $m_0$ generally have values of $\ohsq$ that are too high, and
are therefore unacceptable unless one goes beyond the CMSSM framework used
here. At the ends of the `WMAP lines', smaller values of $m_{1/2}$ are
excluded by either $m_h$~\cite{LEPHWG,Heinemeyer:2000yj} 
and/or $b \to s
\gamma$~\cite{bsg}, and larger values of $m_{1/2}$ have values of $\ohsq$
above the WMAP range and/or the LSP is the charged ${\tilde \tau_1}$. We
note that, as was commented in~\cite{Bench}, most of the proposed
benchmark points for $\mu > 0$ yield a value of $g_\mu - 2$ that lies
within 2~$\sigma$ of the present experimental value based on $e^+ e^-$ 
data, corresponding to the
lighter (pink) regions of the strips in Fig.~\ref{fig:strips}(a). However,
we do not impose this as a requirement on the benchmark points, as
exemplified in Fig.~\ref{fig:strips}(b) for $\mu < 0$.

We see also in Fig.~\ref{fig:strips} that the majority of the benchmark
points lie on the portions of the WMAP lines with lower values of
$m_{1/2}$, making them more interesting for the early stages of LHC
operation, and for any sub-TeV linear $e^+ e^-$ collider. Generally 
speaking, less fine-tuning of parameters is required in this region than 
in the focus-point, funnel and tail regions~\cite{EO}: see also the 
discussion in~\cite{Bench}.

\subsection{Discussion of Spectra}

In order to study the detectability of MSSM particles and the possible
accuracies in measurements for the proposed benchmarks, it is essential to
relate the input CMSSM parameters, defined in Table~\ref{tab:Keith} for
the {\tt SSARD} program, to those necessary to obtain an equivalent MSSM
spectrum in Monte Carlo generators~\footnote{See also the discussion
in~\cite{Bench}.}.  Accordingly, we have tuned the inputs for {\tt
ISASUGRA 7.67} to reproduce as accurately as possible the spectra given in
Table~\ref{tab:Keith}, by scanning $m_0$ and $\tan \beta$ to identify the
parameter pairs minimizing the sum of the relative differences of particle
masses $|m_{\tt SSARD} - m_{\tt ISASUSY}|/M_{\tt SSARD}$. In view of their
importances for the calculation of $\ohsq$, we have treated specially the
mass splitting $|m_{\tilde \tau_1} - m_\chi|$, which is important along
the coannihilation `tail', and also $|m_A - 2 m_\chi|$, where it becomes
important in the `funnel' regions. We have given each of them 30~\% of the 
total weighting when optimizing the {\tt Isasugra 7.67} parameters in the
relevant $(m_{1/2}, m_0)$ regions. The results of the optimization are
given in Table~\ref{tab:isatable}.

\begin{table}[p!]
\centering
\renewcommand{\arraystretch}{0.90}
{\bf Supersymmetric spectra in post-WMAP benchmarks\\ calculated with
{\tt ISASUGRA~7.67}}\\
{~}\\
\begin{tabular}{|c|r|r|r|r|r|r|r|r|r|r|r|r|r|}
\hline
Model          & A'  &  B' &  C' &  D' &  E' &  F' &  G' &  H' &  I' &  J' &  K' &  L' &  M'  \\ 
\hline
$m_{1/2}$      & 600 & 250 & 400 & 525 &  300& 1000& 375 & 935 & 350 & 750 & 1300& 450 & 1840 \\
$m_0$          & 107 &  57 &  80 & 101 & 1532& 3440& 113 & 244 & 181 & 299 & 1001& 303 & 1125 \\
$\tan{\beta}$  & 5   &  10 &  10 & 10  & 10  & 10  & 20  & 20  & 35  & 35  &  46 & 47  & 51   \\
sign($\mu$)    & $+$ & $+$ & $+$ & $-$ & $+$ & $+$ & $+$ & $+$ & $+$ & $+$ & $-$ & $+$ & $+$  \\ 
$m_t$          & 175 & 175 & 175 & 175 & 171 & 171 & 175 & 175 & 175 & 175 & 175 & 175 & 175  \\ \hline
Masses         &     &     &     &     &     &     &     &     &     &     &     &     &      \\ \hline
$|\mu(m_Z) |$  & 773 & 339 & 519 & 663 & 217 & 606 & 485 &1092 & 452 & 891 &1420 & 563 & 1940 \\ \hline
h              & 116 & 113 & 117 & 117 & 114 & 118 & 117 & 122 & 117 & 121 & 123 & 118 &  124 \\
H              & 896 & 376 & 584 & 750 &1544 &3525 & 525 &1214 & 444 & 888 &1161 & 480 & 1623 \\
A              & 889 & 373 & 580 & 745 &1534 &3502 & 522 &1206 & 441 & 882 &1153 & 477 & 1613 \\
H$^{\pm}$      & 899 & 384 & 589 & 754 &1546 &3524 & 532 &1217 & 453 & 892 &1164 & 490 & 1627 \\ \hline
$\chi$         & 242 &  95 & 158 & 212 & 112 & 421 & 148 & 388 & 138 & 309 & 554 & 181 &  794 \\
$\chi_2$       & 471 & 180 & 305 & 415 & 184 & 610 & 286 & 750 & 266 & 598 &1064 & 351 & 1513 \\
$\chi_3$       & 778 & 345 & 525 & 671 & 229 & 622 & 492 &1100 & 459 & 899 &1430 & 568 & 1952 \\
$\chi_4$       & 792 & 366 & 540 & 678 & 302 & 858 & 507 &1109 & 475 & 908 &1437 & 582 & 1959 \\
$\chi^{\pm}_1$ & 469 & 178 & 304 & 415 & 175 & 613 & 285 & 750 & 265 & 598 &1064 & 350 & 1514 \\
$\chi^{\pm}_2$ & 791 & 366 & 541 & 679 & 304 & 846 & 507 &1108 & 475 & 908 &1435 & 582 & 1956 \\ \hline
$\tilde{g}$    &1367 & 611 & 940 &1208 & 800 &2364 & 887 &2061 & 835 &1680 &2820 &1055 & 3884 \\ \hline
$e_L$, $\mu_L$ & 425 & 188 & 290 & 376 &1543 &3499 & 285 & 679 & 304 & 591 &1324 & 434 & 1660 \\
$e_R$, $\mu_R$ & 251 & 117 & 174 & 224 &1534 &3454 & 185 & 426 & 227 & 410 &1109 & 348 & 1312 \\
$\nu_e$, $\nu_{\mu}$
               & 412 & 167 & 274 & 362 &1539 &3492 & 270 & 665 & 290 & 579 &1315 & 423 & 1648 \\
$\tau_1$       & 249 & 109 & 167 & 217 &1521 &3427 & 157 & 391 & 150 & 312 & 896 & 194 &  796 \\
$\tau_2$       & 425 & 191 & 291 & 376 &1534 &3485 & 290 & 674 & 312 & 579 &1251 & 420 & 1504 \\
$\nu_{\tau}$   & 411 & 167 & 273 & 360 &1532 &3478 & 266 & 657 & 278 & 558 &1239 & 387 & 1492 \\ \hline
$u_L$, $c_L$   &1248 & 558 & 859 &1103 &1639 &3923 & 814 &1885 & 778 &1554 &2722 &1001 & 3670 \\
$u_R$, $c_R$   &1202 & 542 & 830 &1064 &1637 &3897 & 787 &1812 & 754 &1497 &2627 & 969 & 3528 \\
$d_L$, $s_L$   &1250 & 564 & 863 &1107 &1641 &3924 & 818 &1887 & 783 &1556 &2723 &1004 & 3671 \\
$d_R$, $s_R$   &1197 & 541 & 828 &1059 &1638 &3894 & 786 &1804 & 752 &1491 &2615 & 967 & 3509 \\
$t_1$          & 958 & 411 & 653 & 860 &1052 &2647 & 617 &1477 & 584 &1207 &2095 & 753 & 2857 \\
$t_2$          &1184 & 576 & 837 &1048 &1387 &3373 & 792 &1753 & 748 &1428 &2366 & 920 & 3231 \\ 
$b_1$          &1147 & 514 & 789 &1015 &1375 &3356 & 737 &1719 & 677 &1377 &2297 & 844 & 3149 \\
$b_2$          &1181 & 535 & 816 &1043 &1602 &3816 & 770 &1761 & 725 &1423 &2349 & 904 & 3217 \\ \hline
\end{tabular}
\caption[]{\it 
Proposed post-WMAP CMSSM benchmark points and mass spectra (in GeV), as
calculated using {\tt ISASUGRA 7.67}~\protect\cite{ISASUGRA} and adapting 
the values of $m_0$ and
$\tan \beta$ (when it is large) to give the best fit to the {\tt SSARD}
spectra shown in Table~\ref{tab:Keith}, as described in the text.}
\label{tab:isatable}
\end{table}

\begin{figure}
\begin{center}
\begin{tabular}{c c}
\mbox{\epsfig{file=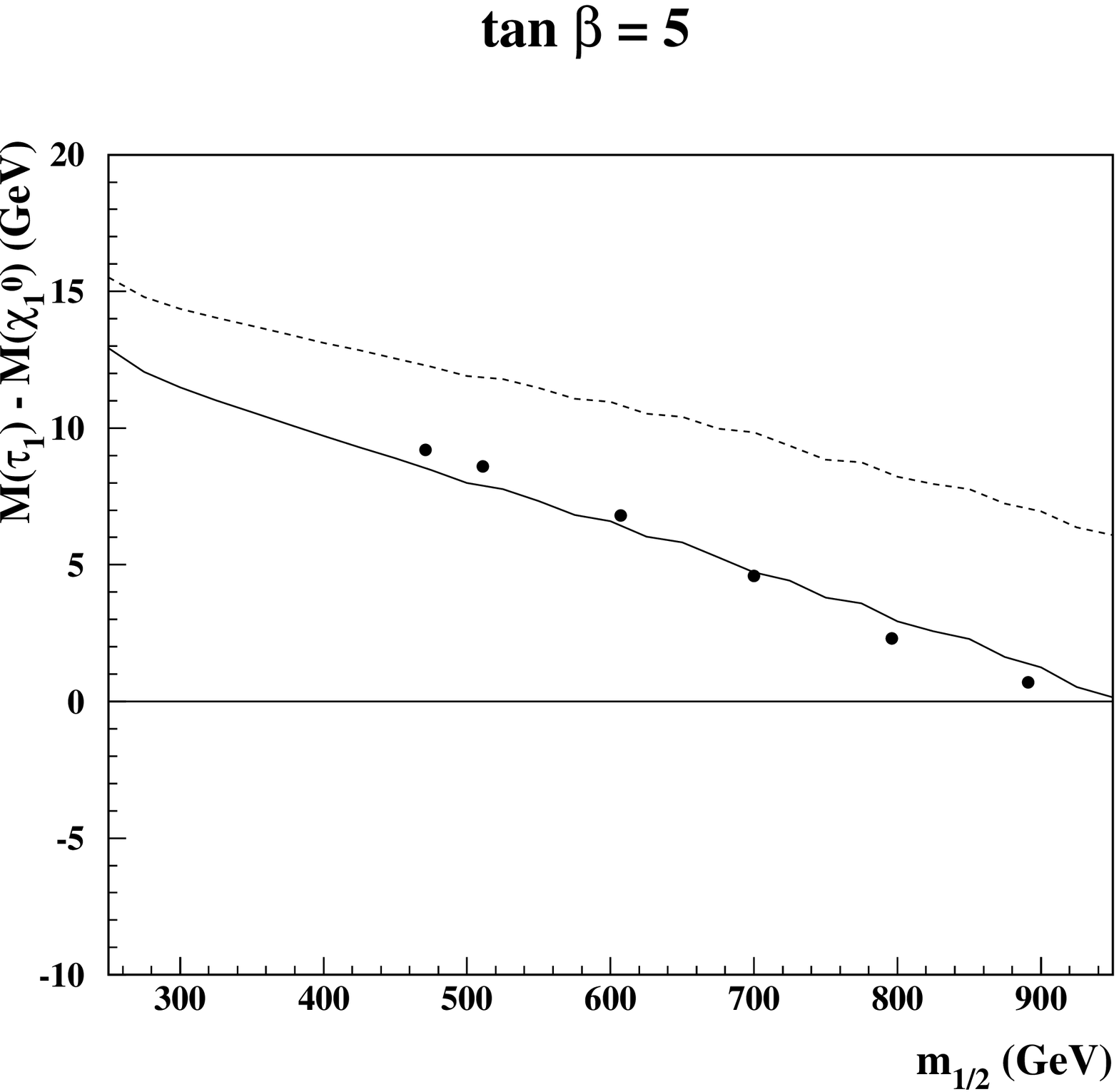,height=6cm,width=7.0cm}} &
\mbox{\epsfig{file=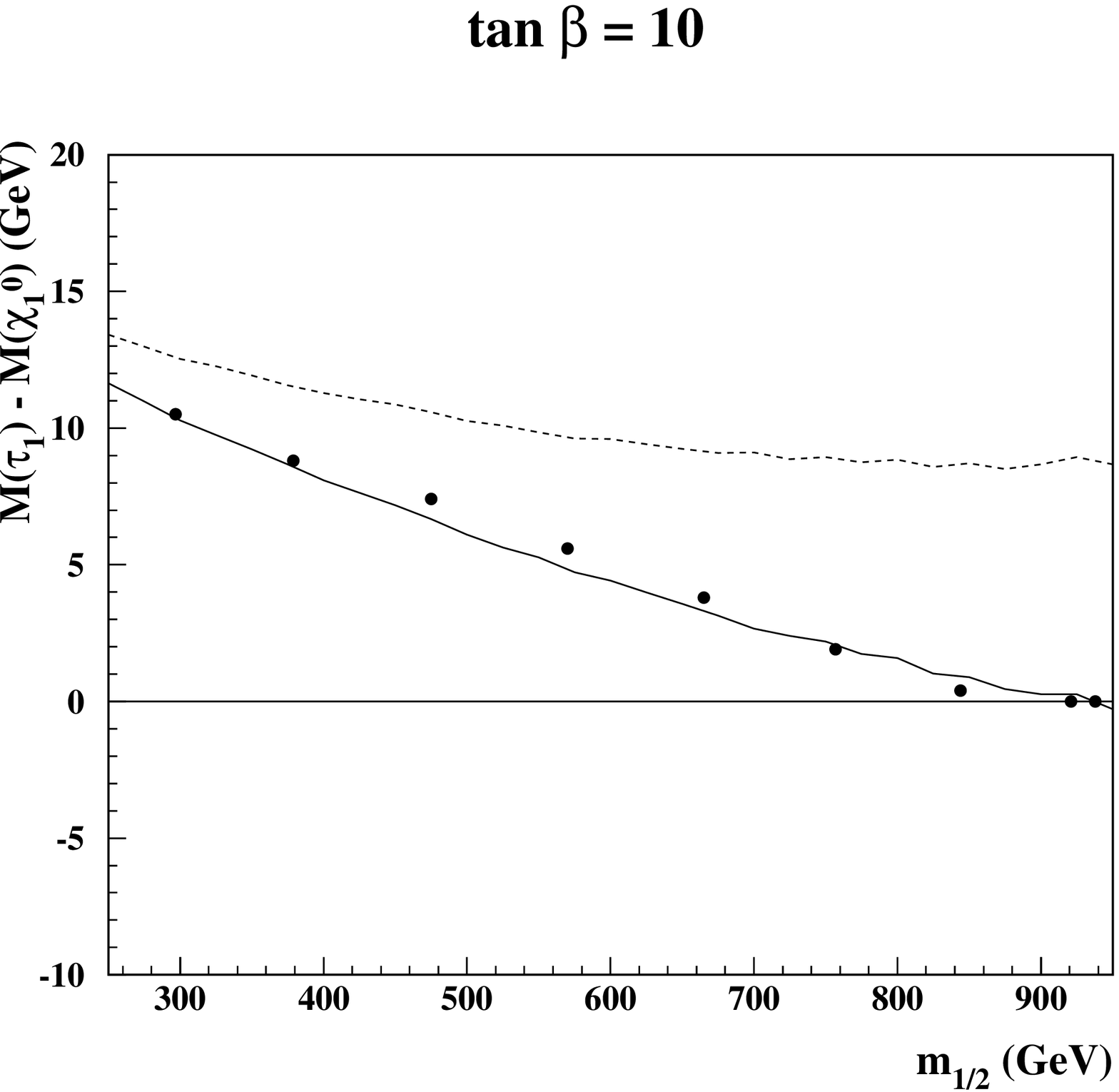,height=6cm,width=7.0cm}} \\
\mbox{\epsfig{file=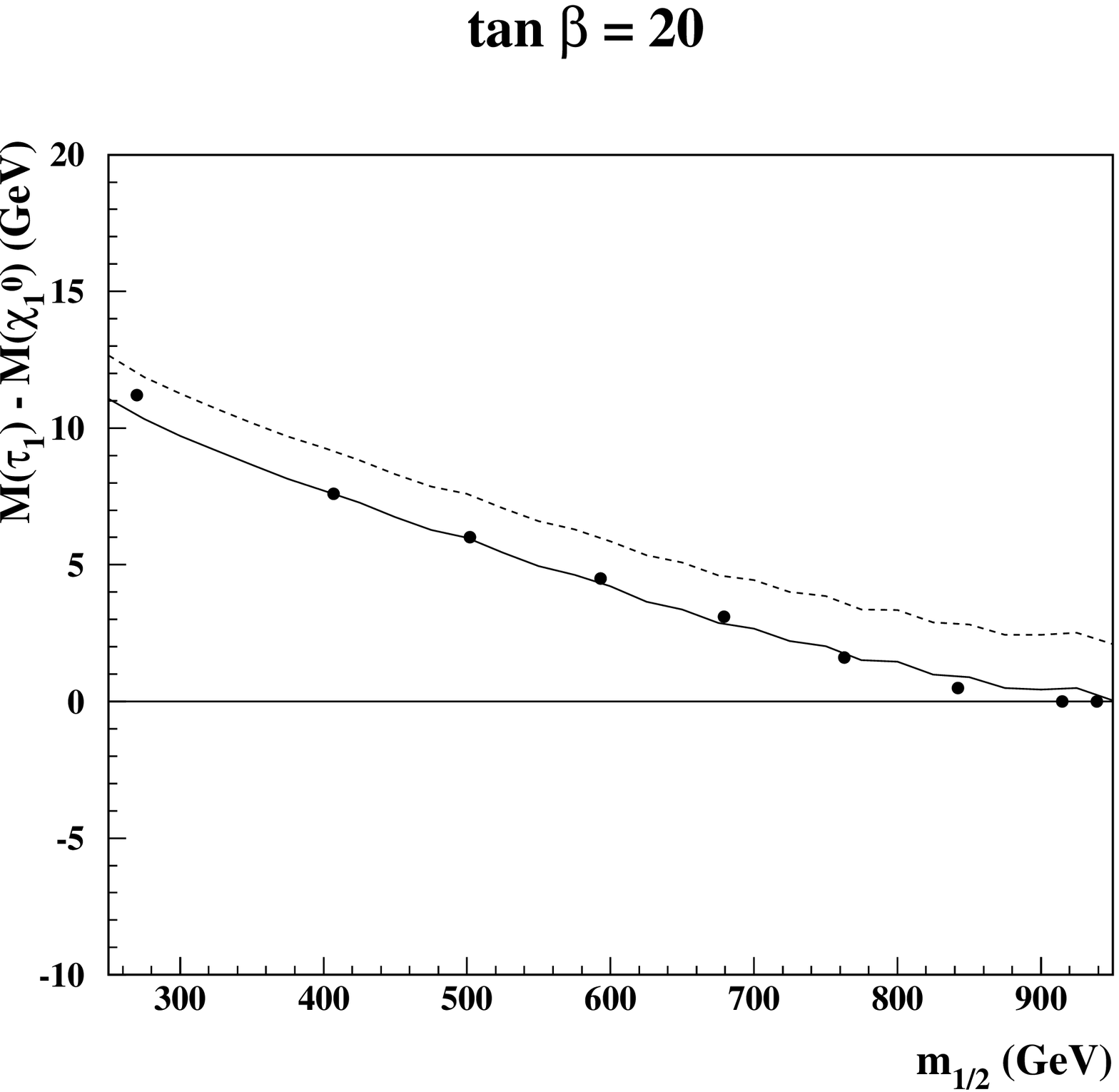,height=6cm,width=7.0cm}} &
\mbox{\epsfig{file=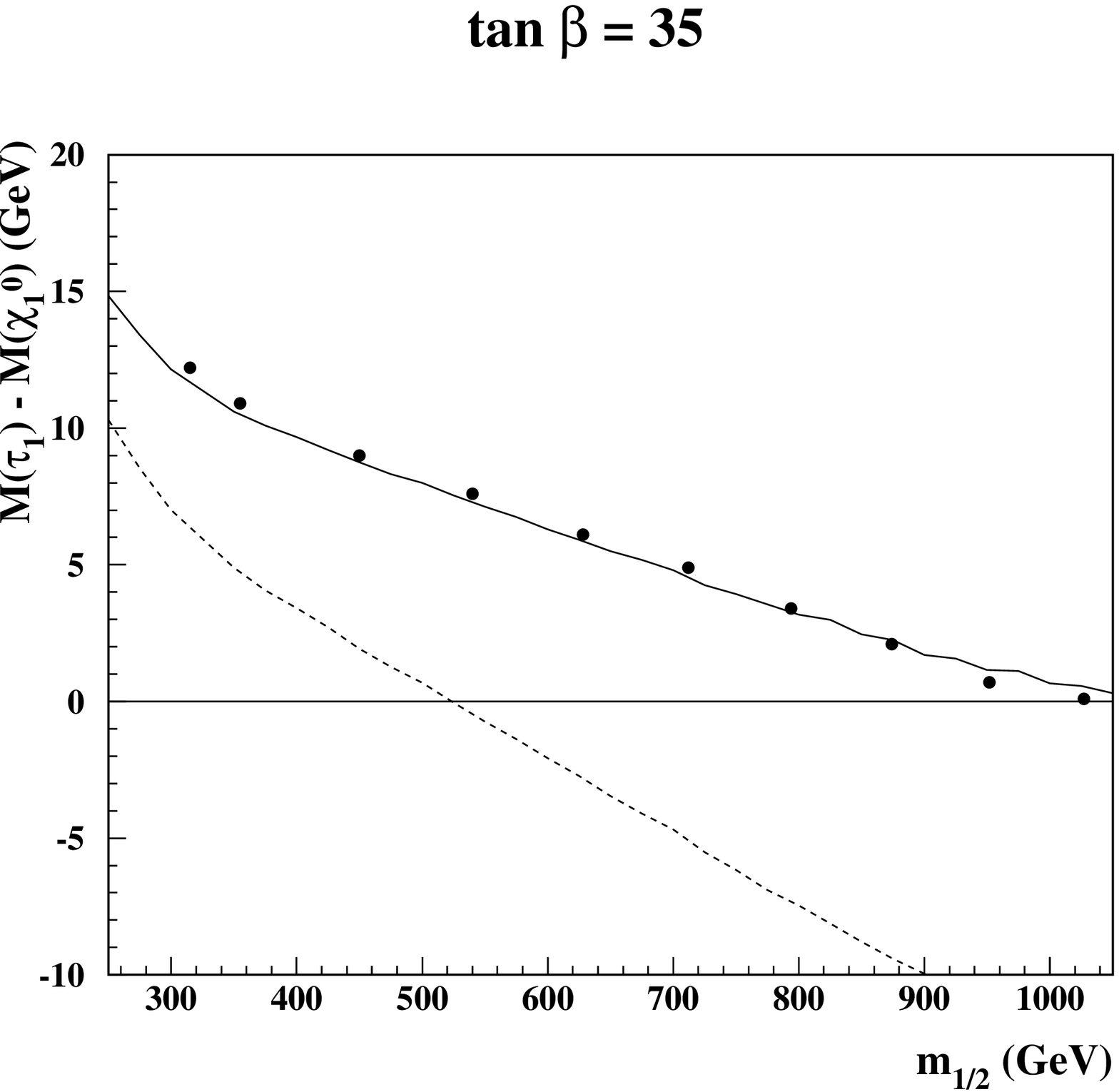,height=6cm,width=7.0cm}} \\
\mbox{\epsfig{file=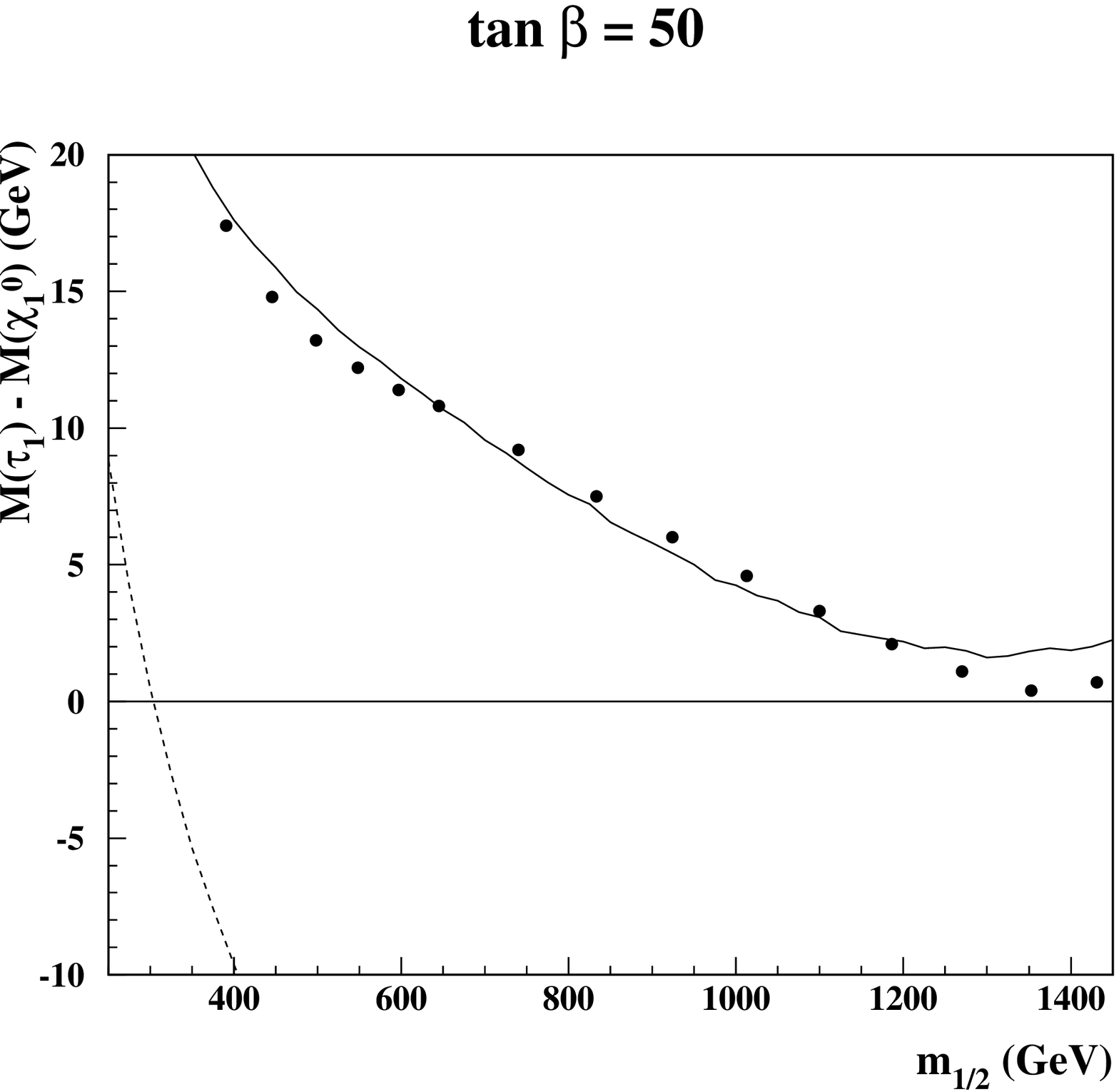,height=6cm,width=7.0cm}} &
\mbox{\epsfig{file=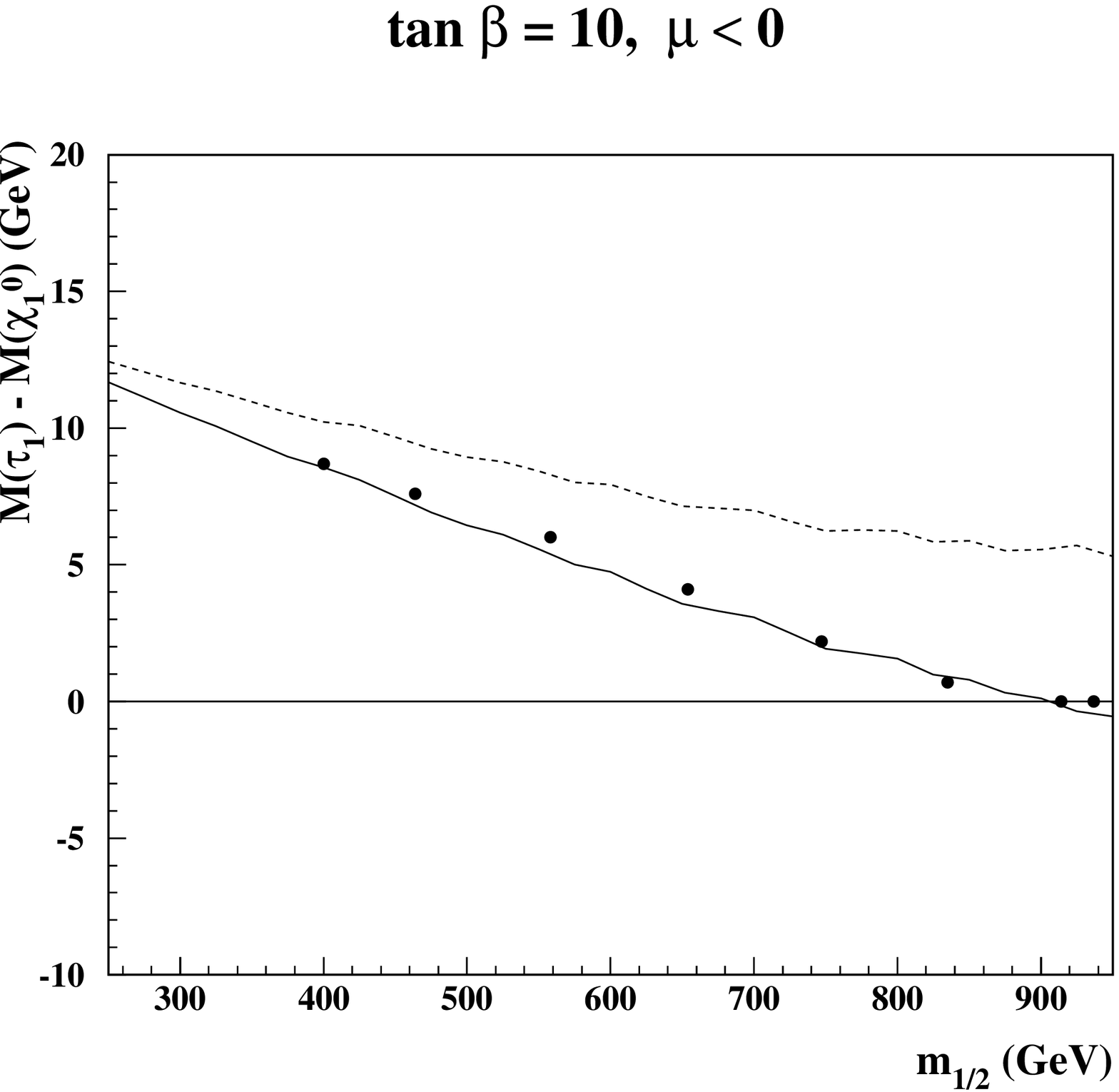,height=6cm,width=7.0cm}} \\
\end{tabular}
\end{center}
\caption{\label{fig:shift}
{\it 
Comparison of the mass differences $\Delta M \equiv
|m_{\tilde \tau_1} - m_\chi|$ along the WMAP lines, as calculated 
in the {\tt SSARD} code 
(dots) and {\tt ISASUGRA 7.67} using the same values of $m_0$ (dashed 
lines), as functions of $m_{1/2}$ for $\mu > 0$ and $\tan 
\beta = 5, 10, 20, 
35, 50$ and $\mu < 0, \tan \beta = 10$. We see the need to shift $m_0$ 
systematically, particularly at larger 
$m_{1/2}$ and $\tan \beta$. This has been done for the solid lines, which 
are improved {\tt ISASUGRA 7.67} fits to the WMAP lines described in the 
text.
}}
\end{figure}

In general, good matching between the predictions of the two codes can be
found with only moderate shifts of the input parameters, and we have
chosen only to adjust $m_0$ in most cases. The typical average relative
difference in the benchmark point masses is of the order of a few percent.  
However, occasionally individual masses may exhibit discrepancies of up to
15\%, and ensuring compatible mass splittings $\Delta M$ requires a
systematic shift in $m_0$ to higher values, particularly at large $\tan
\beta$, as seen in Fig.~\ref{fig:shift}, where the cases $\tan \beta = 5,
10, 20, 35$ and 50 are exhibited for $\mu > 0$, and $\tan \beta = 10$ for
$\mu > 0$. In addition to increasing $m_0$, here we find a better
correspondence between {\tt SSARD} and {\tt ISASUGRA 7.67} if $\tan \beta$
is allowed to be reduced, as done for benchmarks (K, L, M) in
Table~\ref{tab:isatable}.

We have further studied whether the resulting {\tt ISASUGRA 7.67} versions
of the benchmark points yield values of $\ohsq$ that are indeed compatible
with the WMAP data. The first line of Table~\ref{tab:micromega} shows the
values of $\ohsq$ calculated using the {\tt SSARD}~\cite{SSARD} code at
the updated benchmark points shown in Table~\ref{tab:Keith}, which are
well within the WMAP range. The second line shows the corresponding values
of $\ohsq$ calculated using the {\tt Micromegas} code~\cite{Micromegas}
interfaced with the {\tt ISASUGRA 7.67} code that is used to calculate the
spectra in Table~\ref{tab:isatable}. 
The agreement between {\tt SSARD} and {\tt Micromegas/ISASUGRA 7.67} is
generally good in the bulk and coannihilation tail regions, with the
latter falling within the WMAP range, except for point H (which is right
at the tip of the coannihilation tail)  and point J (where the discrepancy
is only marginal). The {\tt Micromegas/ISASUGRA 7.67} code gives
results that are outside the WMAP range in the focus-point and at the upper 
edge of the funnel region (points E, F and M), reflecting the fact that the
calculation of $\ohsq$ is notoriously delicate in these regions~\cite{EO} and 
that the tuning of the mass spectra between the two different codes 
becomes 
difficult. 

\begin{table}
\centering
\renewcommand{\arraystretch}{0.90}
{\bf Relic density $\ohsq, b \to s \gamma$ and $g_\mu - 2$ in post-WMAP 
benchmark scenarios}\\
{~}\\
\begin{tabular}{|c|c|c|c|c|c|c|c|c|c|c|c|c|c|}
\hline
~  & A'   &  B'  &  C'  &  D'  &  E'  &  F'  &  G'  &  H'  &  I'  &  
J'  & K'  & L'  &  M'   \\ 
\hline
$\ohsq$: {\tt SSARD} & .12 & .12 & .12 & .10 & .10 & .10 & .13 & .13 & 
.12 & .10 & .10 & .10 & .14 \\
$\ohsq$: {\tt Micromegas} & .09 & .12 & .12 & .09 & .33 & 2.56 & .12 & .16 
& .12 & .08 & 0.12 & .11 & 0.27 \\
BR($b \to s \gamma$) $\times 10^4$ & 3.9 & 3.1 & 3.8 & 4.5 & 3.7 & 3.7 &
3.3 & 3.6 & 2.5 & 3.4 & 4.2 & 2.5 & 3.5 \\
$\delta a_{\mu}$ $\times 10^{9}$ & 0.3 & 3.2 & 1.3 & -0.8 & 0.2 & 0.03 &
2.7 & 0.4 & 4.5 & 1.1 & -0.3 & 3.4 & 
0.3 \\  
\hline
\end{tabular}
\caption[]{\it Comparison of $\ohsq$ for the benchmark points in 
Table~\ref{tab:Keith}
computed with the {\tt SSARD} code~\protect\cite{SSARD} (top line), and 
the {\tt Micromegas} 
code~\protect\cite{Micromegas} interfaced with {\tt ISASUGRA 
7.67}~\cite{ISASUGRA} (second line) using the fitted 
parameters shown in Table~\ref{tab:isatable}. The third and fourth lines 
show the values of $b \to s \gamma$ and $g_\mu - 2$ calculated using {\tt 
SSARD}.} 
\label{tab:micromega}
\end{table}

For completeness, we also present in Table~\ref{tab:isatable} the values
of the $b \to s \gamma$ branching ratio and the supersymmetric
contribution to $g_\mu - 2$ calculated using {\tt SSARD}. These may be
compared with any evolution in the future experimental values of these
quantities. As already mentioned, the situation with regard to any
possible non-Standard Model contribution to the experimental value of
$g_\mu - 2$ is sufficiently volatile that we do not use this as a
criterion for selecting benchmarks, though it is used in
Fig.~\ref{fig:newM} to order the benchmark points, as described below.

\subsection{Observability at Different Accelerators}

As an important ingredient in the subsequent analyses of sparticle
observability at different colliders, we first report a comparison of the
principal sparticle branching ratios at the updated benchmark points with
those found previously~\cite{Bench} at the original benchmark points.  
The new branching ratios have been calculated using {\tt ISASUGRA 7.67},
whereas previously they were calculated using {\tt ISASUGRA
7.51}~\cite{ISASUGRA}, and we comment on the differences associated with
the improvement in {\tt ISASUGRA}, as opposed to the changes in the
benchmark parameter choices~\footnote{For the purposes of our discussion,
one of the most important upgrades to {\tt ISASUGRA} has been the
inclusion of one-loop radiative corrections to sparticle masses.}. Since
the differences in branching ratios are not large, in general, we limit
ourselves to commenting on instances where the more significant
differences occur.

$\bullet$ In the case of point A, the most notable difference is an
increase in $B({\tilde b_2} \to b \chi)$ from 39 to 62 \%, with
corresponding decreases in $B({\tilde b_2} \to t + \chi^-, {\tilde t_1} +
W^-)$. There is also an increase of $\chi_4 \to h \chi_2$ from 23 to
29 \% at the expense of $\chi_4 \to W \chi^{\pm}_1$. These changes are
not related to the new parameter choice, but to the differences in the
{\tt ISASUGRA} versions.

$\bullet$ In the case of point B, there is again an increase in $B({\tilde
b_2} \to b \chi)$, from 15 to 25 \%, and a decrease in $B({\tilde b_2} \to
{\tilde t_1} + W^-)$. There is also an increase in $B({\tilde e_L} \to e
\chi)$ from 47 to 87 \%, accompanied by decreases in $B({\tilde e_L} \to e
\chi_2, \nu + \chi^\pm_1)$. We also note that the $B({\tilde \nu}_\ell \to
\ell^- \chi^+)$ now vanish, so that the ${\tilde \nu}_\ell$ are {\it
invisible}. There is also an increase in $B({\tilde \tau}_2 \to \tau
\chi)$ from 51 to 84 \%, which decreases $B({\tilde \tau}_2 \to \nu_\tau
+\chi^- , \tau + \chi_2)$.  Finally, there are decreases in $B(\chi_2 \to
{\tilde \tau}_1 + \tau)$ from 84 to 42 \%, which increases the branching
ratio for the invisible channel $B(\chi_2 \to {\tilde \nu} + \nu)$, and in
$B(\chi^\pm \to {\tilde \tau}_1 + \nu_\tau)$ from 96 to 36 \%, which
increases $B(\chi^\pm \to {\tilde \nu} + \ell)$. The changes for the
sleptons are mostly because they are lighter at the updated benchmark
point with smaller $m_0$, and those for $\tilde b_2$ are related to the
differences in the {\tt ISASUGRA} versions.  Both effects influence the
gaugino branching ratios.

$\bullet$ At benchmarks C and D, $B({\tilde b_2} \to b \chi)$ increases
from 13 to 24 \% and from 20 to 44 \%, respectively, with corresponding
decreases in $B({\tilde b_2} \to {\tilde t_1} W^-)$ and $B({\tilde b_2}
\to t + \chi^\pm)$. At point C, decreases are observed of $\chi_2 \to
\stau_1 \tau$ from 23 to 13 \% and of $\chi^{\pm}_1 \to \stau_1 \nu$ from
21 to 11 \%, compensated by increases of $\chi_2 \to \snu \nu$ and
$\chi^{\pm}_1 \to \snu l$, respectively. All changes are mostly related to
differences in the {\tt ISASUGRA} versions.

$\bullet$ At the updated points E and F, the most notable effect is an
increase of the branching ratios of squarks to gluinos, due to the
increase in $m_0$ and hence of the squark masses at the updated benchmark
points. For instance, the $B(\sbot_1 \to b \tilde{g})$ changes from 45 to
52 \% at E and from 22 to 38 \% at F. They are compensated by a reduction
of the decays to charginos and neutralinos. Moreover, at point E the
decays of sleptons into $\chi^{\pm}_1$ tend to decrease and those into
$\chi^{\pm}_2$ to increase correspondingly.

$\bullet$ At the updated benchmark G, there are increases in $B({\tilde
b_1} \to t \chi^{\pm}_2)$, from 13 to 22 \%, with accompanying decreases
in other modes, mainly in $B({\tilde b_1} \to t + \chi^{\pm}_1, b +
\chi_2)$. A decrease is observed of $\chi_2 \to \stau_1 \tau$ from 82 to
62 \% and of $\chi^{\pm}_1 \to \stau_1 \nu$ from 81 to 57 \%, compensated
by an increase of $\chi_2 \to \snu \nu$ and $\chi^{\pm}_1 \to \snu l$,
respectively. The changes are mostly due to the differences in the {\tt
ISASUGRA} versions.

$\bullet$ Although the updated point H has a significantly lighter
spectrum than previously, no major changes ($>$ 5-6 \%) are observed in
the decay branching ratios.

$\bullet$ For the old point I the decay $\sbot_1 \to t \chi^{\pm}_2$ was
kinematically forbidden. For the updated point it is allowed and has a
branching ratio of 11 \%. Also, $B(\snu_e \to \nu \chi)$ increases
from 61 to 73 \% and $B(\snu_e \to e \chi^{\pm}_1)$ decreases
correspondingly, mostly due to the change in the {\tt ISASUGRA} version 
used.

$\bullet$ At point J the main changes are a decrease of $B(\chi_2 \to
\stau_1 \tau)$ from 82 to 66 \% and of $B(\chi^{\pm}_1 \to \stau_1 \nu)$
from 82 to 64 \%, mostly compensated by an increase of $B(\chi_2 \to
\snu \nu)$ and $B(\chi^{\pm}_1 \to \snu l)$, respectively. The changes are
mainly due to the differences in the {\tt ISASUGRA} versions.

$\bullet$ At the updated benchmark K, there is an increased branching
ratio for the direct decay of $\sel_L$ and $\snu_e$ to $\chi$ from 26 to
35 \% and 27 to 37 \% respectively, at the expense of the decays to
$\chi^{\pm}_1$. The increase of $B(\snu_{\tau} \to W \stau_1)$ from 28 to
40 \% is accompanied by a decrease of $B(\snu_{\tau} \to \tau
\chi^{\pm}_1)$. Also, the decays of $\chi_2$ and $\chi^{\pm}_1$ to
$\stau_1$, which were previously forbidden kinematically, are now allowed
by the new parameter values and reach 43 and 41 \%, respectively. A
reduction of $B(\chi_4 \to W \chi^{\pm}_1)$ from 56 to 47 \% is also
observed, associated with the differences in the {\tt ISASUGRA} versions.

$\bullet$ The updated point L is characterized by a decrease of the
branching ratios of $\snu_e$ and $\snu_{\tau}$ to the $\chi^{\pm}_1$,
compensated by an increase of $B(\snu_e \to \nu \chi)$ from 35 to 43 \%
and of $B(\snu_{\tau} \to W \stau_1)$ from 59 to 77 \%. The decay
branching ratios of $\chi_3$ and $\chi_4$ to $\stau_1 \tau$ are increased
from 19 to 27 \% and 13 to 20 \%, respectively, with a corresponding
reduction of the decays to $\chi^{\pm}_1$. These changes are associated
with the updated CMSSM parameter values.

$\bullet$ At the updated point M, the main change is a smaller value of
$m_0$, which affects many branching fractions. We only mention some of the
major changes. The direct decay to $\chi$ is increased for $\sel_L$
and $\snu_e$ from 34 to 74 \% and from 34 to 77 \%, repectively, at the
expense of a reduced decay to the chargino and the heavier neutralino. The
$B(\chi_2 \to h + \chi)$ decreases from 45 to 12 \%, whereas $B(\chi_2 \to
{\tilde \tau_1} + \tau)$ increases, and $B(\chi^\pm_1 \to W^\pm + \chi)$
decreases from 48 to 14 \%, whereas $B(\chi^\pm_1 \to {\tilde \tau_1} +
\nu_\tau)$ increases.

The branching ratios at the updated benchmark points not mentioned above
are not significantly different from those at the original versions of the
points. Among all the above effects, perhaps the most significant is the
invisibility of the ${\tilde \nu_\ell}$ at the updated version of point B.

\begin{figure} 
\epsfig{file=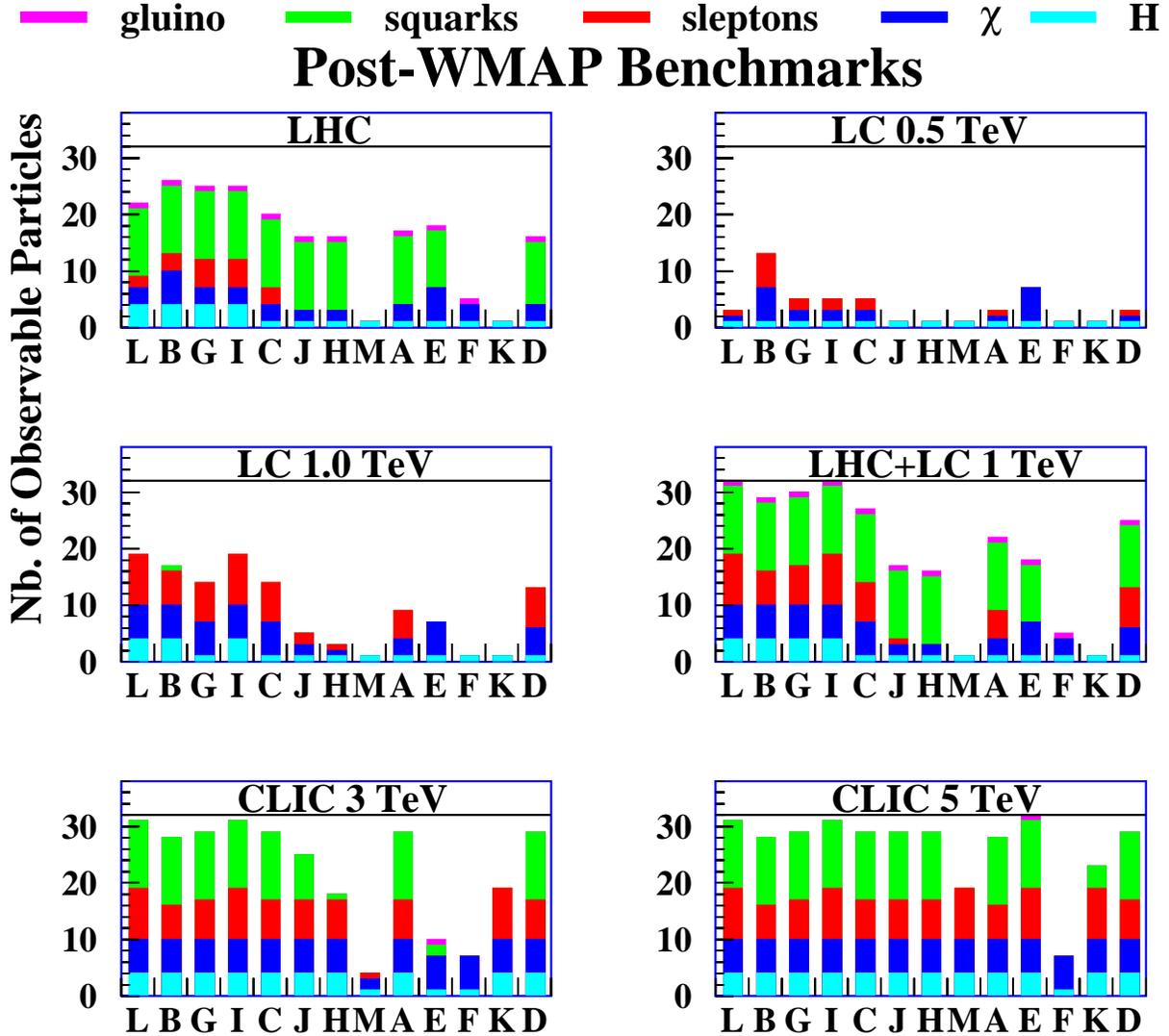,height=6.5in}
\vspace*{-0.5in}
\caption{\label{fig:newM}
{\it 
Summary of the numbers of MSSM particles that may be detectable at various
accelerators in the updated benchmark scenarios. As in~\protect\cite{Bench}, 
we see that the capabilities of the LHC and of linear $e^+ e^-$ colliders 
are largely complementary. We re-emphasize that mass and coupling measurements
at $e^+ e^-$ colliders are usually much cleaner and more precise than at
hadron-hadron colliders such as the LHC, where, for example, it is not
known how to distinguish the light squark flavours. 
}}
\end{figure}

Combining the above information, we now present in Fig.~\ref{fig:newM} an
updated comparison of the numbers of different MSSM particles that should
be observable at different accelerators in the various benchmark
scenarios~\cite{Bench}, ordered by their consistency with $g_\mu -2$ as
calculated using $e^+ e^-$ data for the Standard Model
contribution~\cite{g-2}. We re-emphasize that the qualities of the
prospective sparticle observations at hadron colliders and linear $e^+
e^-$ colliders are often very different, with the latters' clean
experimental environments providing prospects for measurements with better
precision.  Nevertheless, Fig.~\ref{fig:newM} already restates the clear
message that hadron colliders and linear $e^+ e^-$ colliders are largely
complementary in the classes of particles that they can see, with the
former offering good prospects for strongly-interacting sparticles such as
squarks and gluinos, and the latter excelling for weakly-interacting
sparticles such as charginos, neutralinos and sleptons. We discuss later
the detailed criteria used for assessing the detectabilities of different
particles at different colliders.

\section{WMAP Lines}

As has already been mentioned, in view of the reduction in dimensionality
of the CMSSM parameter space enforced by WMAP~\cite{EOSS}, one may
progress beyond the previous approach of sampling, more or less sparsely,
the CMSSM parameter space. In particular, as proposed in~\cite{SPS} but
with a different attitude towards the cosmological density constraint, one
may explore, more or less systematically, the CMSSM phenomenology along
the lines in parameter space shown in Fig.~\ref{fig:strips}. Again as
proposed in~\cite{SPS}, one may then focus special attention on particular
points along these lines. The updated benchmark points discussed in the
previous section are examples and, as we discuss below, they manifest some
of the distinctive possibilities that may appear along the WMAP lines.
However, for certain purposes it may be helpful to select a few additional
points that exemplify other generic possibilities, as we discuss later.

\subsection{Parametrizations of WMAP Lines}

As seen in Fig.~\ref{fig:strips}, the WMAP lines have a relatively simple 
form for $\tan \beta \lappeq 35$. It is therefore possible to provide 
simple quadratic parametrizations for them that are a convenient summaries
of the 
correlations between $m_{1/2}$ and $m_0$ that are enforced by WMAP within 
the CMSSM framework adopted here. Using the {\tt SSARD} code, we find the 
following convenient parametrizations for the $\mu > 0$ lines:
\begin{eqnarray}
\tan \beta = ~5: \; & m_0 & \; = \; -5.46 + 0.206 m_{1/2} + 4.02 \times 
10^{-6} m_{1/2}^2, \; (475 \le m_{1/2} \le 960) \\
\tan \beta = 10: \; & m_0 & \; = \; 11.97 + 0.163 m_{1/2} + 4.90 \times 
10^{-5} m_{1/2}^2, \; (300 \le m_{1/2} \le 940) \\
\tan \beta = 20: \; & m_0 & \; = \; 50.89 + 0.144 m_{1/2} + 6.77 \times 10^{-5} 
m_{1/2}^2, \; (270 \le m_{1/2} \le 940), \\
\tan \beta = 35: \; & m_0 & \; = \; 90.24 + 0.219 m_{1/2} +
5.70 \times 10^{-5} m_{1/2}^2, \; (315 \le m_{1/2} \le 1070)
\label{parameters}
\end{eqnarray}
where the masses are expressed in units of GeV, and the numbers in 
parentheses specify the allowed ranges of $m_{1/2}$. When $\tan \beta = 50$, 
a rapid-annihilation funnel appears, 
and parametrizing the regions allowed by WMAP becomes more 
complicated. In the case of $\tan \beta = 50$, one may conveniently use:
\begin{eqnarray}
m_0 \; & = & \; 140.6 + 0.33 m_{1/2} + 6.3 \times 10^{-5} m_{1/2}^2 \; 
(350 \le m_{1/2} \le 1550), \nonumber \\
m_0 \; & = & \; - 1941 + 1.65 m_{1/2} \; (1700 \le m_{1/2} \le 1800),
\label{tb50}
\end{eqnarray}
where the masses are again expressed in GeV units. For $\mu < 0$, we 
propose the following convenient parametrizations:
\begin{eqnarray}
\tan \beta = 10: \; & m_0 & \; = \; 15.0 + 0.16 m_{1/2} + 4.2 \times
10^{-5} m_{1/2}^2, \; (400 \le m_{1/2} \le 940) \\
\tan \beta = 35: \; & m_0 & \; = \; 39.6 + 0.76 m_{1/2} - 1.9 \times 
10^{-5} m_{1/2}^2, \; (790 \le m_{1/2} \le 1650),
\label{negparameters}
\end{eqnarray}
where for $\tan \beta = 35$ we parametrize only the top branch allowed by
WMAP in the $(m_{1/2}, m_0)$ plane shown in panel (b) of 
Fig.~\ref{fig:strips}, which includes the updated point K.

As mentioned earlier when discussing the specific benchmark points, the
appropriate values of $m_0$ as functions of $m_{1/2}$ must be
re-evaluated for the {\tt ISASUGRA 7.67} code. A similar minimization
procedure to that discussed earlier has been repeated,
parametrizing the shifts w.r.t. the values obtained
with the {\tt SSARD} code by a linear function of $m_{1/2}$, which have
then been added to (\ref{parameters}, \ref{tb50}, \ref{negparameters}). 
The resulting
parametrizations are:
\begin{eqnarray}
\tan \beta = ~5: \; & m_0 & \; = \; -9.96 + 0.197 m_{1/2} + 4.02 \times 
10^{-6} m_{1/2}^2, \; (475 \le m_{1/2} \le 960) \\
\tan \beta = 10: \; & m_0 & \; = \; 13.57 + 0.142 m_{1/2} + 4.90 \times 
10^{-5} m_{1/2}^2, \; (300 \le m_{1/2} \le 940) \\
\tan \beta = 20: \; & m_0 & \; = \; 50.39 + 0.142 m_{1/2} + 6.77 \times 10^{-5} 
m_{1/2}^2, \; (270 \le m_{1/2} \le 940), \\
\tan \beta = 35: \; & m_0 & \; = \; 89.56 + 0.239 m_{1/2} +
5.70 \times 10^{-5} m_{1/2}^2, \; (300 \le m_{1/2} \le 1050).
\label{isapar}
\end{eqnarray}
These have been used to produce the solid lines in Fig.~\ref{fig:shift}.

We note in passing that, as the mass difference $\Delta M \equiv
|m_{\tilde \tau_1} - m_\chi| \to 0$ towards the high-$m_{1/2}$ ends
of the WMAP lines, these have portions at high $m_{1/2}$ where $\Delta M <
m_\tau$. In these portions, the ${\tilde \tau_1}$ NLSP decays via a
virtual $\tau$, resulting in the dominance of four-body decay modes such
as ${\tilde \tau_1} \to \nu \chi \ell {\bar \nu}$ and $\nu \chi q {\bar
q}$. In this case, the ${\tilde \tau_1}$ is stable on the scale of the
size of the detector. This observation has important implications for
${\tilde \tau_1}$ observability at different colliders, as we discuss
below.

\subsection{Discussion of $\chi_2$ Decay Branching Ratios}

One of the key particles appearing in sparticle decay chains is the second
neutralino $\chi_2$, whose branching ratios are quite model-dependent and
have significant impact on sparticle detectability at future colliders.  
Moreover, $\chi_2$ decays play crucial roles in reconstructing sparticle
masses via cascade decays. Therefore, we now use {\tt ISASUGRA 7.67} to
discuss how the principal branching ratios of the $\chi_2$ vary along the
WMAP lines, noting several significant features that are important for
phenomenology.

\begin{figure}
\begin{center}
\begin{tabular}{c c}
\mbox{\epsfig{file=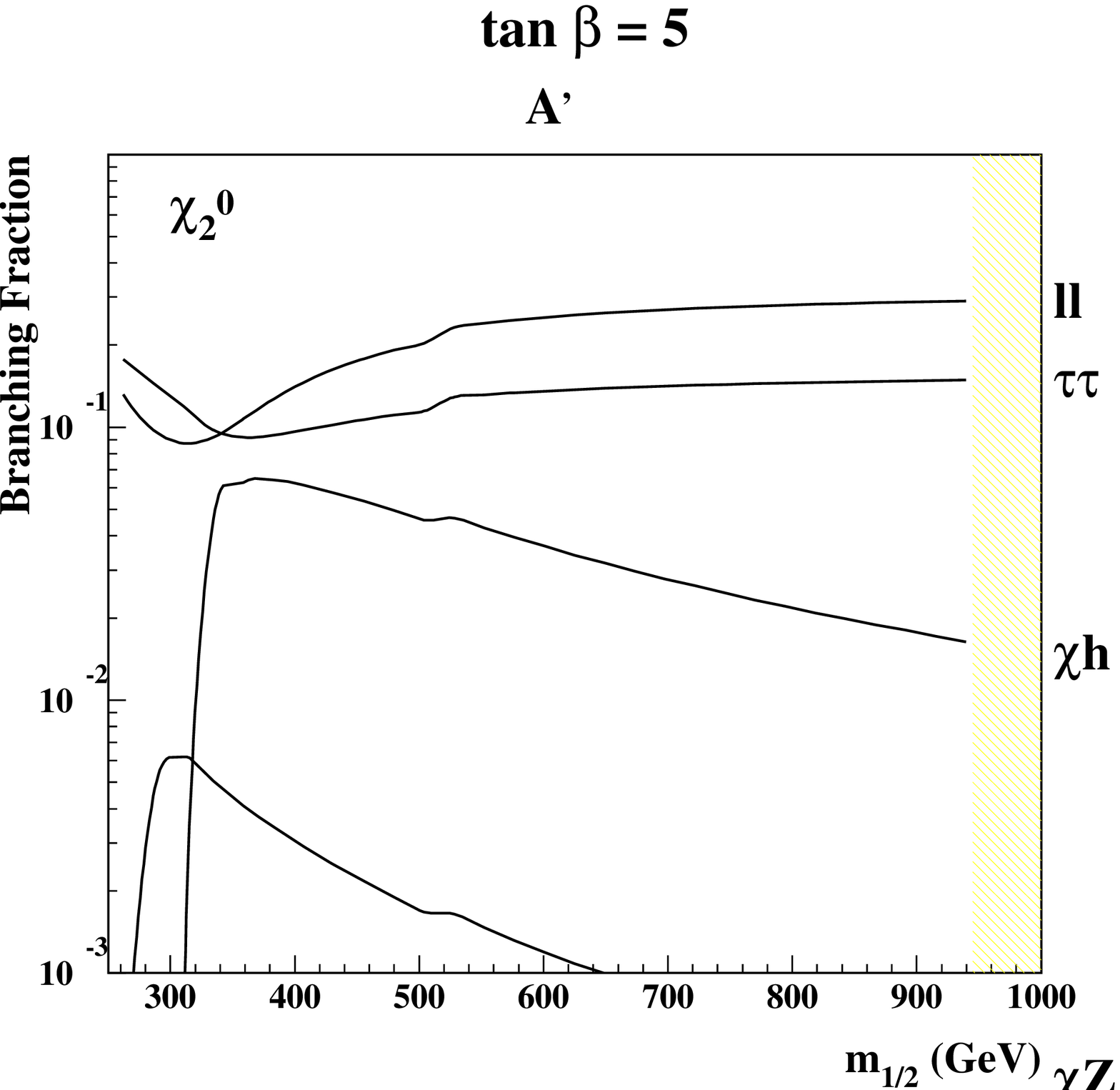,height=5cm}} &
\mbox{\epsfig{file=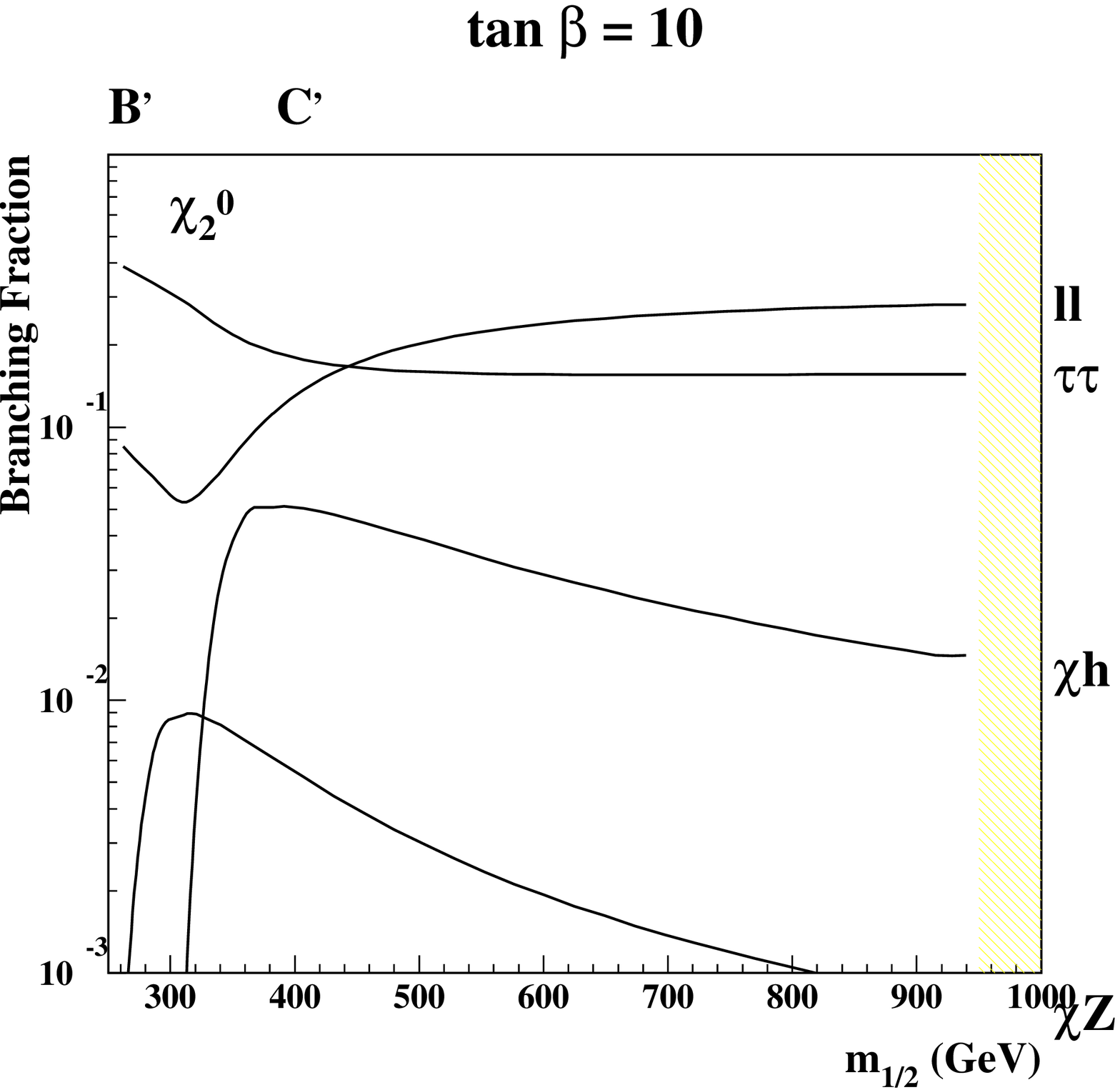,height=5cm}} \\
\mbox{\epsfig{file=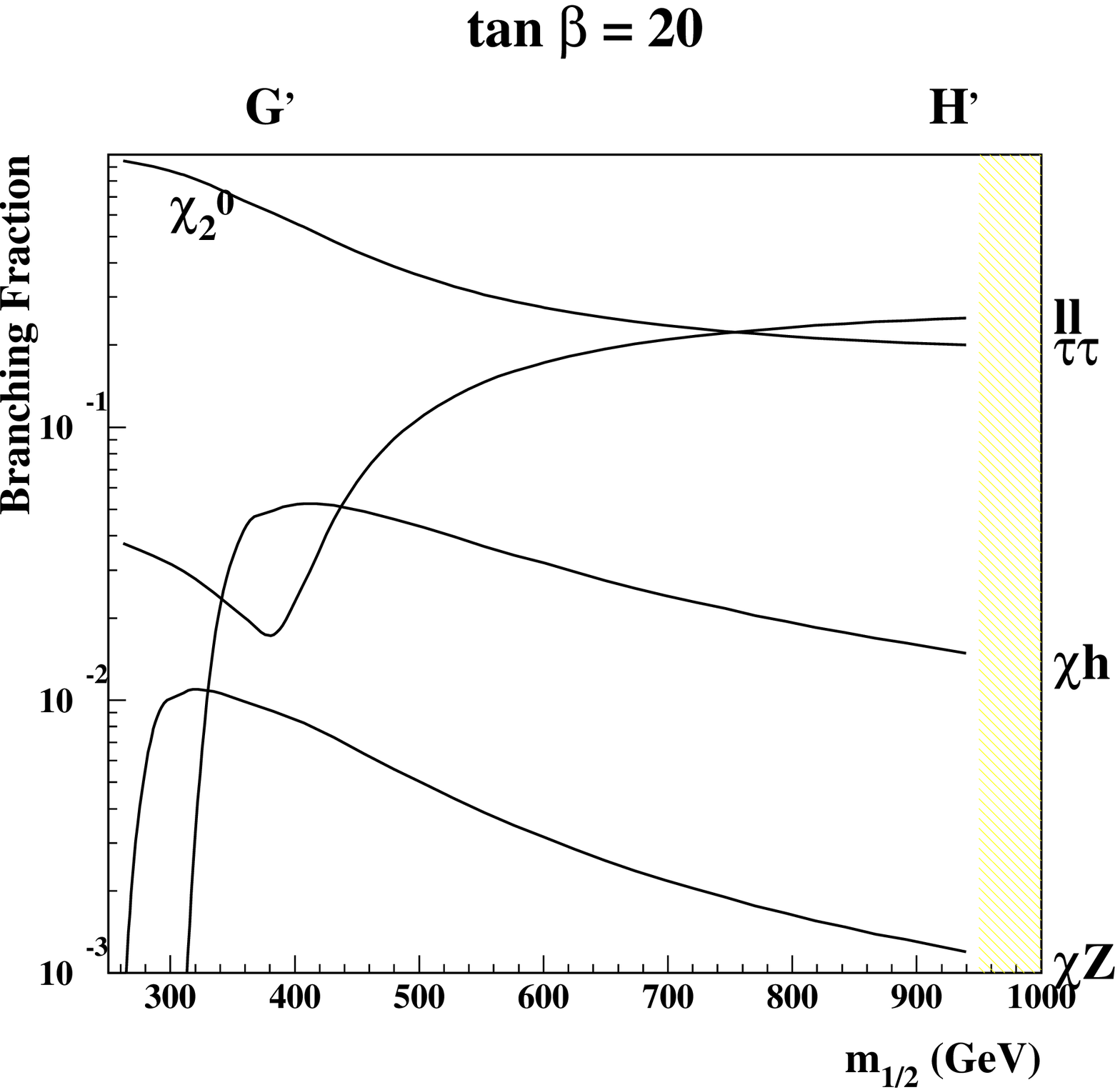,height=5cm}} &
\mbox{\epsfig{file=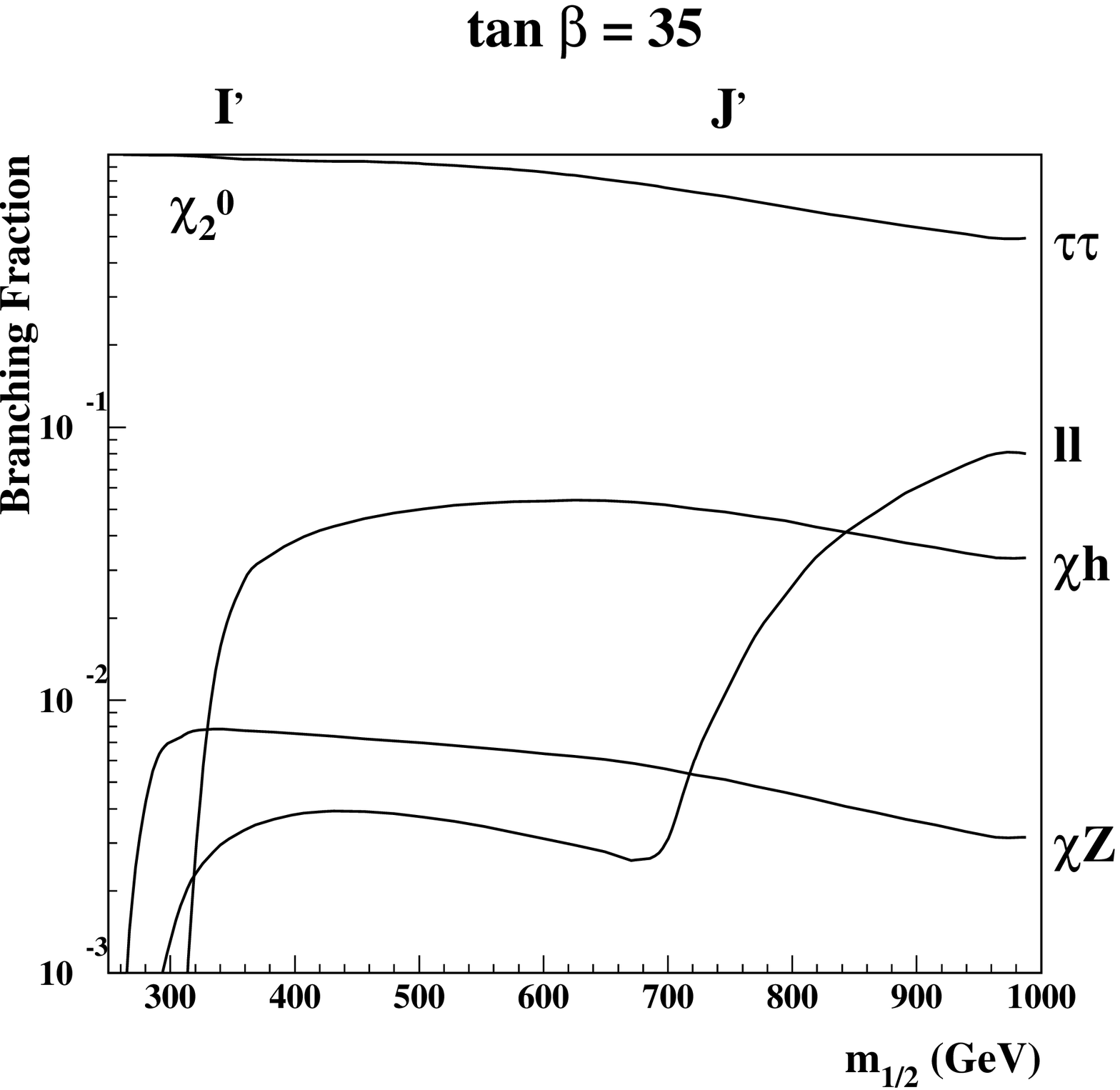,height=5cm}} \\  
\mbox{\epsfig{file=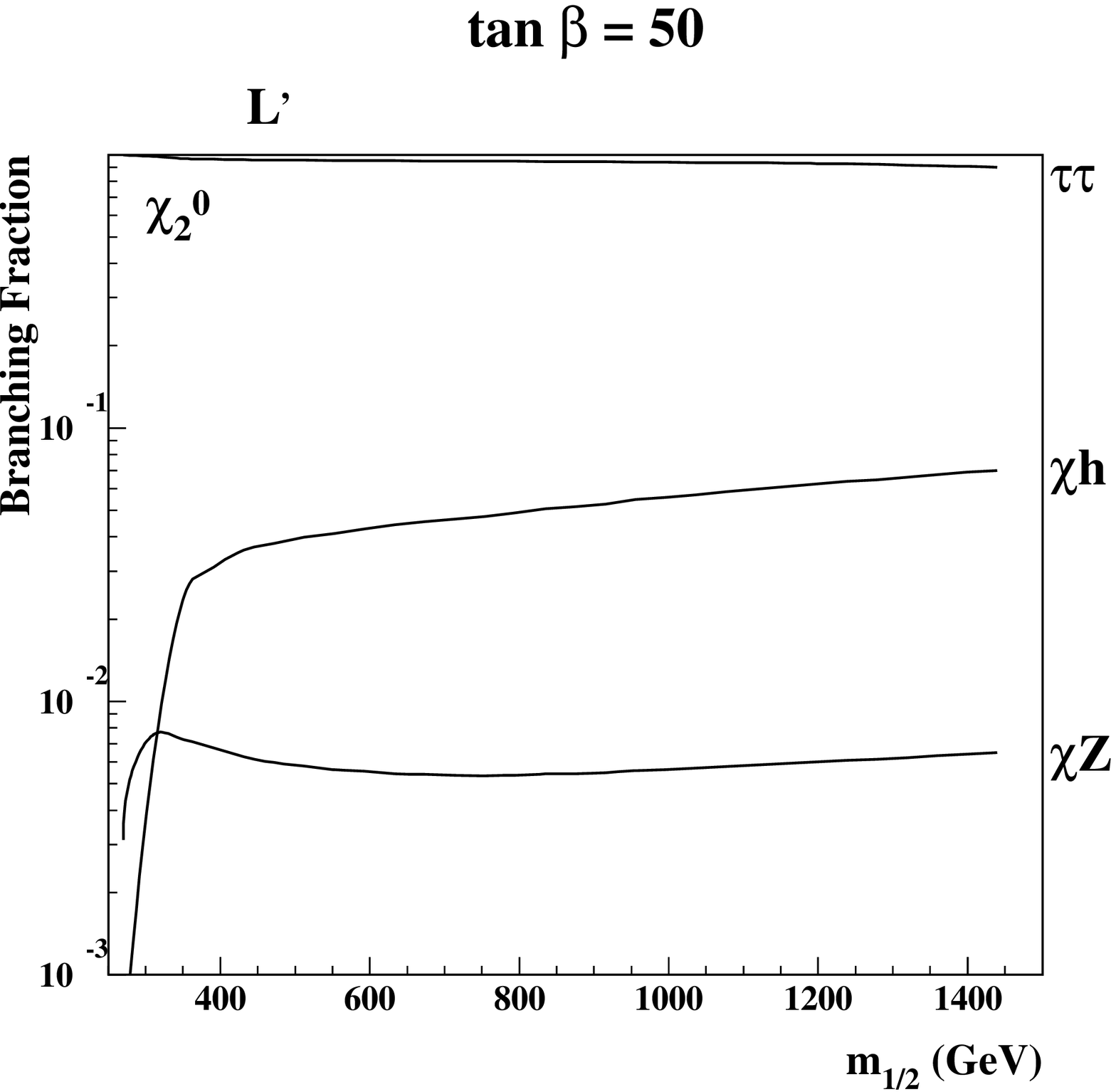,height=5cm}} &
\mbox{\epsfig{file=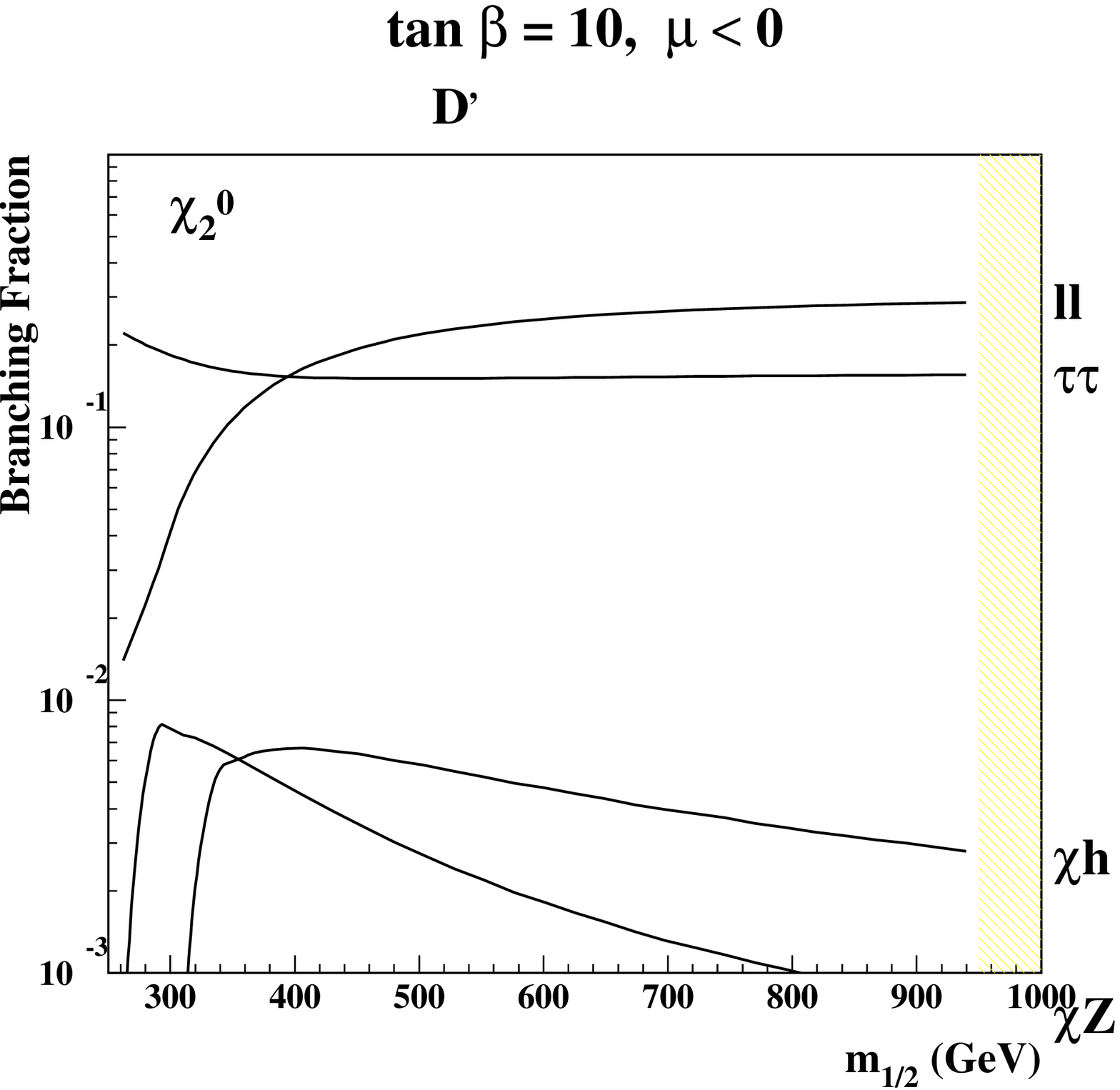,height=5cm}} \\
\end{tabular}
\end{center}
\caption{\label{fig:br}\it
Dominant branching ratios of the next-to-lightest neutralino $\chi_2$
as functions of $m_{1/2}$ along the WMAP lines for (a) $\tan \beta = 
5$, (b) $\tan \beta = 10$, 
(c) $\tan \beta = 20$, (d) $\tan \beta = 35$ and (e) $\tan \beta = 50$, 
all for $\mu > 0$, and (f) $\tan \beta = 10$ for $\mu < 0$. The locations 
of updated benchmark points 
along these WMAP lines are indicated, as are the upper limits on 
$m_{1/2}$ in panels (a, b, c) and (f).} 
\end{figure}

As seen in Fig~\ref{fig:br}(a) for $\tan \beta = 5$ and $\mu > 0$,
$\chi_2$ decays into ${\tilde \tau} \tau$ and other ${\tilde \ell} \ell$
modes dominate among the decay modes of interest. Both ${\tilde \tau_1}
\tau$ and ${\tilde \tau_2} \tau$ contribute, the latter increasing with
$m_{1/2}$ while the former decreases. The dip in the the ${\tilde \ell}
\ell$ branching ratio when $m_{1/2} \sim 300$~GeV reflects a similar
switch between the ${\tilde \ell_R} \ell$ modes at low $m_{1/2}$ and the
${\tilde \ell_L} \ell$ modes at large $m_{1/2}$~\footnote{We note in
passing that the branching ratio for ${\tilde \ell} \to \chi \ell$ exceeds
50~\% for all the WMAP range of $m_{1/2}$.}. The most important other
decay modes are the invisible $\chi_2 \to {\tilde \nu} \nu$ decays. The
decay modes $\chi_2 \to \chi Z$ and $\chi_2 \to \chi h$ exhibit clear
thresholds at $m_{1/2} \sim 270$~GeV and 310~GeV, respectively, reflecting
the opening up of these two channels. We note that benchmark A is in the
region of $m_{1/2}$ where the branching ratio for $\chi_2 \to \chi h$ is
already declining from its peak value $\simeq 6$ \%, while that for
$\chi_2 \to \chi Z$ has already been reduced from its peak value $\simeq
0.6$ \%.

Similar features are exhibited in Fig~\ref{fig:br}(b) for $\tan \beta =
10$ and $\mu >0$, with the peak in the branching ratio for $\chi_2 \to
\chi h$ reduced to $\simeq 5$ \% and that for $\chi_2 \to \chi Z$
increased to $\simeq 0.9$ \%. The updated benchmark point B with $m_{1/2}
= 250$~GeV is below both thresholds, whilst benchmark C has a near-maximal
branching ratio for $\chi_2 \to \chi h$.

Panel (c) of Fig.~\ref{fig:br} shows the corresponding $\chi_2$ decay
branching ratios for $\tan \beta = 20$ and $\mu > 0$. We see that the
${\tilde \tau} \tau$ mode is larger than for $\tan \beta = 10$, reflecting
the larger $\tau$ Yukawa coupling. The dip in the ${\tilde \ell} \ell$ has
moved to larger $m_{1/2} \sim 400$~GeV. The thresholds in $\chi_2 \to \chi
Z$ and $\chi_2 \to \chi h$ are again prominent, with the former branching
ratio now reaching slightly more than 1~\%. Benchmark G is located near
the peak in $\chi_2 \to \chi h$ and the dip in $\chi_2 \to {\tilde \ell}
\ell$, whilst benchmark H has almost equal branching ratios for ${\tilde
\ell} \ell$ and ${\tilde \tau} \tau$, and smaller branching ratios for
$\chi h$ and particularly $\chi Z$.

When $\tan \beta = 35$ and $\mu > 0$, shown in panel (d) of
Fig.~\ref{fig:br}, $\chi_2 \to {\tilde \tau} \tau$ becomes the dominant
branching ratio for all values of $m_{1/2}$ allowed by WMAP. Apart from
this, the most noticeable feature is the relative suppression of the
$\chi_2 \to {\tilde \ell} \ell$ branching ratio, whose dip has now moved
up to $m_{1/2} \sim 700$~GeV. The branching ratio for $\chi_2 \to \chi h$
exhibits a broad peak of similar height to the lower values of $\tan
\beta$, whereas the branching ratio for $\chi_2 \to \chi Z$ is somewhat
smaller. Benchmark I is located near the peaks of the $\chi_2 \to \chi h, 
\chi
Z$ branching ratios, and benchmark J is located in the region of $m_{1/2}$
where the $\chi_2 \to {\tilde \ell} \ell$ starts to rise.

Panel (e) of Fig.~\ref{fig:br} shows the branching ratios for $\tan \beta
= 50$ and $\mu > 0$. Here we note the strong dominance of $\chi_2 \to
{\tilde \tau} \tau$ decays, the approximate constancies (at relatively low
levels) of the branching ratios for $\chi_2 \to \chi h, Z$ above their 
respective thresholds,
and the strong suppression of $\chi_2 \to {\tilde \ell} \ell$ decays.
Benchmark L has branching ratios that are typical for this value of $\tan
\beta$.

Finally, panel (f) of Fig.~\ref{fig:br} shows the branching ratios for
$\tan \beta = 10$ when $\mu < 0$. In this case, there is dip structure in
the branching ratio for $\chi_2 \to {\tilde \ell} \ell$, which rises
monotonically with $m_{1/2}$, while the branching ratio for $\chi_2 \to
{\tilde \tau} \tau$ is always large. In this case, the peak branching
ratio for $\chi_2 \to \chi h$ is below that for $\chi_2 \to \chi Z$.

The $\chi_2$ branching ratios are rather different at the tip of the
funnel for $\tan \beta = 50$ - which has BR($\chi_2 \to \chi h$) = 0.11 at
the location of point~M, by point~K on the side of the rapid-annihilation
funnel for $\tan \beta = 35$ and $\mu <0$ - where the mode $\chi_2 \to
\chi h$ gives half of the total $\chi_2$ decay rate, and at the focus
points E and F - where the $\chi_2$ decay rate is saturated by $\chi q
\bar{q}$ and by $\chi h$, respectively.

\section{Production and Detectability along WMAP lines}

We now study the reaches at different accelerators in the number of
observable supersymmetric particles, and the changes in experimental
signatures and topologies along the WMAP lines for different values of
$\tan \beta$.

\subsection{LHC}

\subsubsection{Sparticle Signatures}

\begin{figure}  
\begin{center}
\mbox{\epsfig{file=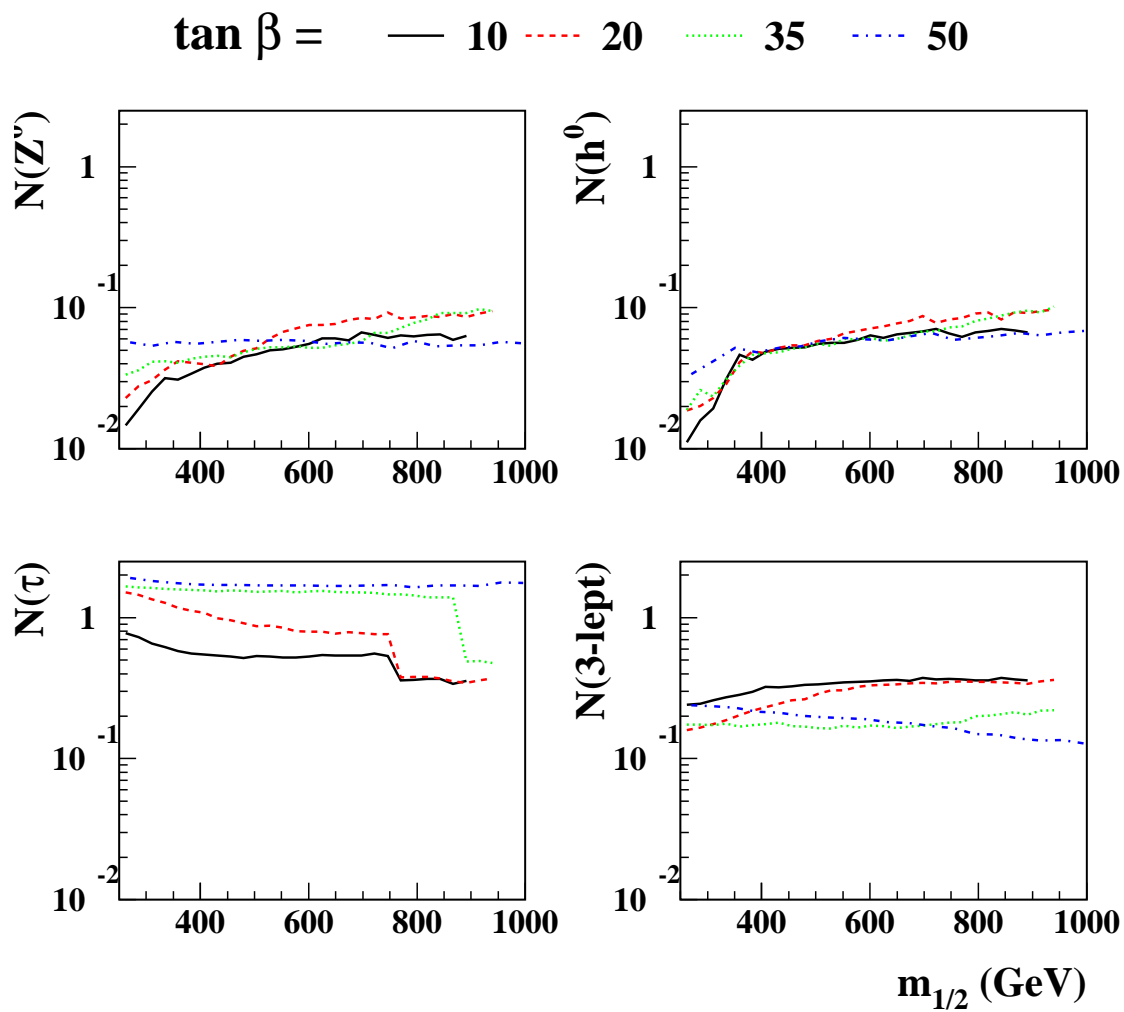,height=15cm}}
\end{center}
\caption{\it Supersymmetric event signatures at the LHC: the numbers of
$Z$ bosons (upper left), light Higgs bosons $h$ (upper right),
$\tau$ leptons (lower left) and three-lepton final states per 
supersymmetric event are shown as functions of $m_{1/2}$ along the WMAP 
lines for four values of $\tan \beta$, all with $\mu > 0$. These plots 
were obtained with {\tt PYTHIA 6.215}~\protect\cite{PYTHIA} interfaced to 
{\tt ISASUGRA 7.67}~\protect\cite{ISASUGRA}.}
\label{fig:dec} 
\end{figure}

In order to visualize the changes in the signatures of sparticle decays
along the WMAP lines, we first compute the mean numbers of different
particle species produced in sparticle decay chains at LHC. Large samples
of inclusive supersymmetric events have been generated along the WMAP
lines, using {\tt PYTHIA 6.215}~\cite{PYTHIA} interfaced to {\tt ISASUGRA
7.67}~\cite{ISASUGRA} to compute the average numbers of different particle
species produced per supersymmetric event, and the production cross
sections of exclusive and inclusive sparticle production reactions.
Fig.~\ref{fig:dec} shows the average numbers of (a) $Z$, (b) $h$ bosons,
(c) $\tau$ leptons and (d) trilepton events with three $\ell$ ($\ell =
e,~\mu$) in inclusive sparticle decays at the LHC. The plots are all for
$\mu > 0$ and display results for $\tan \beta = 10, 20, 35$ and 50 along
the corresponding WMAP lines.

We see in panel (a) of Fig.~\ref{fig:dec} that the fractions of sparticle
decays with a $Z$ boson in the cascade is never large in the regions of
the parameter space traversed by the WMAP lines, reaching about
0.1~$Z^0$/sparticle decay at large $m_{1/2}$ and $\tan \beta$.

The decays of sparticles into the lightest Higgs boson $h$, followed by
$h \to b \bar{b}$ decay, provide another important signature at the LHC.
These appear above the threshold for $\chi_2 \to \chi h$ decay at
about $m_{1/2}$=350~GeV, and may reach 7-10\% per sparticle event at 
larger $m_{1/2}$, as seen 
in panel (b) of Fig.~\ref{fig:dec}.

As seen in panel (c), the numbers of $\tau$ leptons produced in sparticle
events are always large, particularly at large $\tan \beta$ where there
are between one and two per event. Finally, we see in panel (d) of
Fig.~\ref{fig:dec} that the fraction of trilepton events is always above
10\%, and may attain $\sim 40$\% at large $m_{1/2}$ and small $\tan
\beta$.

These plots reinforce the importance of the $\tau$ signature for sparticle
detection at the LHC, within the CMSSM framework used here. Clearly the
$\chi_2 \to \chi h, Z$ signatures are also interesting, but they may be
quite challenging to exploit. However, we would like to emphasize that
other decay patterns should not be neglected, and may be preferred in
other supersymmetric models. For example, if the soft
supersymmetry-breaking scalar masses for the Higgs multiplets are
non-universal, $\chi_2 \to \chi h, Z$ decays may be more copious.

\subsubsection{Detectability at the LHC}

We now estimate the numbers of different species of supersymmetric
particles that may be detectable at the LHC as functions of $m_{1/2}$
along the WMAP lines for different values of $\tan \beta$, using the 
following criteria.

$\bullet$ {\it Higgs bosons}: We generally follow the ATLAS and CMS
studies of the numbers of observable bosons as functions of $M_A$ and
$\tan \beta$~\cite{ACHiggs}. In contrast to~\cite{Bench}, here we also
consider $H/A \to \chi_2 \chi_2$ decays.

$\bullet$ {\it Gauginos}: Our criteria for the observability of heavier
neutralinos at the LHC have been refined compared to our previous
publication~\cite{Bench}. In particular, we first compute the total
numbers of $\chi_2$ produced in all sparticle events and the branching
fraction corresponding to a dilepton final state. We consider this to be
observable at the LHC if the product is at least 0.01~pb, corresponding to
1000 events produced with 100~fb$^{-1}$ of integrated luminosity. The
lightest neutralino $\chi$ is considered always to be detectable via the
cascade decays of observed supersymmetric particles.

$\bullet$ {\it Gluinos}: These are considered to be observable for masses
below 2.5~TeV~\cite{squarks}.

$\bullet$ {\it Squarks}: The spartners of the lighter quark flavours $u,
d, s, c$ are considered to be observable if $m_{\tilde{q}} <$
2.5~TeV~\cite{squarks}, but we recall that it is not known how to
distinguish the spartners of different light-quark flavours at the LHC. In
general, we assume that the stops and sbottoms ${\tilde t}, {\tilde b}$
are observable only if they weigh below 1~TeV, unless the gluino weighs $<
2.5$~TeV and the stop or sbottom can be produced in its two-body decays.

$\bullet$ {\it Charged Sleptons}: These are considered to be observable
when the mass splitting $m_{\tilde{\ell}}-m_{\chi} >
30$~GeV~\footnote{This relatively conservative cut is to ensure that the
$m_{\tilde{\ell}}$ decay lepton should be observable, and causes sleptons
to be lost in some scenarios, particularly with the lower values of $m_0$
at the updated benchmark points.} and the inclusive production cross
section, which includes their direct production and that in the decays of
other supersymmetric particles, exceeds 0.1~pb, giving at least 10000
events with 100~fb$^{-1}$ of integrated luminosity~\footnote{This 
criterion is illustrative: in a more detailed study, one should also 
include backgrounds such as $W^+ W^-$, that may be very important.}.

$\bullet$ {\it Sneutrinos}: These have not been considered as observable
at the LHC, due to their large yield of invisible $\tilde{\nu} \to \nu
\chi$ decays.

We first note the key differences between the expected numbers of
detectable MSSM particles at the updated benchmark points, shown in
Fig.~\ref{fig:newM}, and the analogous LHC analysis in~\cite{Bench}.
Howver, we would like to caution the reader that it is impossible to be 
precie about the capabilities of the LHC (or other accelerator) without 
detailed simulations that go beyond the scope of this paper.

$\bullet$ At point B, we now consider the
heavier neutral Higgs bosons to be detectable via their decays into the
$\chi_2$ and into $\tau$ leptons, and the $H^\pm \to \tau \nu$ decays
should also be detectable.

$\bullet$ We no longer consider the $H^\pm$ to be observable at point C.

$\bullet$ At point F, we now observe that the ${\tilde g}$ has branching
ratios of 16~\% and 17~\%, respectively, for decays into $\chi_{2,3} {\bar
t} t$, followed by branching ratios of 97~\% and 99~\% for $\chi_2 \to
\chi h$ and $\chi_3 \to \chi Z$, respectively. The ${\tilde g}$ also has a
branching ratio of 33~\% for the decay into $\chi^\pm {\bar t} b$. We now 
consider that the $\chi_{2,3}$ should be detectable at this point, but not 
the $\chi^\pm$.

$\bullet$ At point H, supersymmetric particles are now well within the
reach of the LHC, essentially because of the reduction in $m_{1/2}$. We
now consider the ${\tilde q}$, ${\tilde g}$, $\chi$ and $\chi_2$ to be
detectable at this point.

$\bullet$ At point J, we now consider the $\chi_2$ to be observable 
because of its large production rate, even though its decays 
generally include $\tau$ leptons.

$\bullet$ At point K, we now find that no sparticles are observable with 
our present criteria, although the lightest Higgs boson $h$ is observable.

A corollary of these changes is that supersymmetric particles appear to be
detectable at the LHC at all the updated benchmark points except K and M,
which are located in rapid-annihilation funnels.

\begin{figure}[t]
\centerline{\epsfxsize = 0.4\textwidth \epsffile{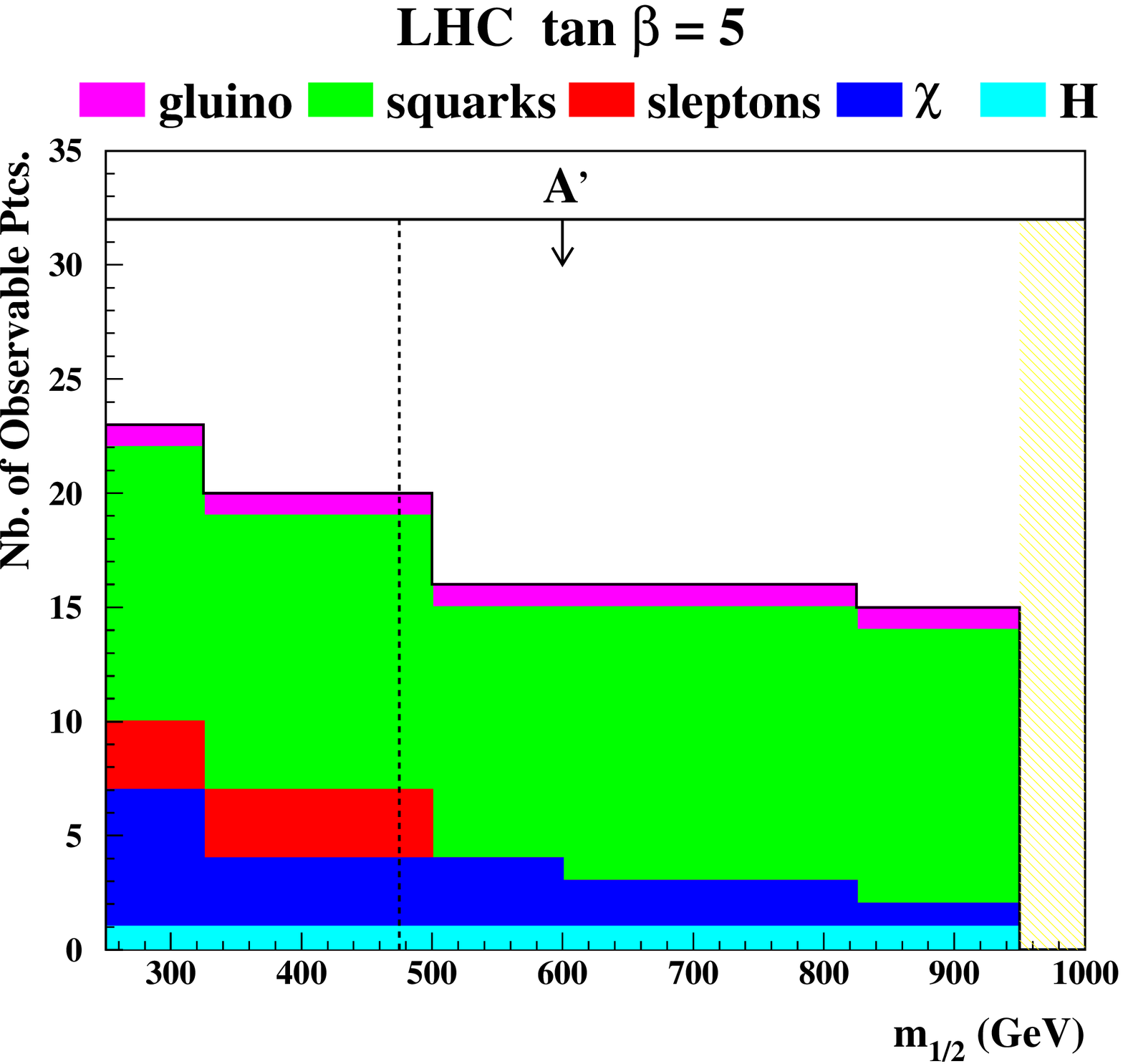}
\hfill \epsfxsize = 0.4\textwidth \epsffile{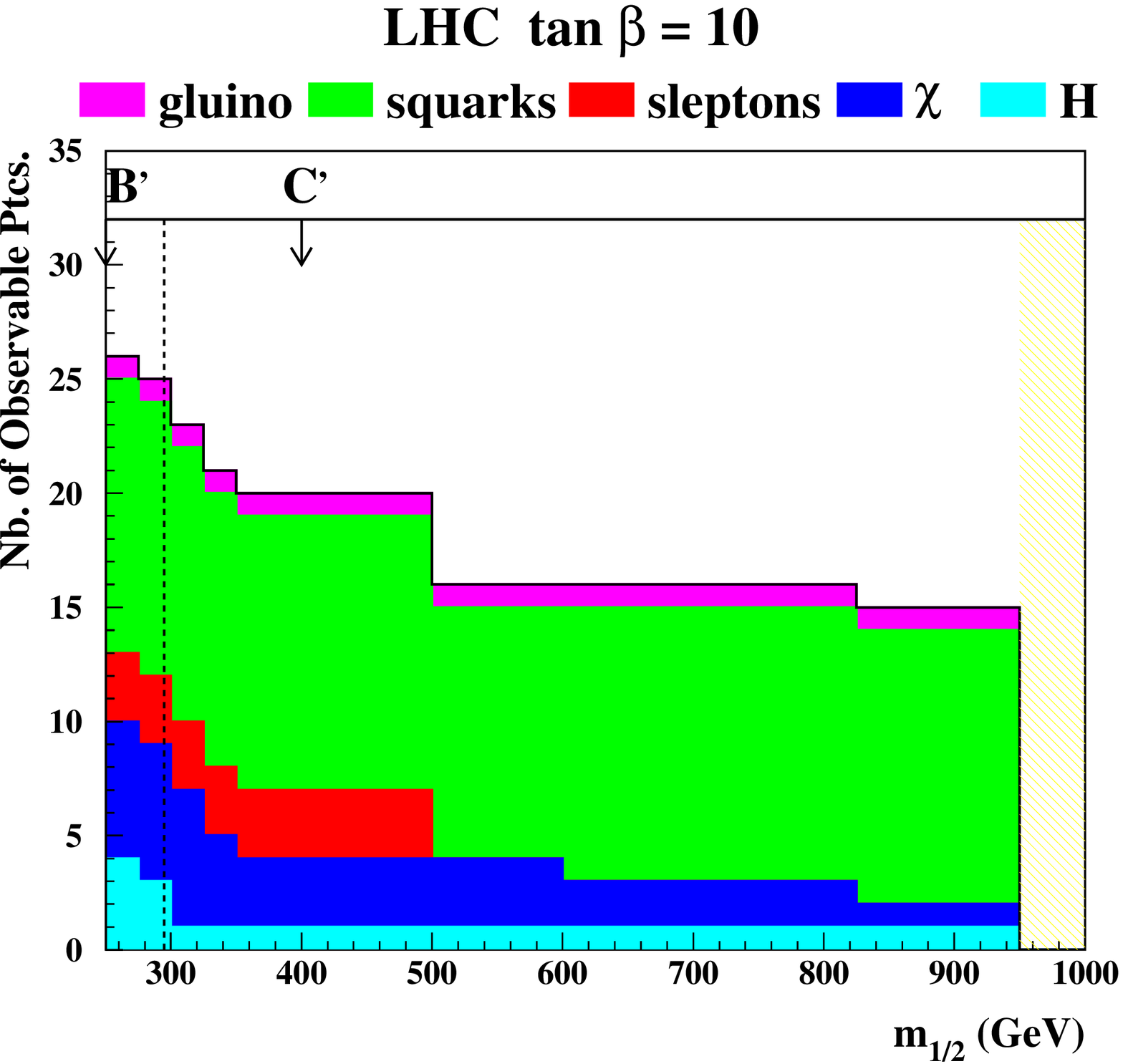}}
\centerline{\epsfxsize = 0.4\textwidth \epsffile{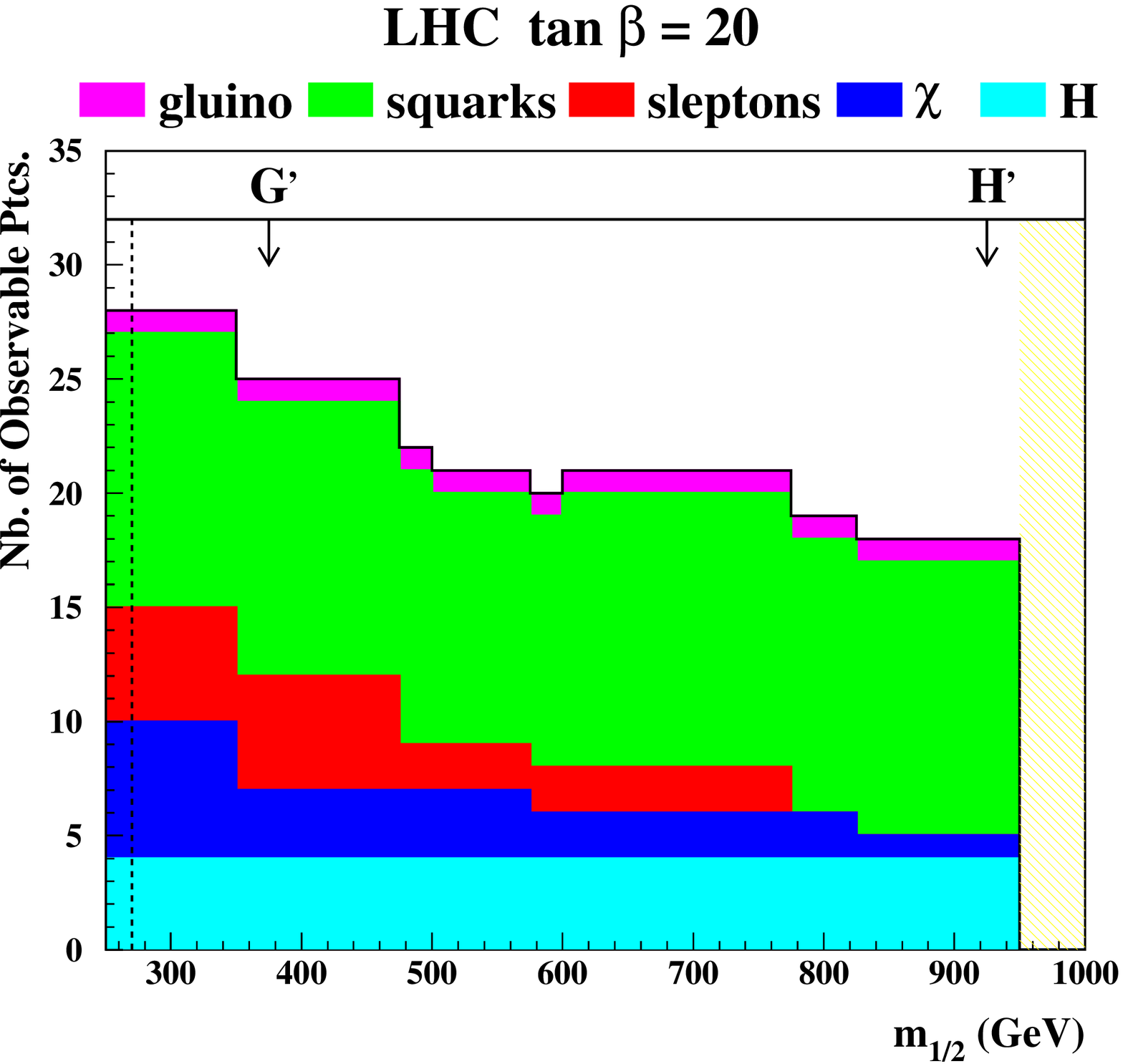}
\hfill \epsfxsize = 0.4\textwidth \epsffile{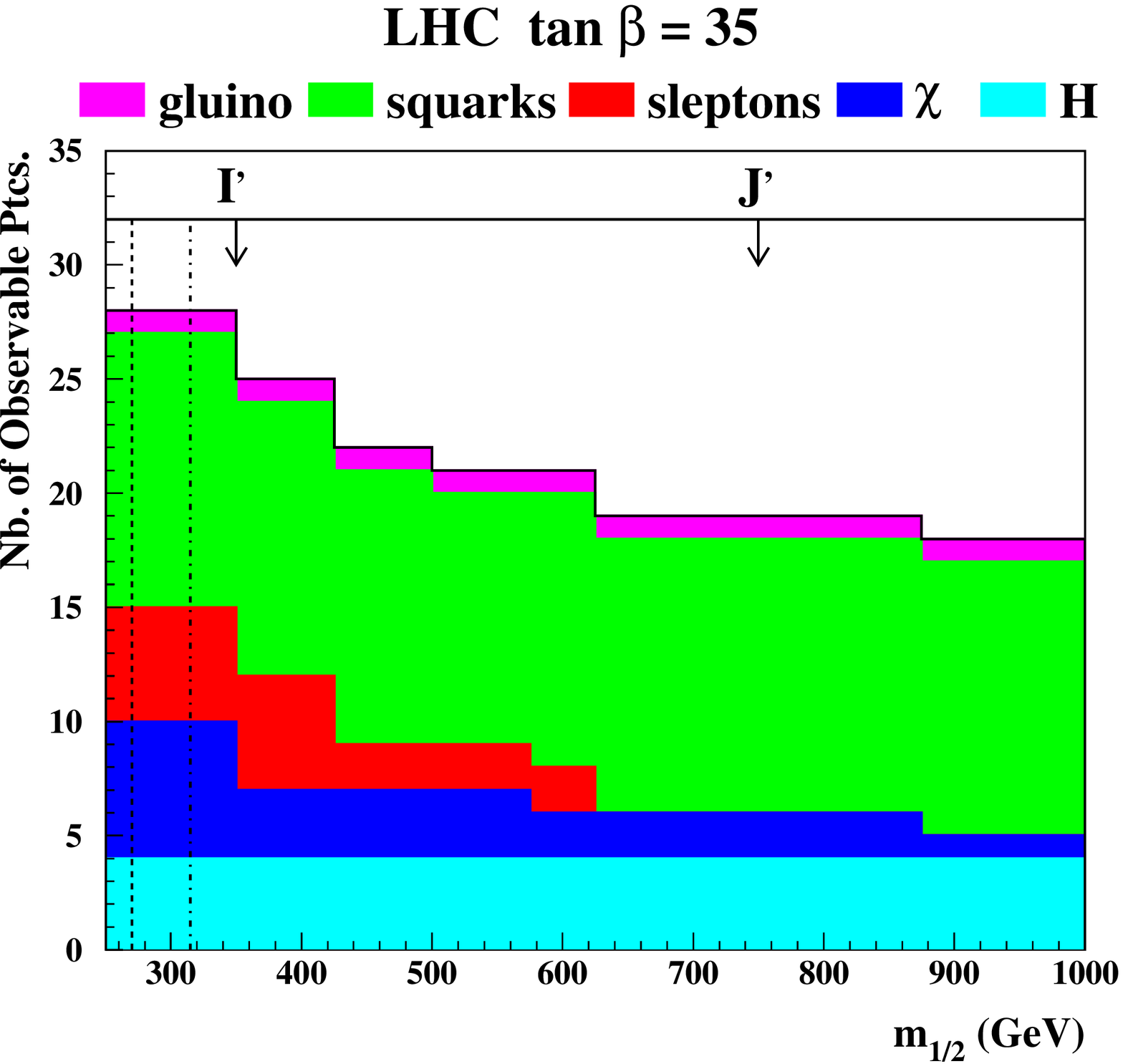}}
\centerline{\epsfxsize = 0.4\textwidth \epsffile{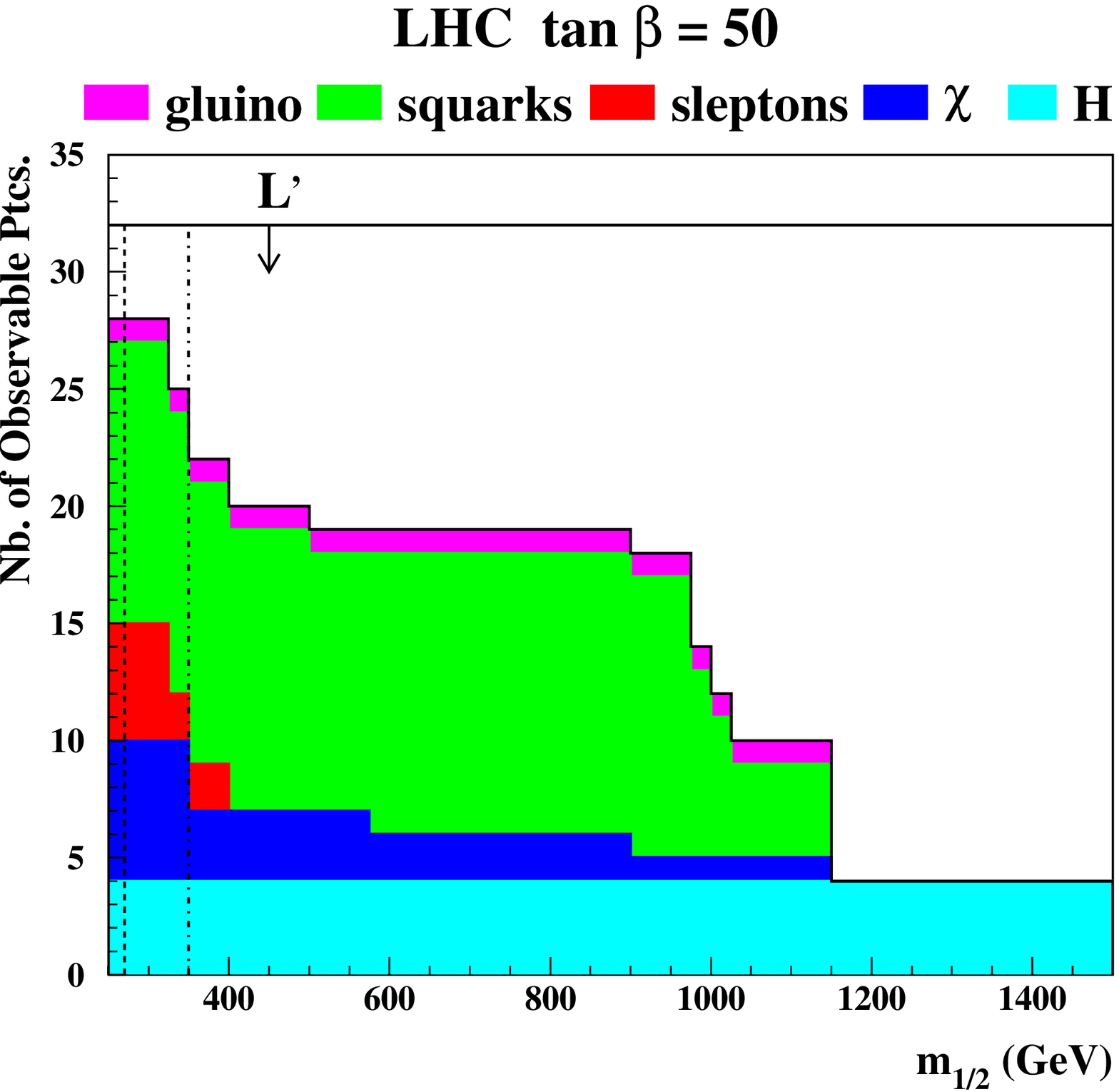}
\hfill \epsfxsize = 0.4\textwidth \epsffile{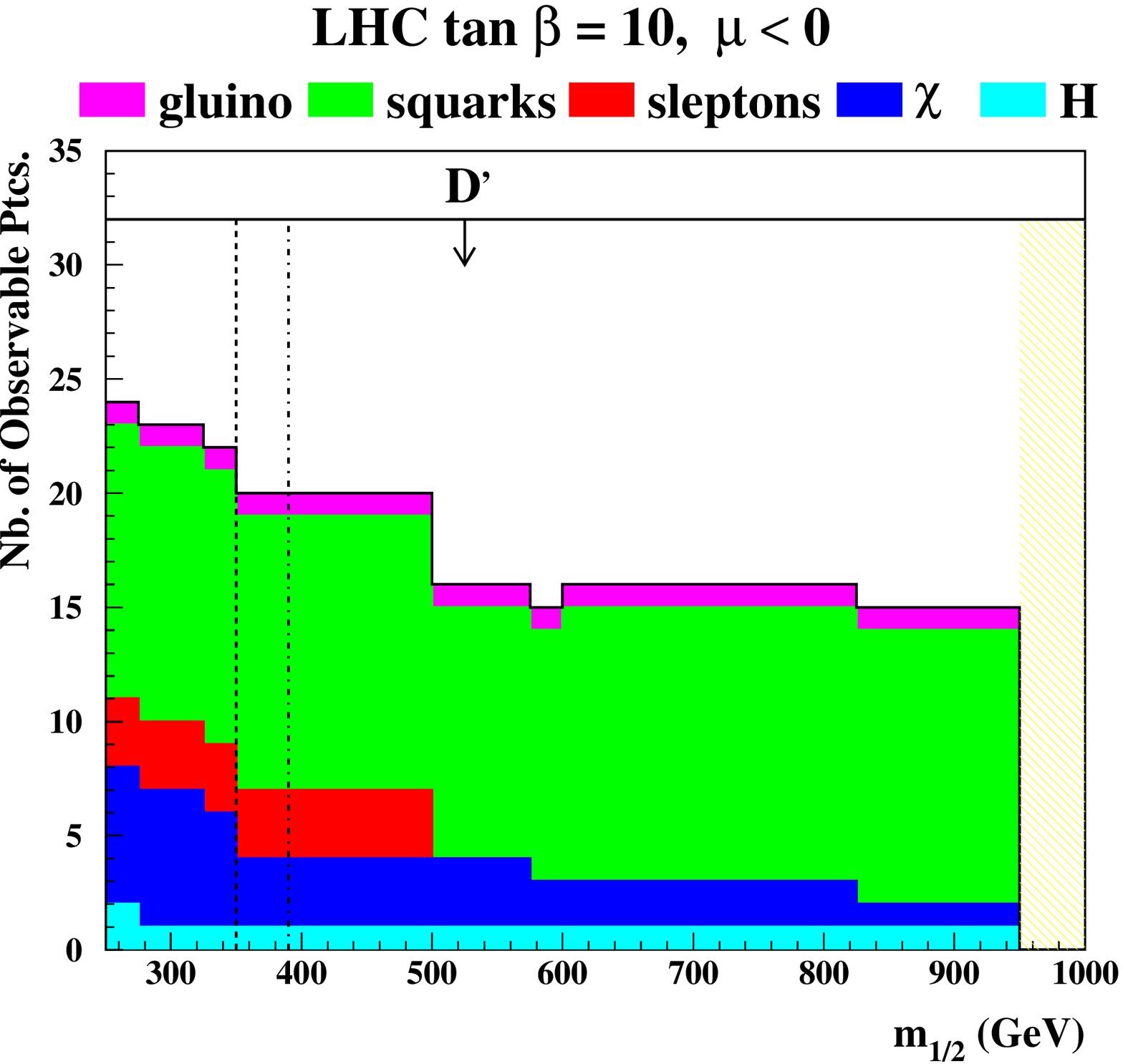}}
\caption{\it
Estimates of the numbers of MSSM particles that may be detectable at the 
LHC as functions of $m_{1/2}$ along the WMAP lines for $\mu > 0$ and $\tan 
\beta = 5, 10, 20, 35$ and $50$, and for $\mu < 0$ and $\tan \beta = 10$. 
The locations of updated benchmark points along these WMAP lines are 
indicated, as are the nominal lower bounds on $m_{1/2}$ imposed by $m_h$ 
(dashed lines) and $b \to s \gamma$ (dot-dashed lines). 
\vspace*{0.5cm}}
\label{fig:lhc}
\end{figure}

We now display in Fig.~\ref{fig:lhc}, as functions of $m_{1/2}$ along the
WMAP lines for different values of $\tan \beta$, the numbers of different
types of MSSM particles that should be observable at the LHC with
100~fb$^{-1}$ of integrated luminosity in each of ATLAS and CMS. The
nominal lower bounds on $m_{1/2}$ imposed by $m_h$ (dashed lines) and $b
\to s \gamma$ (dot-dashed lines) are also shown. These each have some
uncertainties, for example {\tt FeynHiggs} has a quoted error of $\sim
2$~GeV in the calculation of $m_h$ so that (specifically) benchmark B in
the top right panel should not be regarded as excluded.

We use the criteria explained above and generalize the results for the
indicated benchmark scenarios that were shown in Fig.~\ref{fig:newM}. In
the top left panel, for $\tan \beta = 5$, we note first that only one MSSM
Higgs boson is expected to be visible. In the top right panel for $\tan
\beta = 10$, we see that the other Higgs bosons are also expected to be
observable at low $m_{1/2}$ in the neighbourhood of point B, for example
via decays into $\chi_2 \chi_2$. All of the Higgs bosons
are expected to observable over the entire range of $m_{1/2}$ for $\tan
\beta = 35, 50$ and $\mu > 0$, but not for $\tan \beta = 10$ and $\mu < 0$
(bottom right panel). All of the charginos and neutralinos are expected to
be observable at low $m_{1/2}$, but only the lightest neutralino $\chi$ at
high $m_{1/2}$. Since the sneutrinos decay invisibly and the rates for
${\tilde \ell_2}$ production are inadequate, only the ${\tilde \ell_1}$
are considered to be observable at the LHC for this value of $\tan \beta$,
and only at low $m_{1/2}$. However, all the squarks and gluinos are
expected to be observable anywhere along any of the WMAP lines, except
that for $\tan \beta = 50$ and $\mu > 0$, where sparticles would be
unobservable at the end of the funnel, as exemplified by benchmark point M
in Fig~\ref{fig:newM}.

In the cases of larger values of $\tan \beta$, shown in the other panels
of Fig.~\ref{fig:lhc}, all the MSSM Higgs bosons are expected to be
observable for all the WMAP ranges of $m_{1/2}$. However, the
observabilities of the different charginos and neutralinos resemble those
for $\tan \beta = 10$. Some of the ${\tilde \ell_2}$ may be observable, at
least for small $m_{1/2}$. As for $\tan \beta = 10$, all the squarks and
gluinos are expected to observable at the LHC anywhere along the parts of
the WMAP lines that are shown, with the exception of the tip of the line 
for $\tan \beta = 50$.

We comment finally on the case of the metastable ${\tilde \tau_1}$ that
appears towards the end of the WMAP lines where $\Delta M =
|m_{\tilde \tau_1} - m_\chi| < m_\tau$. There is no specific study of
this scenario at the LHC, but it has been estimated in an analogous
gauge-mediated supersymmetry-breaking scenario that the efficiency for
observing a metastable particle with $m = 640$~GeV would be 25~\%
(increasing for larger masses), and that a signal should be detectable if
the overall production cross-section exceeds 1~fb~\cite{CMS-CR1999-019}.
In view of the large sparticle production cross sections at the LHC, and
the fact (see Fig.~\ref{fig:dec}) that each sparticle event produces on
average 0.1 or more ${\tilde \tau_1}$ particles per event towards the ends
of the WMAP lines, we believe that a metastable ${\tilde \tau_1}$ could be
detected out to the ends of the coannihilation tails. On the other hand,
it may not be easy to determine the mass accurately, as the peak measured
from the relativistic $1/\beta$ factor broadens with increasing mass. For
example, when $m = 640$~GeV the FWHM of the mass distribution is estimated
to be about 250 GeV~\cite{CMS-CR1999-019}. The error in the mass estimate
may be reduced in a large statistical sample, but such an analysis would
need to assume a realistic experimental environment in order to evaluate
systematic errors.

\subsection{Detectability at $e^+ e^-$ Linear Colliders}

Our criteria for the observability of supersymmetric particles at linear
colliders are based on their pair-production cross sections. 

$\bullet$ Particles
with cross sections in excess of 0.1~fb are considered as observable,
thanks to their production in more than 100 events with an integrated
luminosity of 1~ab$^{-1}$. 

$\bullet$ The lightest neutralino $\chi$ is considered to
be observable only through its production in the decays of heavier
supersymmetric particles. 

$\bullet$ Sneutrinos are considered to be detectable when
the sum of the branching fractions for decays which lead to clean
experimental signatures, such as $\tilde{\nu}_{\ell} \to \chi^{\pm}
\ell^{\mp}$ ($\ell$ = $e$, $\mu$, $\tau$) and $\tilde{\nu}_{\tau} \to W^+
\tilde{\tau_1}^-$, exceeds 15\%. 

$\bullet$ The $\gamma \gamma$ collider option at a linear collider would
allow one to produce heavy neutral Higgs bosons via the $s$-channel
processes $\gamma \gamma \to A$ and $\gamma \gamma \to H$, extending the
reach up to 750~GeV for 0.5~TeV $e^\pm$ beams and up to 1.5-2.0~TeV for
1.5~TeV $e^\pm$ beams. A $\gamma \gamma$ collider may also be used to look 
for gluinos, but we do not include this possibility in our analysis.

$\bullet$ Finally, we assume that a metastable ${\tilde \tau_1}$ could be
detected at any linear $e^+ e^-$ collider with more than 100 events, and
note that the mass could be measured more accurately than at the LHC, by
measuring the production threshold as well as $1/\beta$.

We consider $e^+ e^-$ collision energies $\sqrt{s}$ = 0.5~TeV, 1.0~TeV,
3~TeV and 5~TeV, and also the combined capabilities of the LHC and a 1-TeV
linear collider. Comparing our present estimates of the physics reaches of
linear $e^+ e^-$ colliders of different energies with those
in~\cite{Bench}, we observe changes due both to the criteria adopted and
to the mass spectra. We note briefly the principal changes.

$\bullet$ The updated point F has no supersymmetric particles observable
at 1~TeV centre-of-mass energy. This is in part due to differences in the
{\tt ISAJET} spectrum optimisation, where it now reproduces more closely
that from {\tt SSARD}.

$\bullet$ On the other hand, the reductions in the parameters $m_{1/2},
m_0$ for point H make slepton-pair production possible already below
1~TeV. The same point now has one squark accessible at CLIC with
centre-of-mass energy 3~TeV, and all of the squarks at 5~TeV.

$\bullet$ Point M now has no squarks accessible even to CLIC operating
with a centre-of-mass energy of 5~TeV.

\subsubsection{TeV-Class $e^+ e^-$ Linear Colliders}

\begin{figure}[t]
\centerline{\epsfxsize = 0.4\textwidth
\epsffile{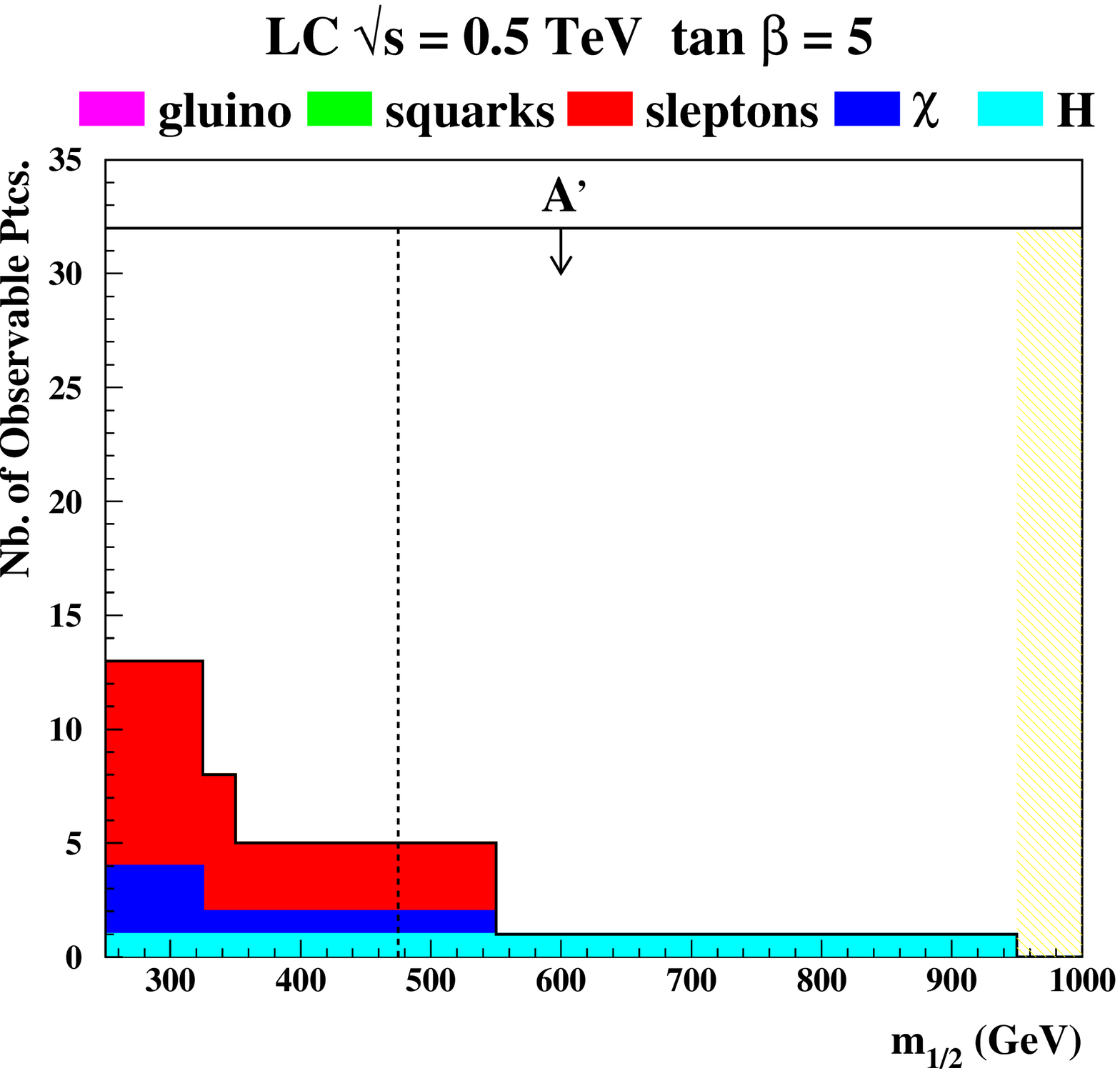}
\hfill \epsfxsize = 0.4\textwidth \epsffile{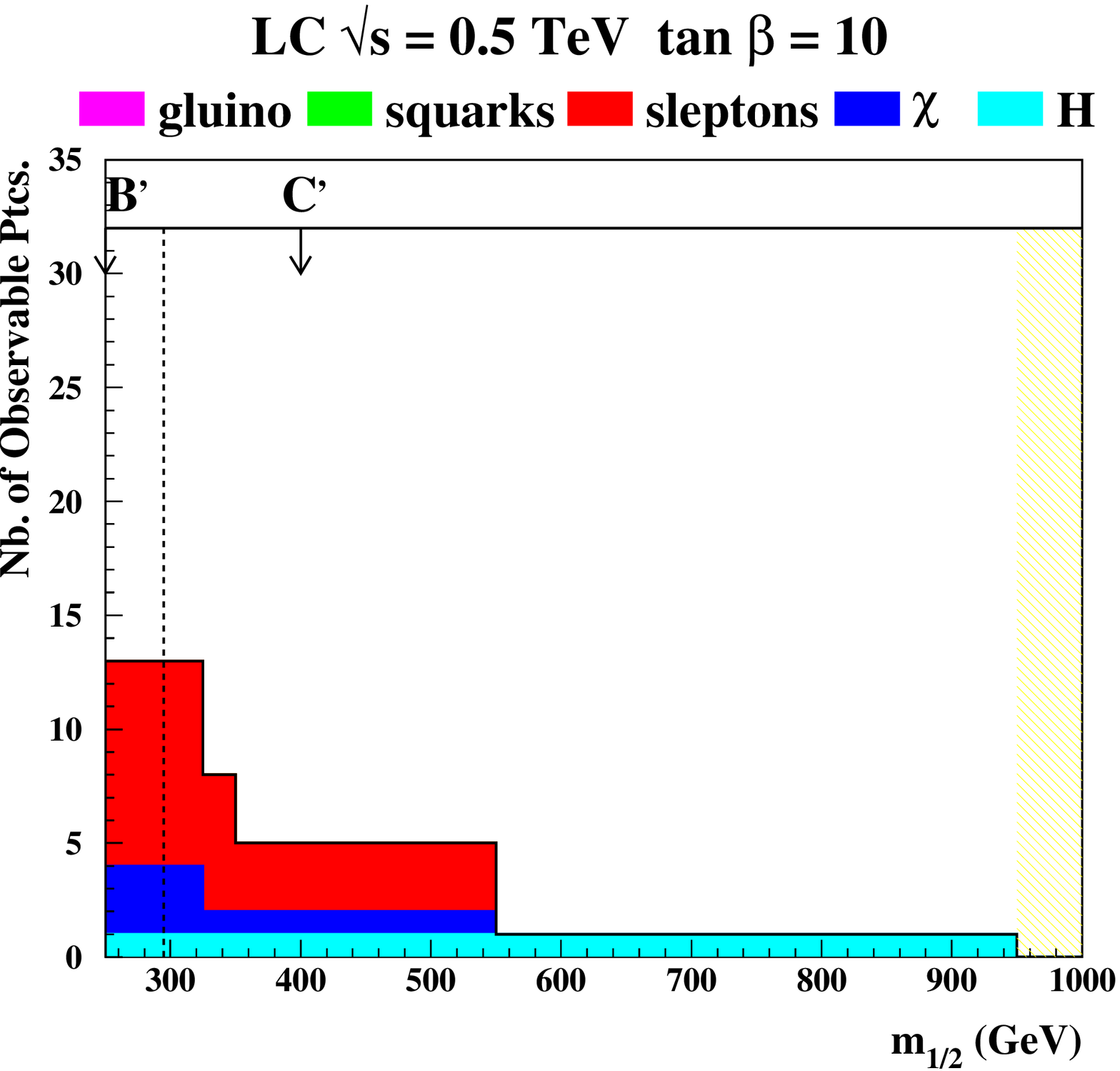}}
\centerline{\epsfxsize = 0.4\textwidth
\epsffile{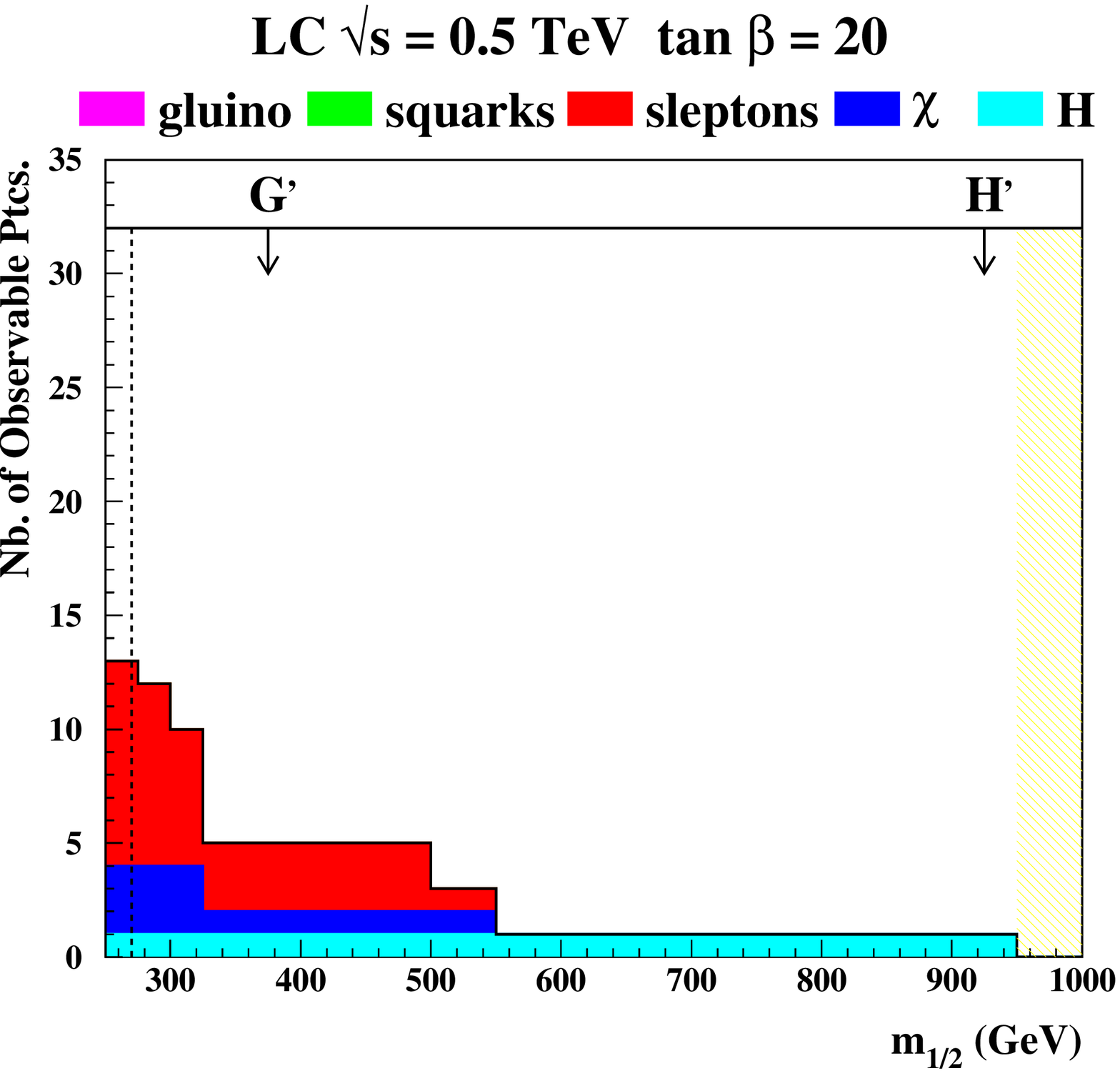}
\hfill \epsfxsize = 0.4\textwidth \epsffile{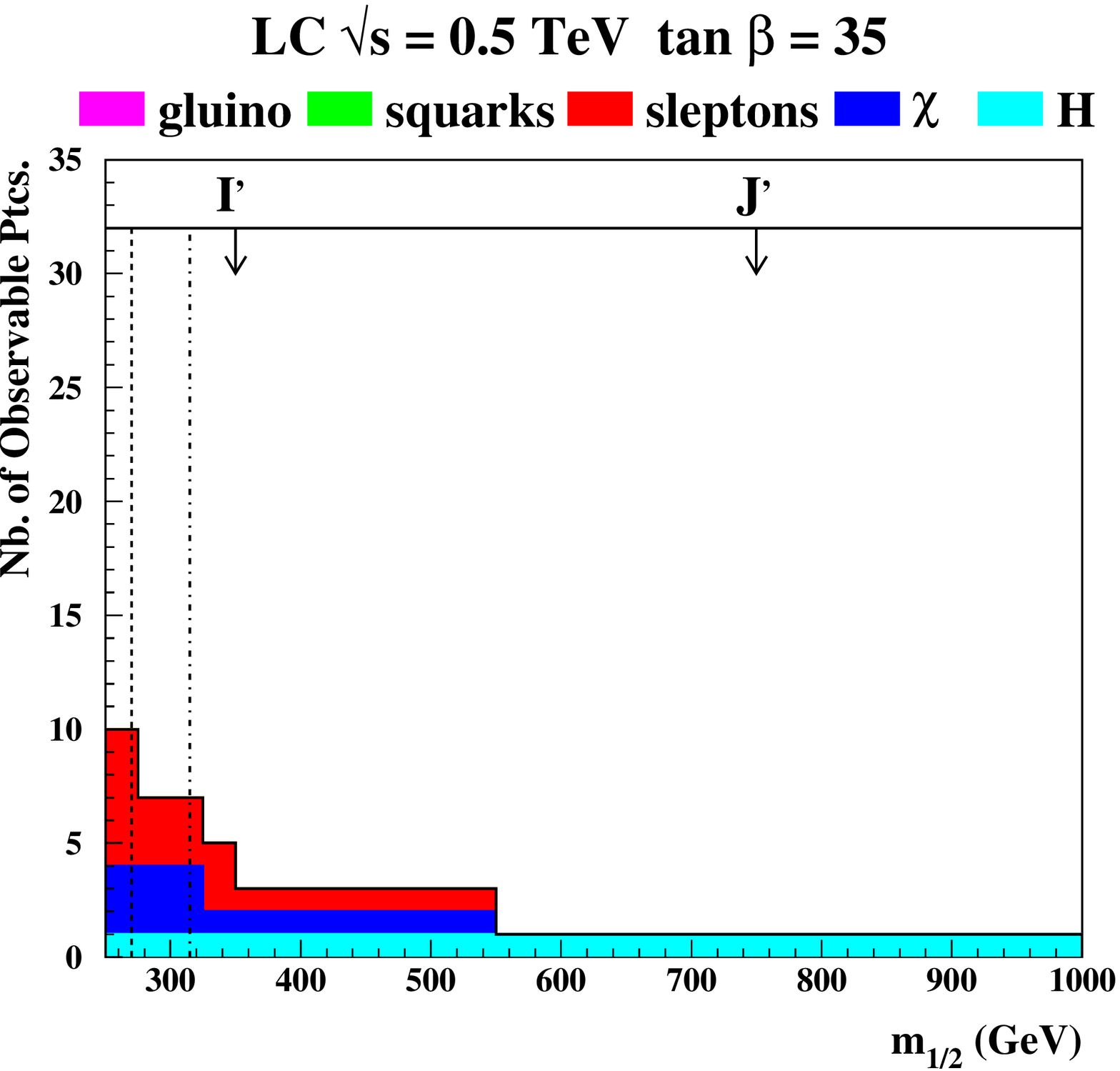}}

\centerline{\epsfxsize = 0.4\textwidth
\epsffile{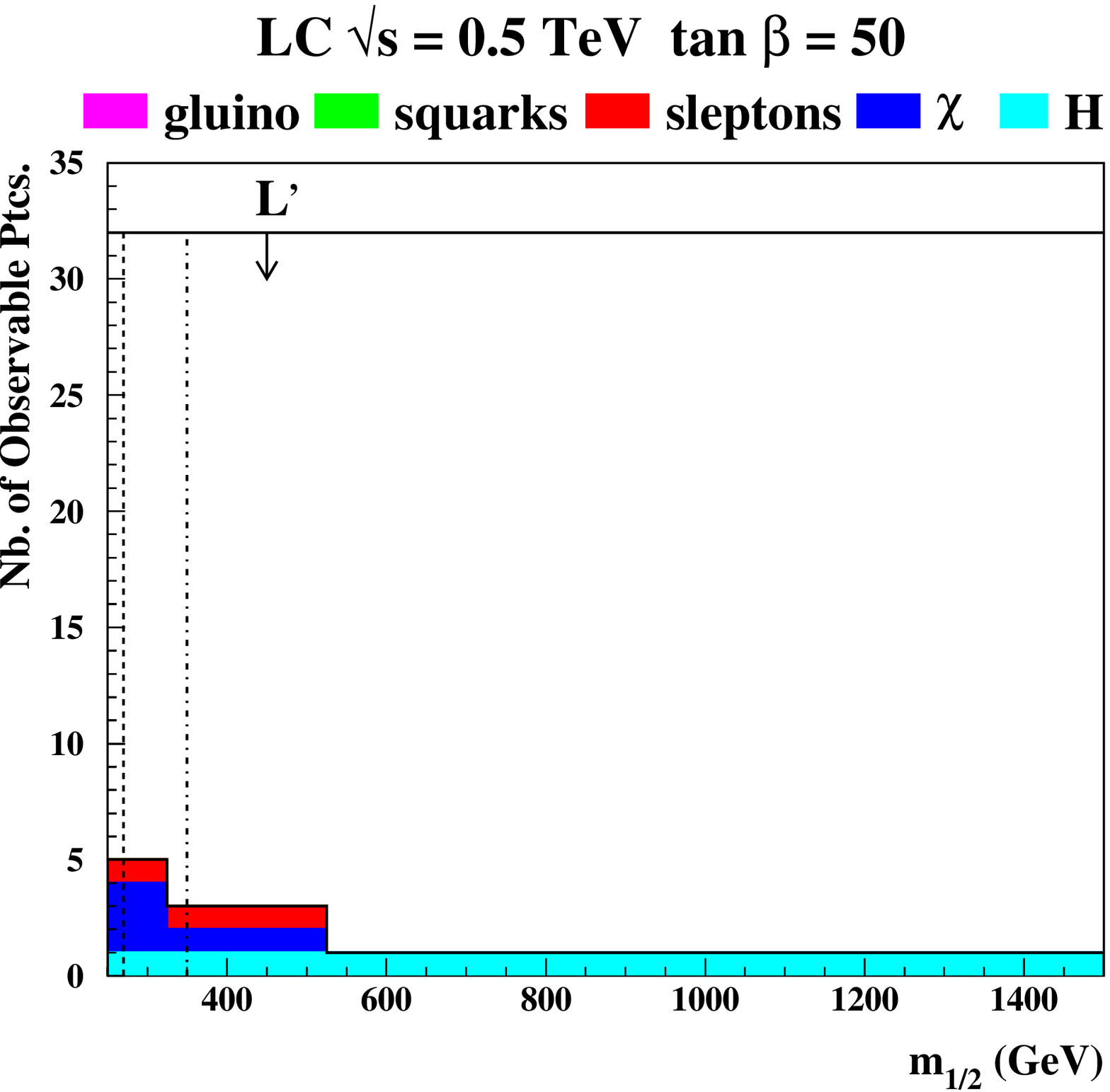}
\hfill \epsfxsize = 0.4\textwidth \epsffile{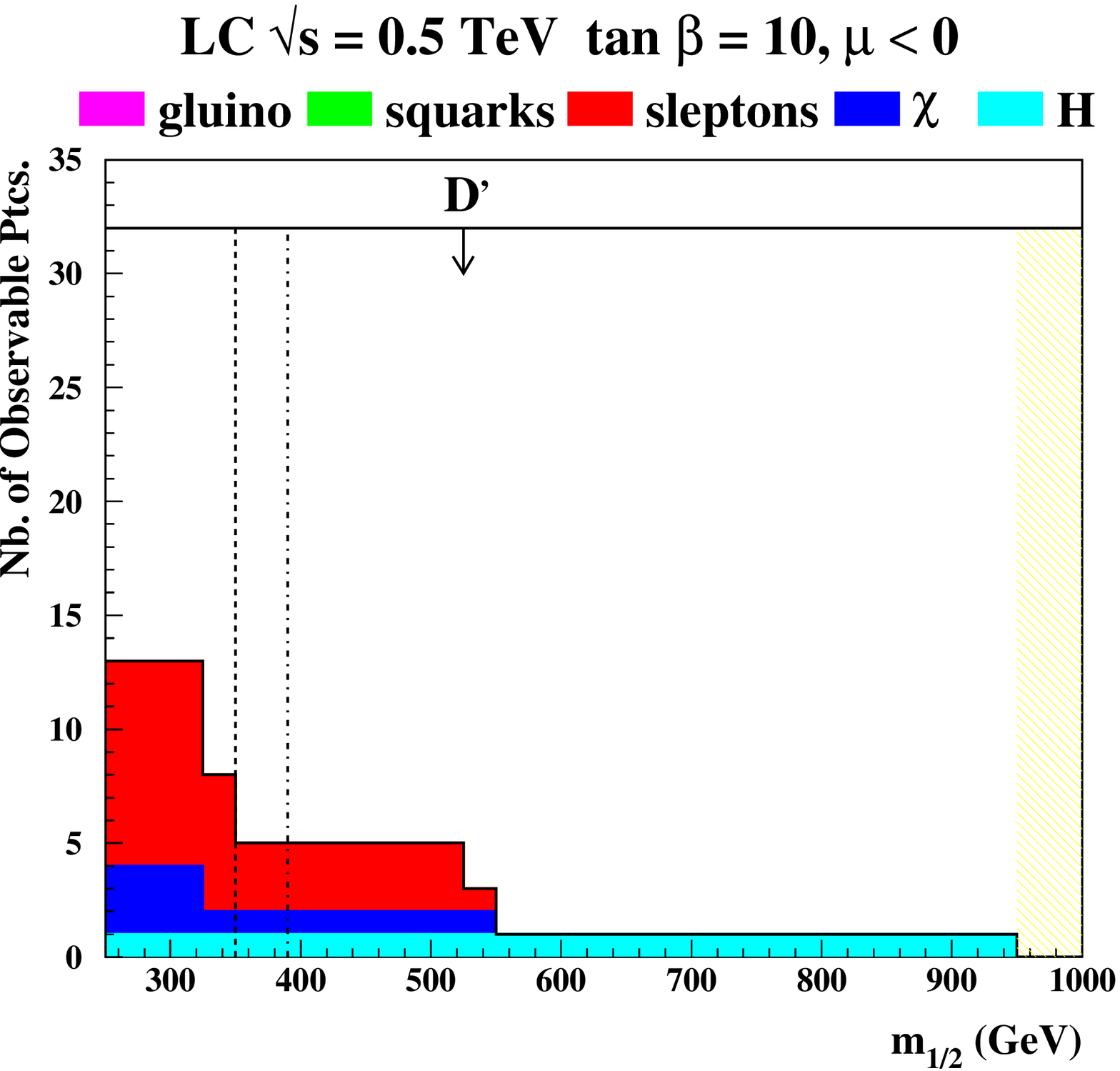}}
\caption{\it
Estimates of the numbers of MSSM particles that may be detectable at a
0.5-TeV linear $e^+ e^-$ collider
as functions of $m_{1/2}$ along the WMAP lines for $\mu > 0$ and $\tan
\beta = 5, 10, 20, 35$ and $50$, and for $\mu < 0$ and $\tan \beta = 10$. 
The locations of updated benchmark points along these WMAP lines are 
indicated, as are the nominal lower bounds on $m_{1/2}$ imposed by $m_h$  
(dashed lines) and $b \to s \gamma$ (dot-dashed lines).
\vspace*{0.5cm}}
\label{fig:lc0500}
\end{figure}

The capabilities of a linear $e^+e^-$ collider with $\sqrt{s}$ = 0.5~TeV
are illustrated in Fig.~\ref{fig:lc0500}~\cite{LC}. It can see some
gauginos and sleptons if $m_{1/2} \lappeq 500$~GeV. Its capabilities are
therefore well suited to the $g_\mu - 2$-friendlier scenarios displayed in
the left columns of Fig.~\ref{fig:newM}. Moreover, the sparticles that it
would see would complement those detectable at the LHC. Also, we recall
that such a linear collider would be able to measure sparticle properties
much more accurately than the LHC~\cite{Bench,SPS}.

\begin{figure}[t]
\centerline{\epsfxsize = 0.4\textwidth 
\epsffile{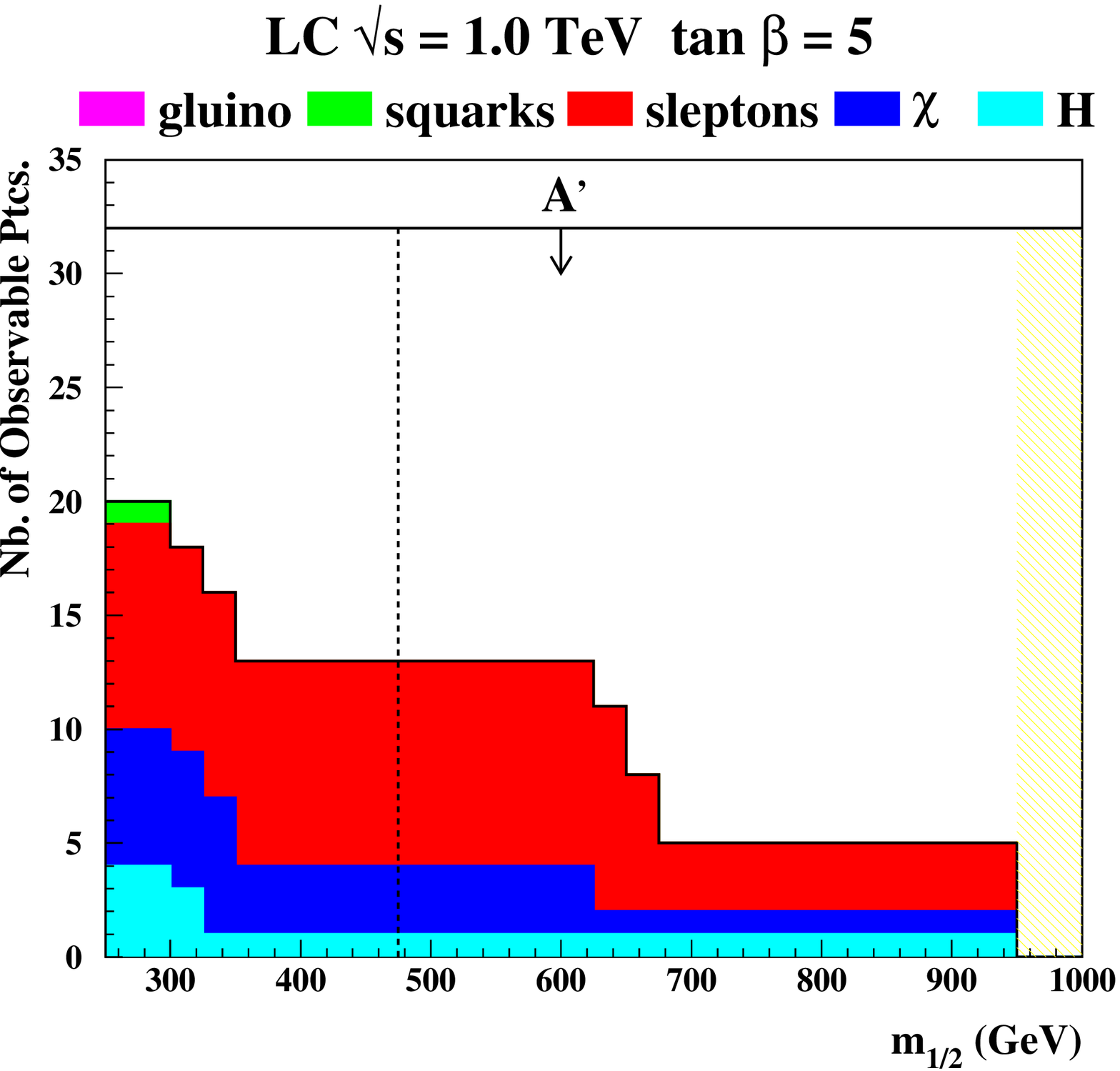}
\hfill \epsfxsize = 0.4\textwidth \epsffile{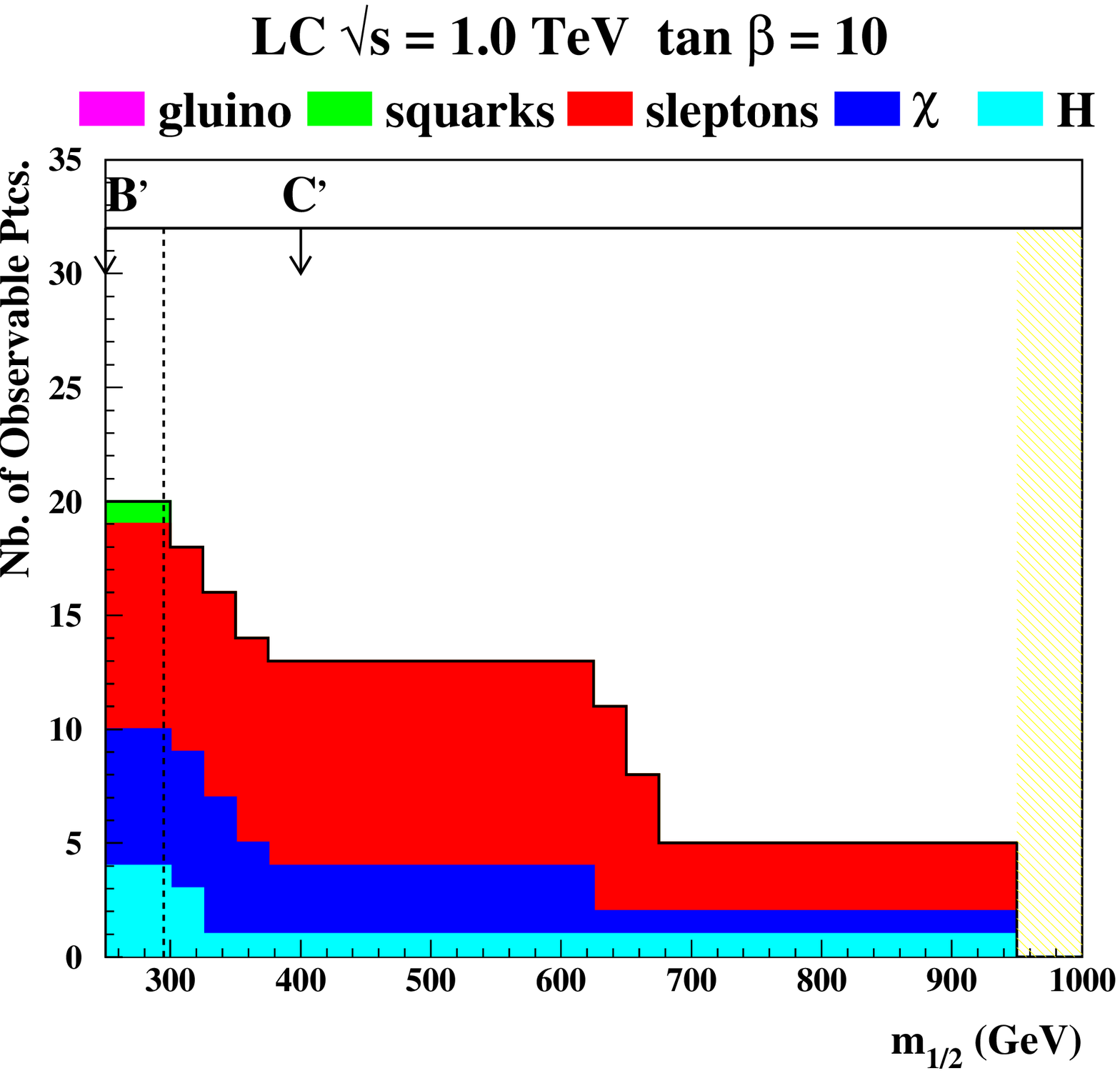}}
\centerline{\epsfxsize = 0.4\textwidth 
\epsffile{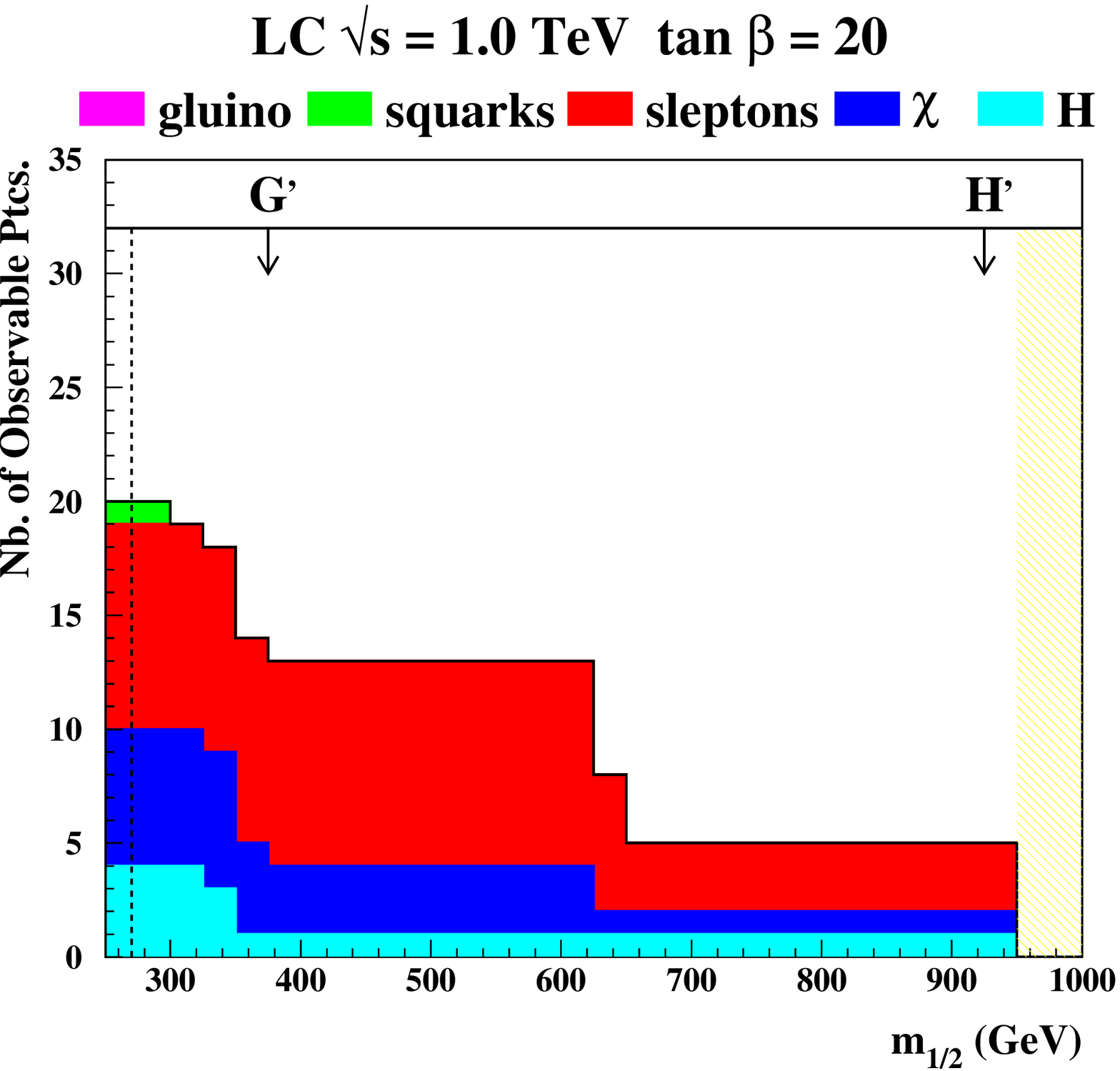}
\hfill \epsfxsize = 0.4\textwidth \epsffile{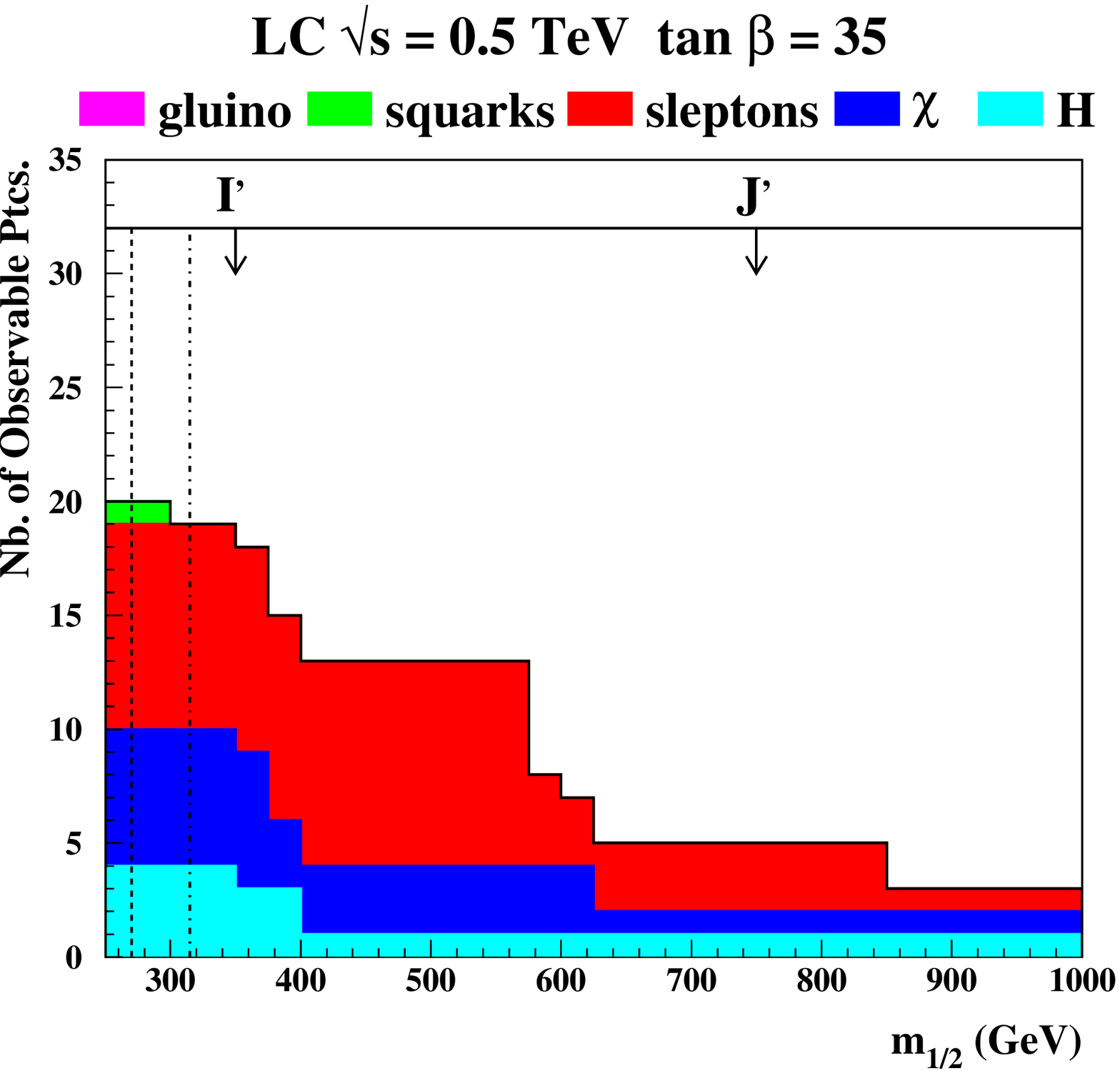}}
\centerline{\epsfxsize = 0.4\textwidth 
\epsffile{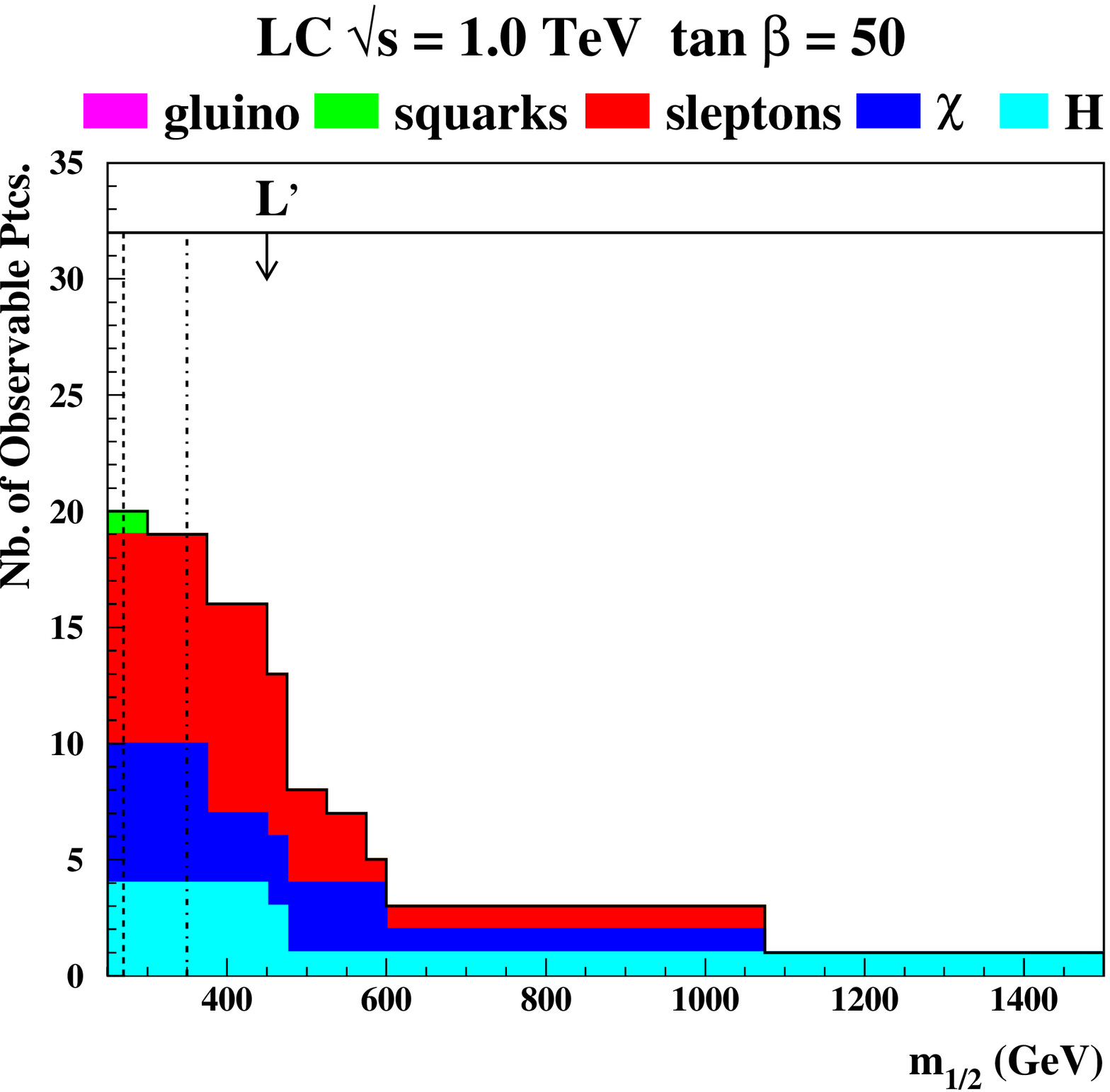}
\hfill \epsfxsize = 0.4\textwidth \epsffile{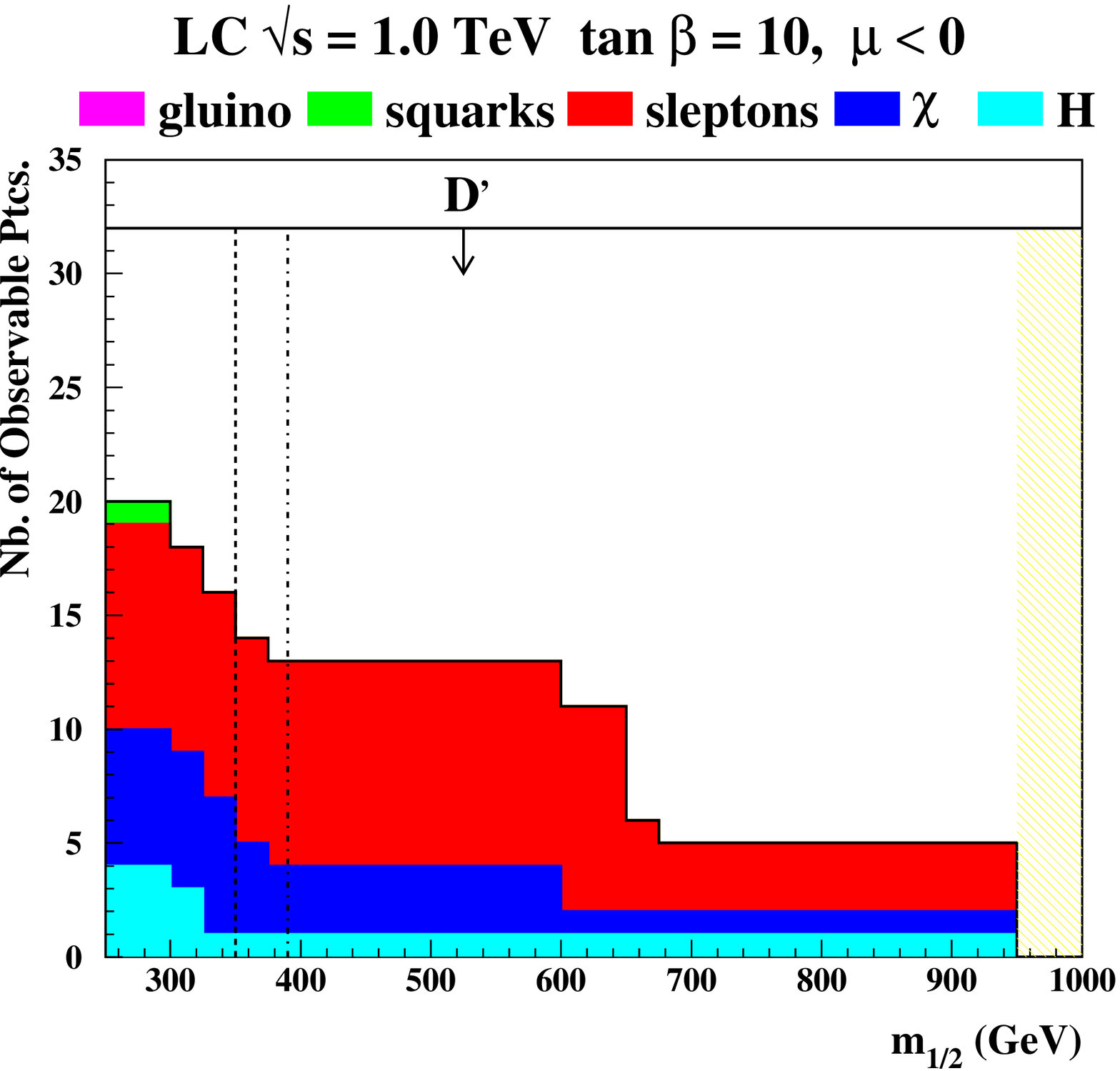}}
\caption{\it
Estimates of the numbers of MSSM particles that may be detectable at a  
1-TeV linear $e^+ e^-$ collider
as functions of $m_{1/2}$ along the WMAP lines for $\mu > 0$ and $\tan   
\beta = 5, 10, 20, 35$ and $50$, and for $\mu < 0$ and $\tan \beta = 10$.
The locations of updated benchmark points along these WMAP lines are
indicated, as are the nominal lower bounds on $m_{1/2}$ imposed by $m_h$
(dashed lines) and $b \to s \gamma$ (dot-dashed lines).
\vspace*{0.5cm}}  
\label{fig:lc1000}   
\end{figure}

A linear $e^+e^-$ collider able to deliver collisions at a centre of mass
energy $\sqrt{s} = 1$~TeV has the potential to complement significantly
the LHC in studying the supersymmetric spectrum, as seen in
Fig.~\ref{fig:lc1000}. In particular, such a linear collider would observe
most sleptons and gauginos as long as $m_{1/2}$ is below about 600~GeV, as
exemplified by benchmarks B, C, G, I and L. Also, it would detect at least
one supersymmetric particle over the whole range of the WMAP lines, even
up to their upper $m_{1/2}$ limits exemplified by benchmark point H,
except in the funnel cases: $\mu > 0, \tan \beta = 50$ and $\mu < 0, \tan
\beta = 35$. The accuracy it would provide in the measurements of several
sparticle masses would enable GUT mass relations to be tested, thereby
completing the exploitation of the LHC data. We note also that a 1-TeV
linear $e^+e^-$ collider would be able to observe the lightest
squark, the ${\tilde t_1}$, at low values of $m_{1/2}$. This possibility
is exemplified by benchmark point B, as also seen in Fig.~\ref{fig:newM}.

\begin{figure}[t]
\centerline{\epsfxsize = 0.4\textwidth
\epsffile{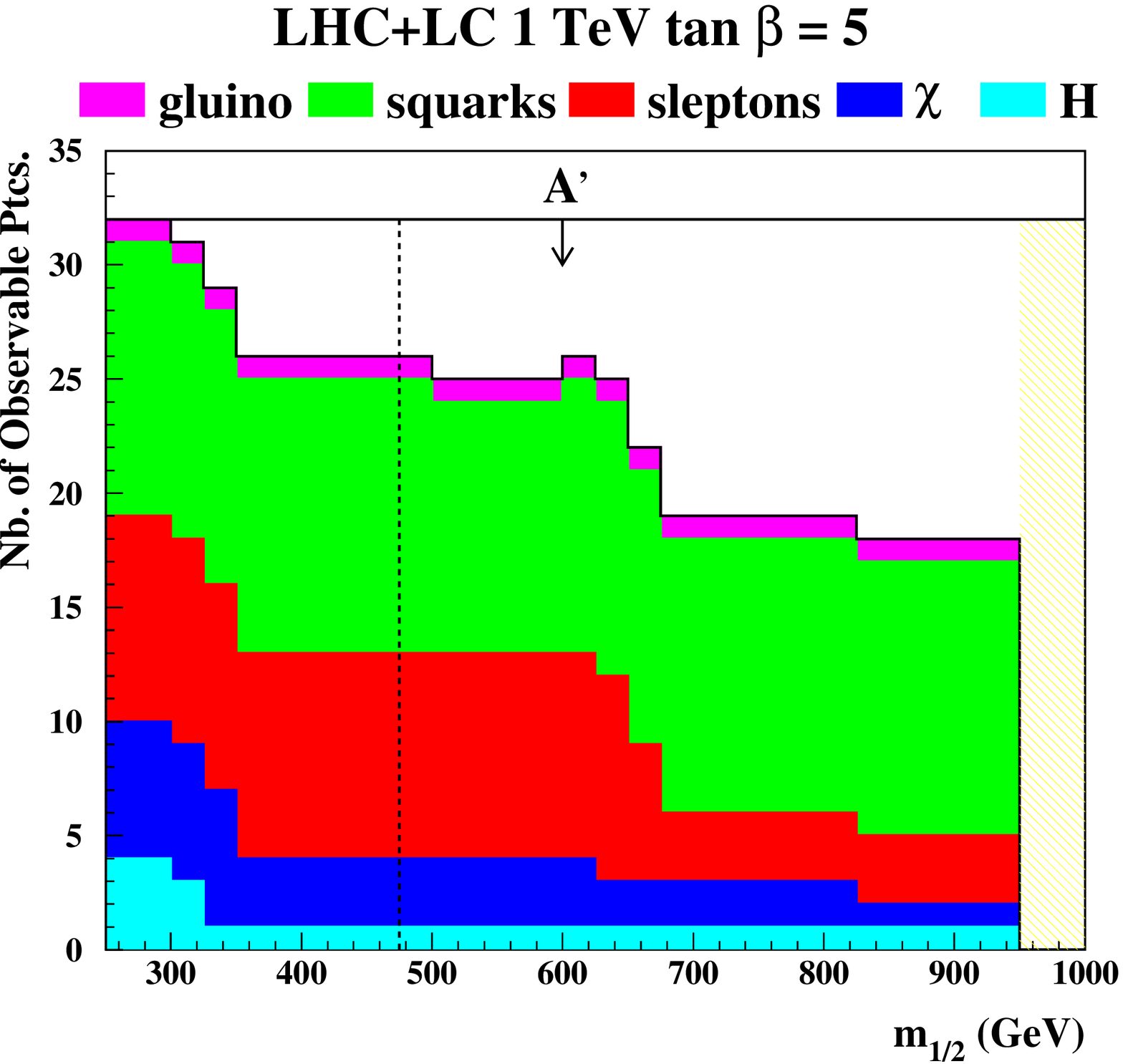}
\hfill \epsfxsize = 0.4\textwidth \epsffile{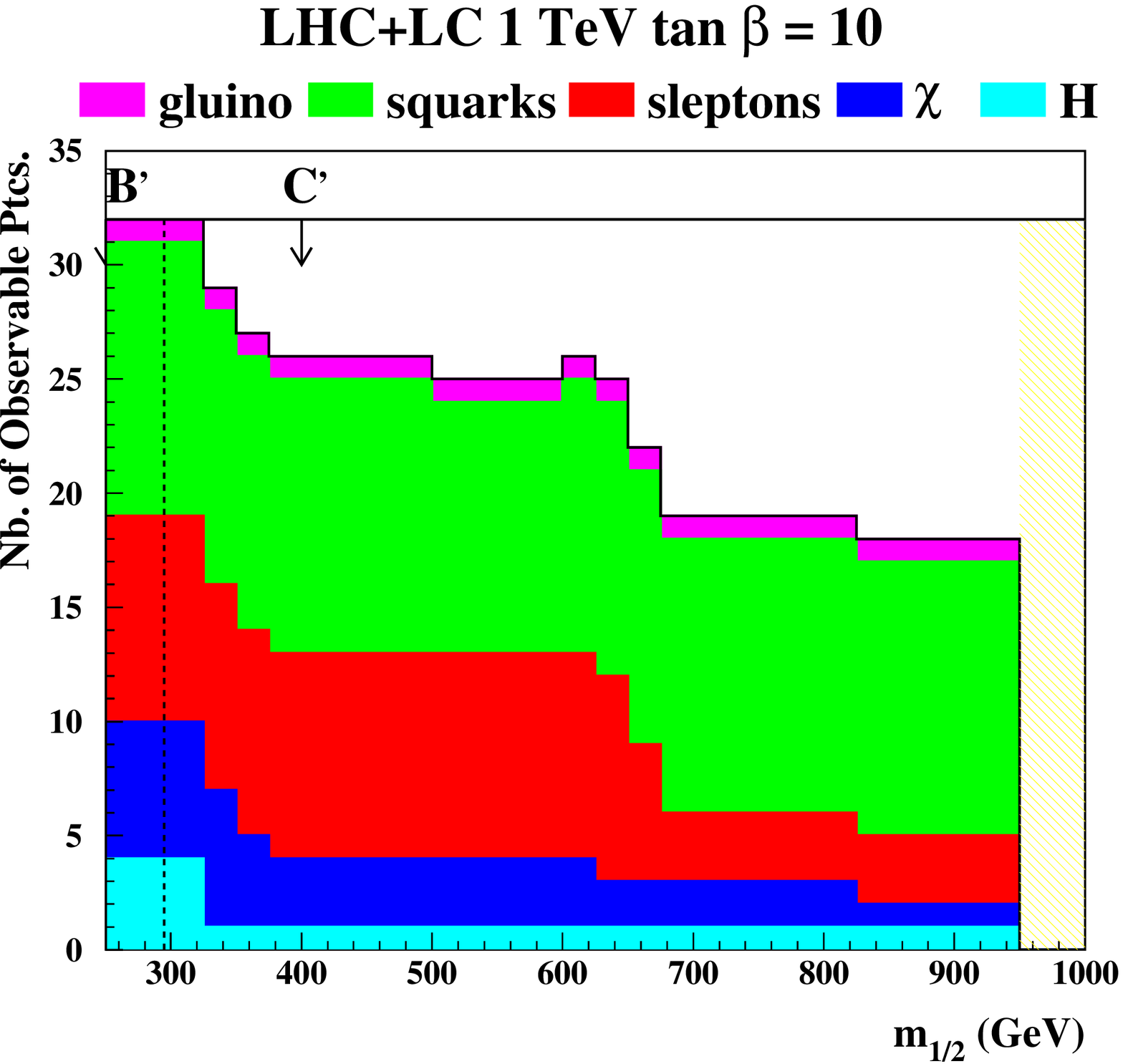}}
\centerline{\epsfxsize = 0.4\textwidth
\epsffile{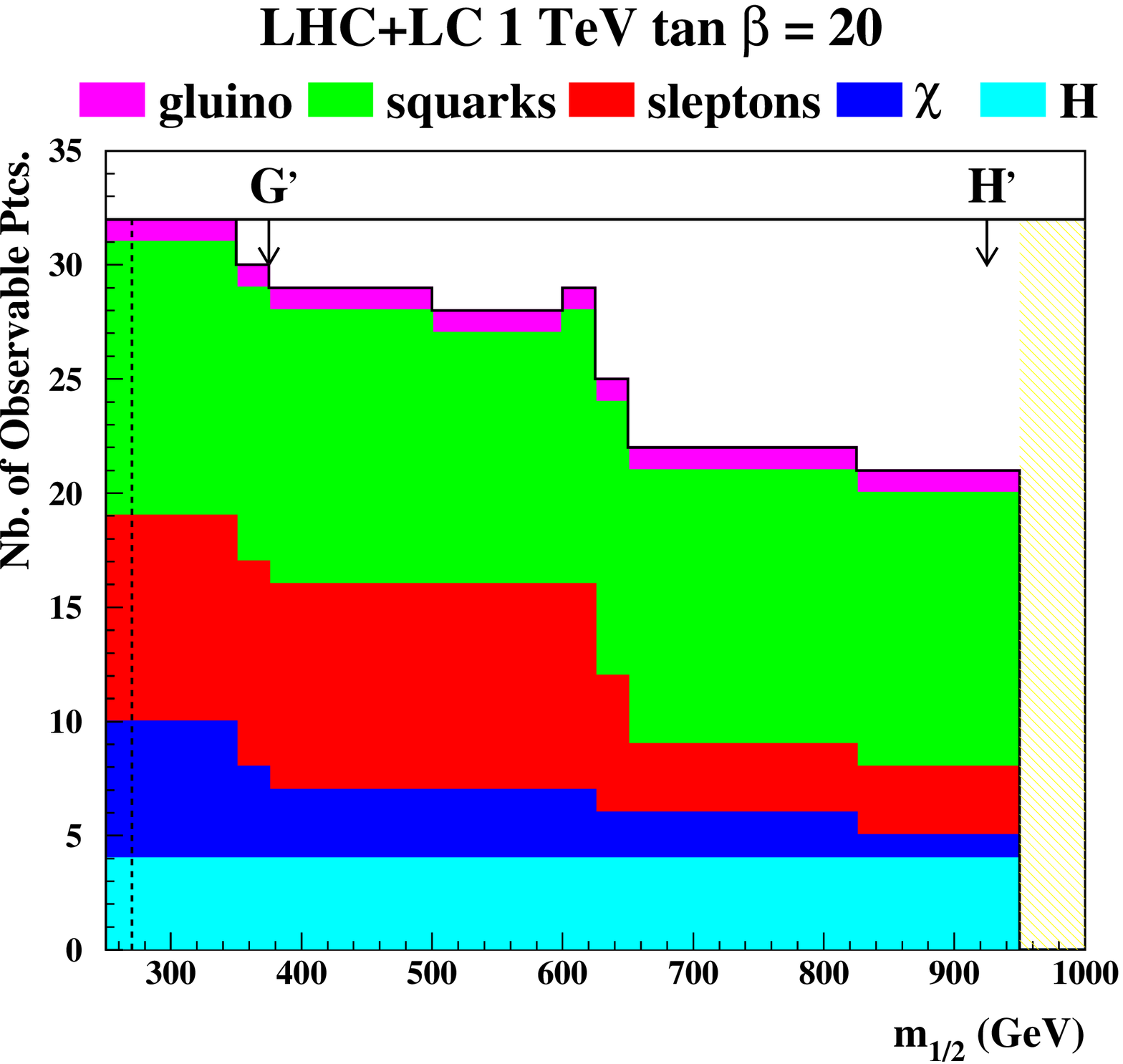}
\hfill \epsfxsize = 0.4\textwidth \epsffile{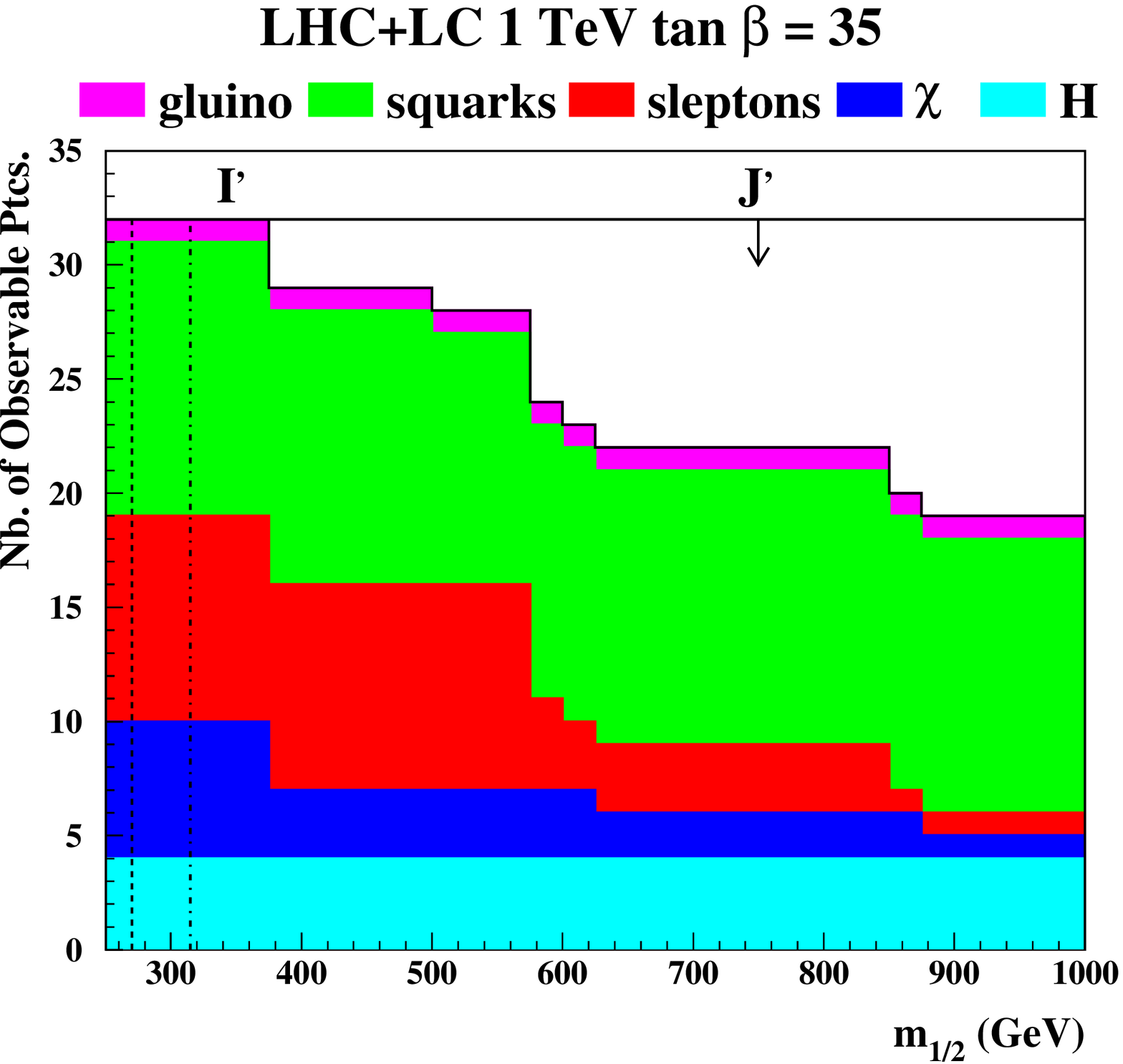}}
\centerline{\epsfxsize = 0.4\textwidth
\epsffile{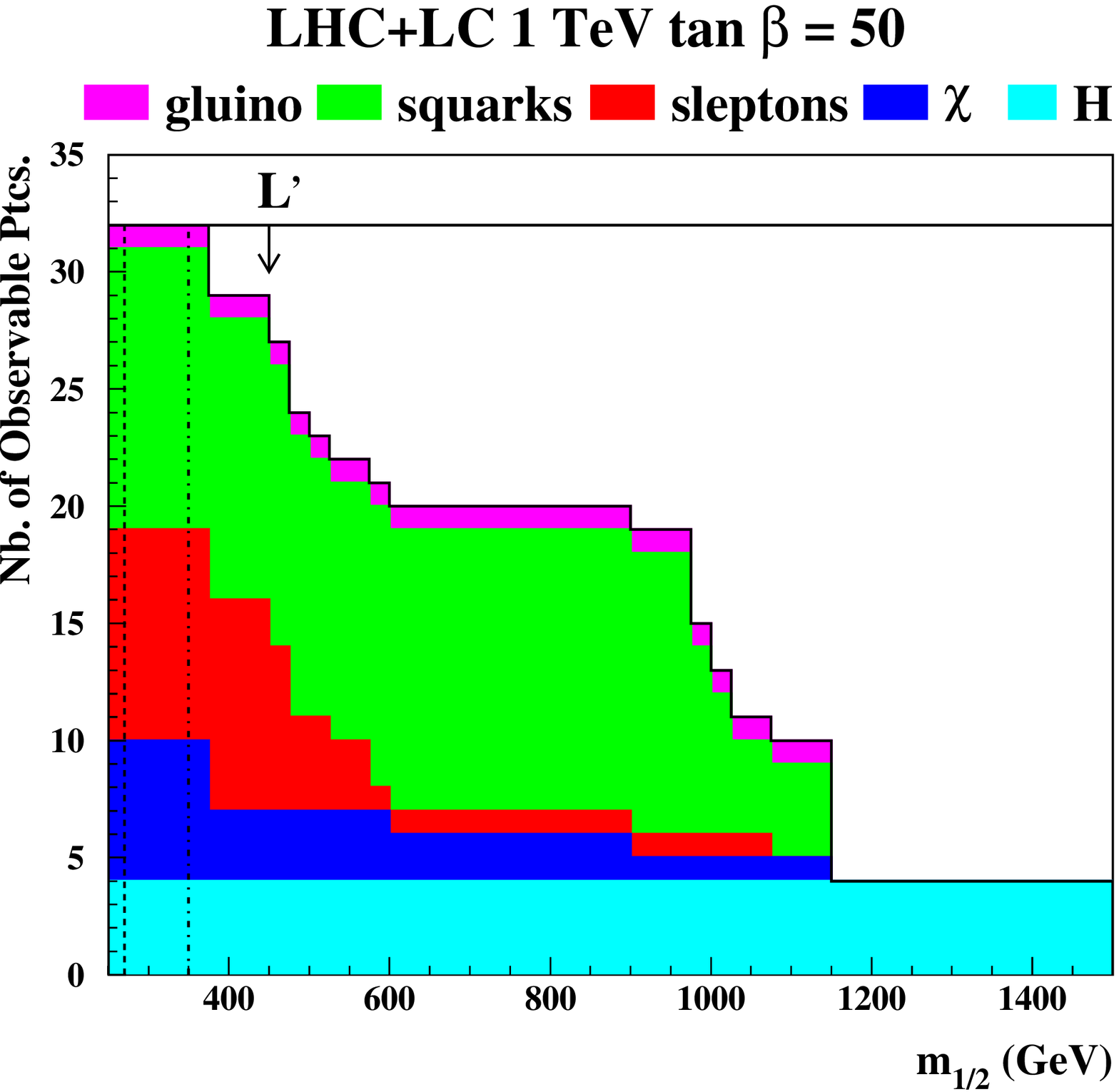}
\hfill \epsfxsize = 0.4\textwidth \epsffile{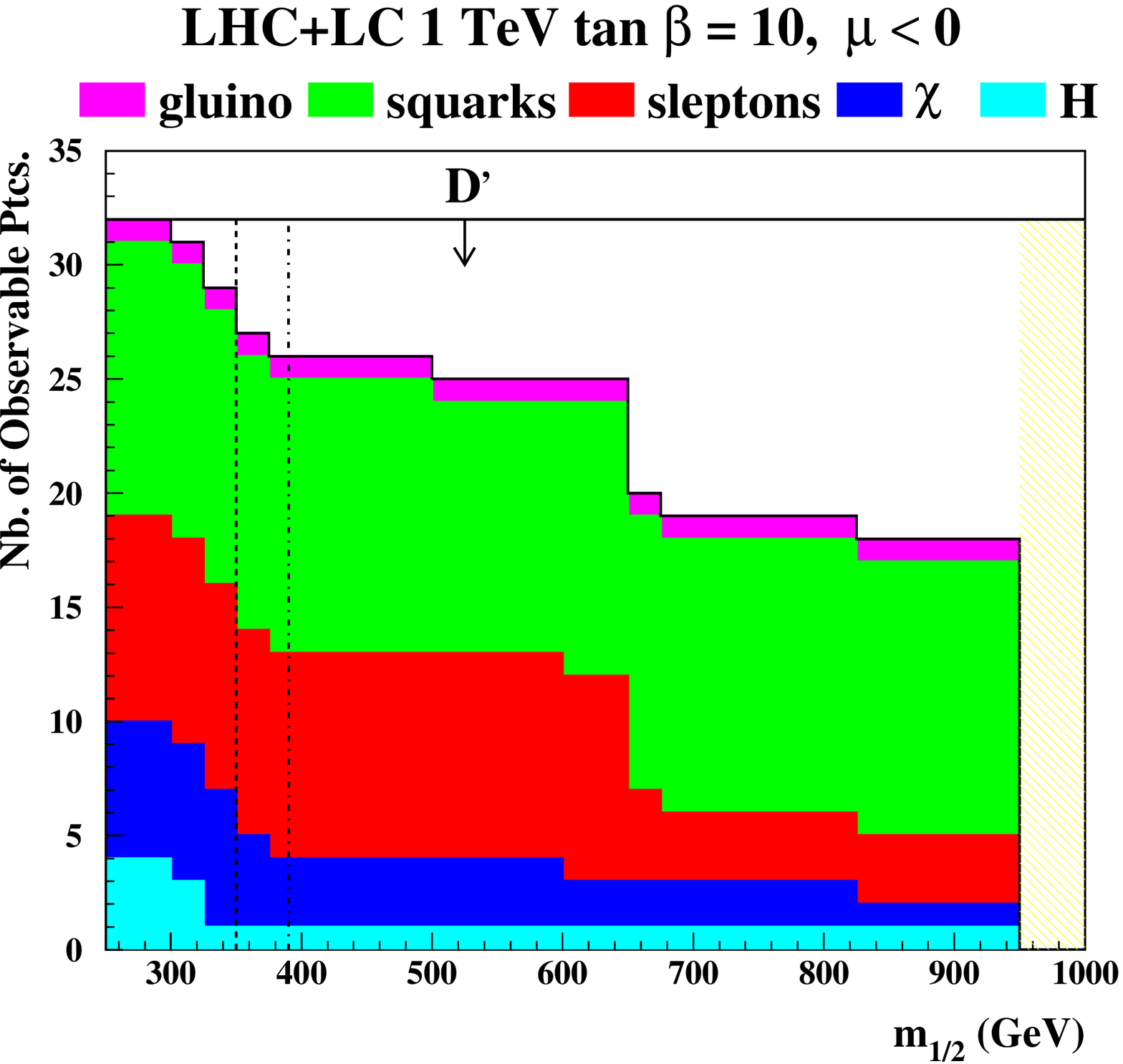}}
\caption{\it
Estimates of the combined numbers of MSSM particles that may be detectable 
at a combination of the LHC and a 1-TeV linear $e^+ e^-$ collider
as functions of $m_{1/2}$ along the WMAP lines for $\mu > 0$ and $\tan   
\beta = 5, 10, 20, 35$ and $50$, and for $\mu < 0$ and $\tan \beta = 10$.
The locations of updated benchmark points along these WMAP lines are
indicated, as are the nominal lower bounds on $m_{1/2}$ imposed by $m_h$
(dashed lines) and $b \to s \gamma$ (dot-dashed lines).
\vspace*{0.5cm}}
\label{fig:lhclc}
\end{figure}

The complementarity between the LHC and  1-TeV linear $e^+ e^-$ collider 
is displayed clearly in Fig.~\ref{fig:lhclc}, where we see that, together, 
they cover the majority of the MSSM spectrum over most of the $m_{1/2}$ 
ranges covered by the different WMAP lines.

\subsubsection{CLIC}

\begin{figure}[t]
\centerline{\epsfxsize = 0.4\textwidth 
\epsffile{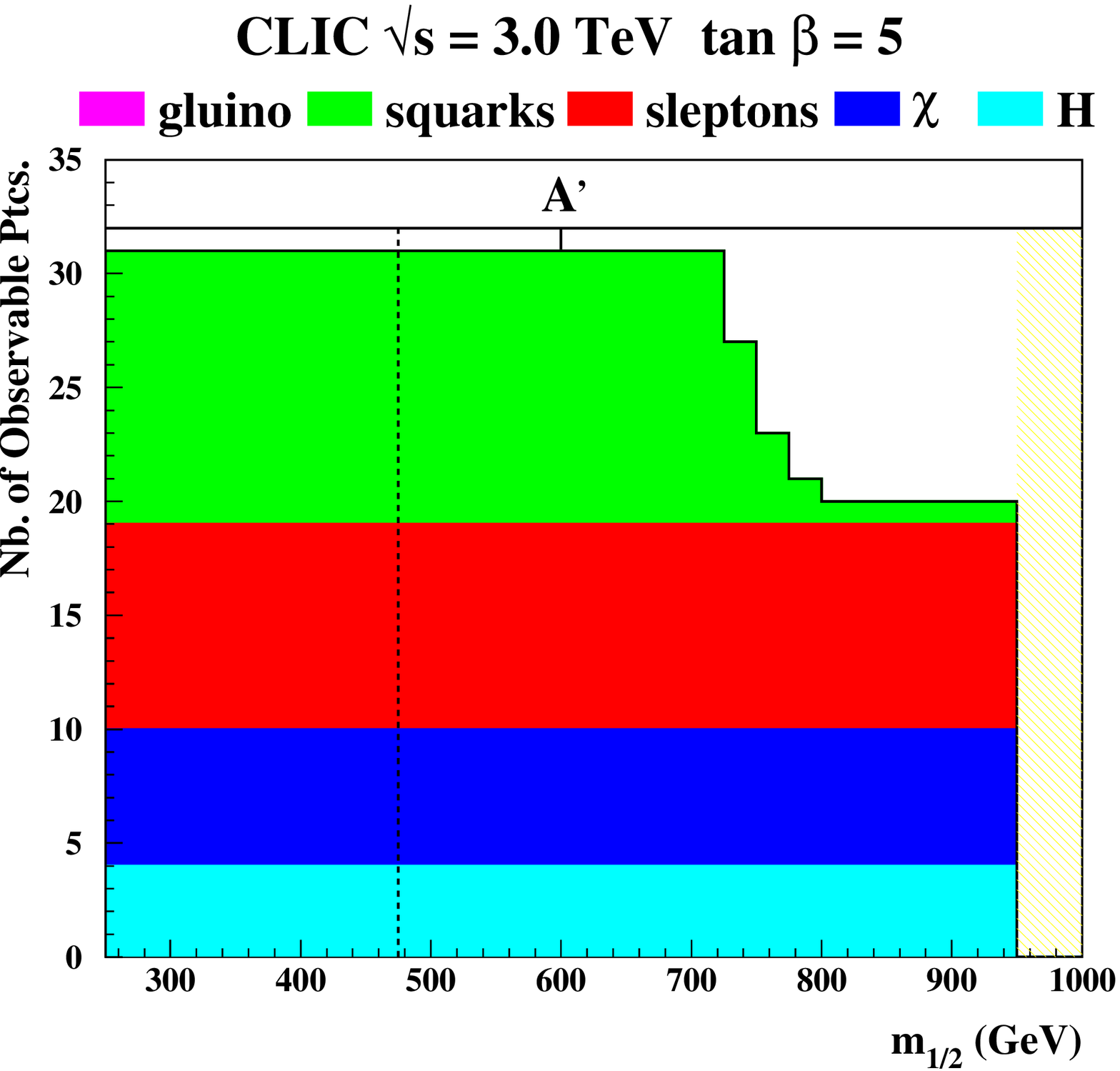}
\hfill \epsfxsize = 0.4\textwidth \epsffile{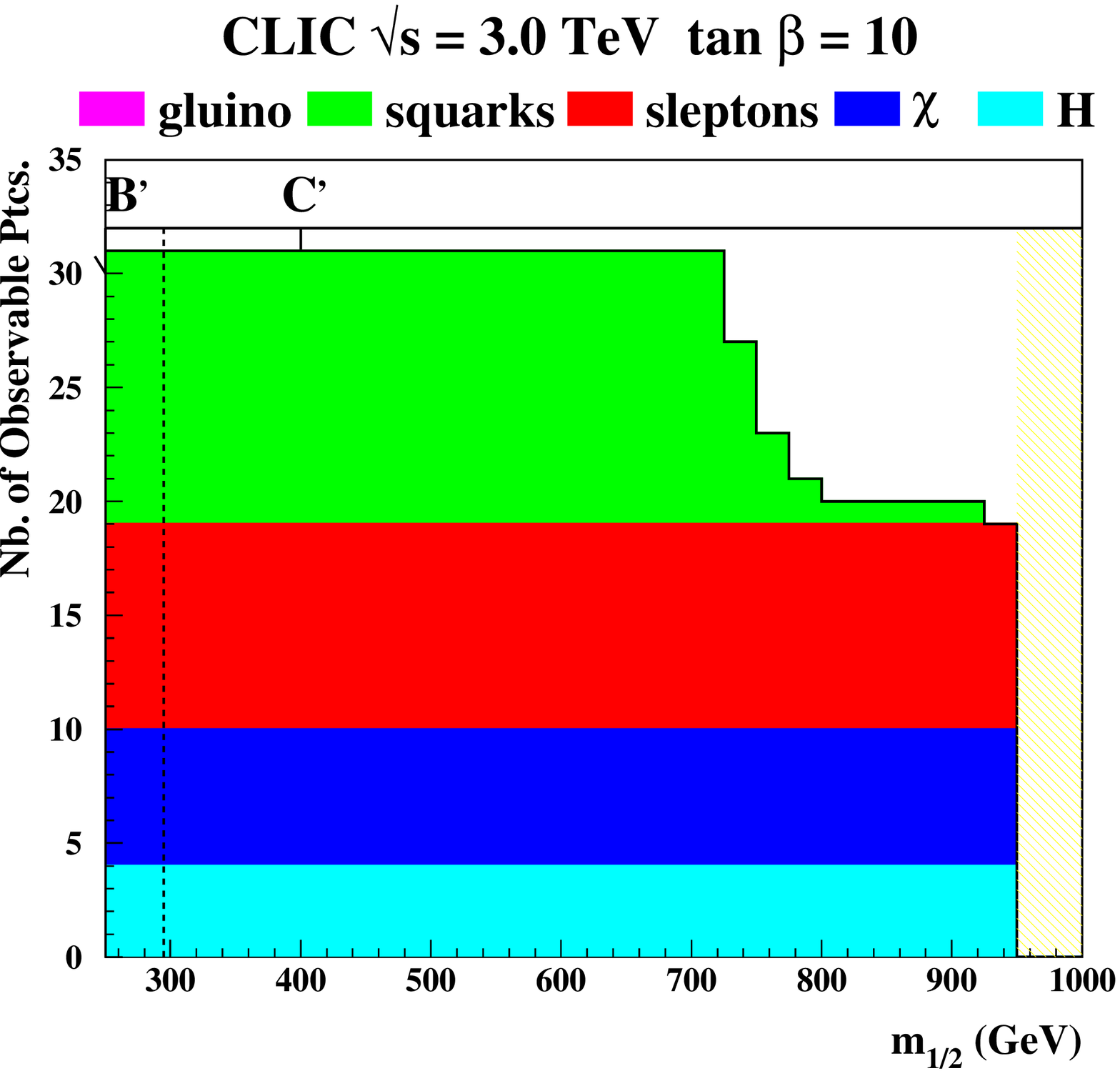}}
\centerline{\epsfxsize = 0.4\textwidth 
\epsffile{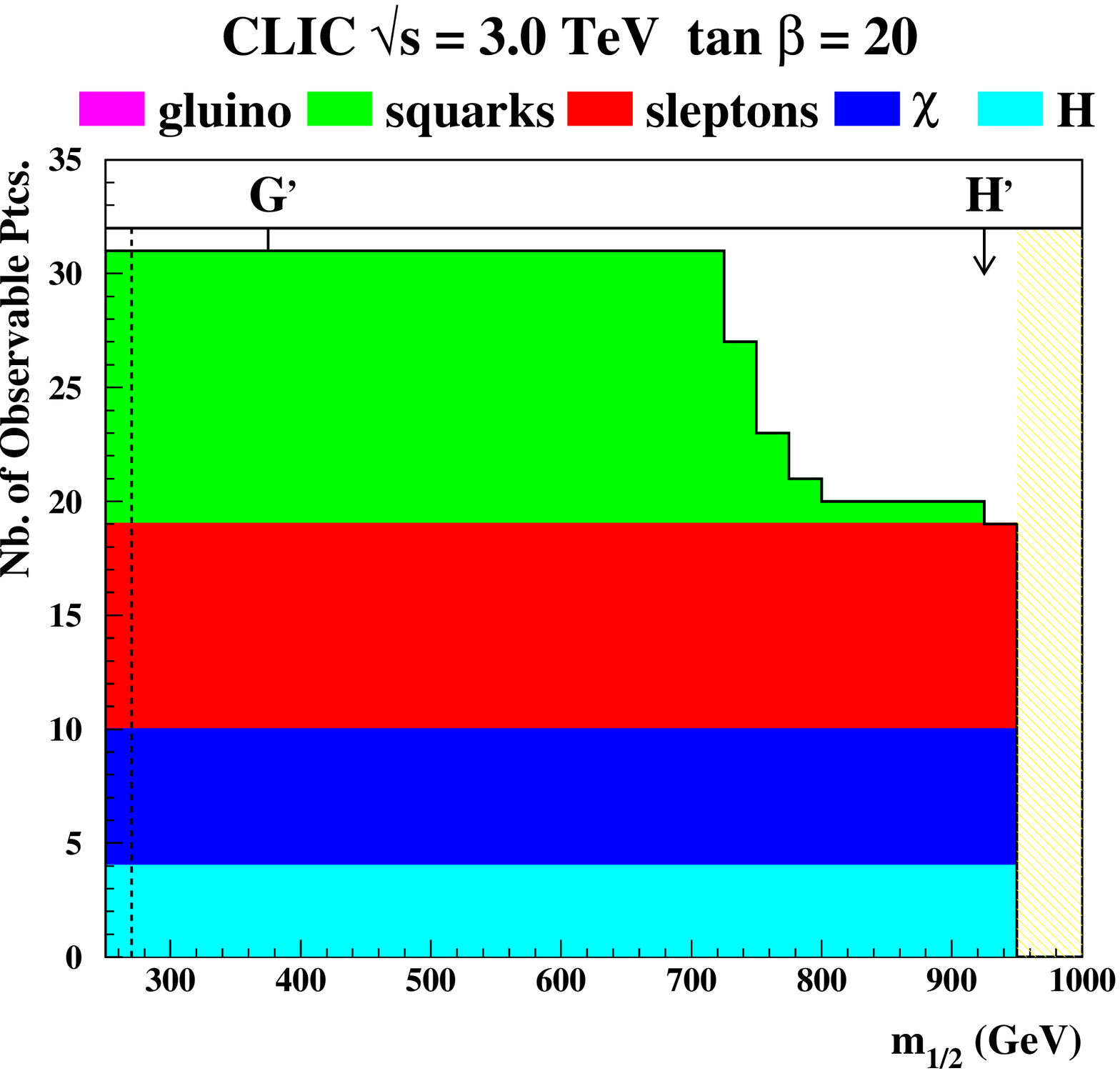}
\hfill \epsfxsize = 0.4\textwidth \epsffile{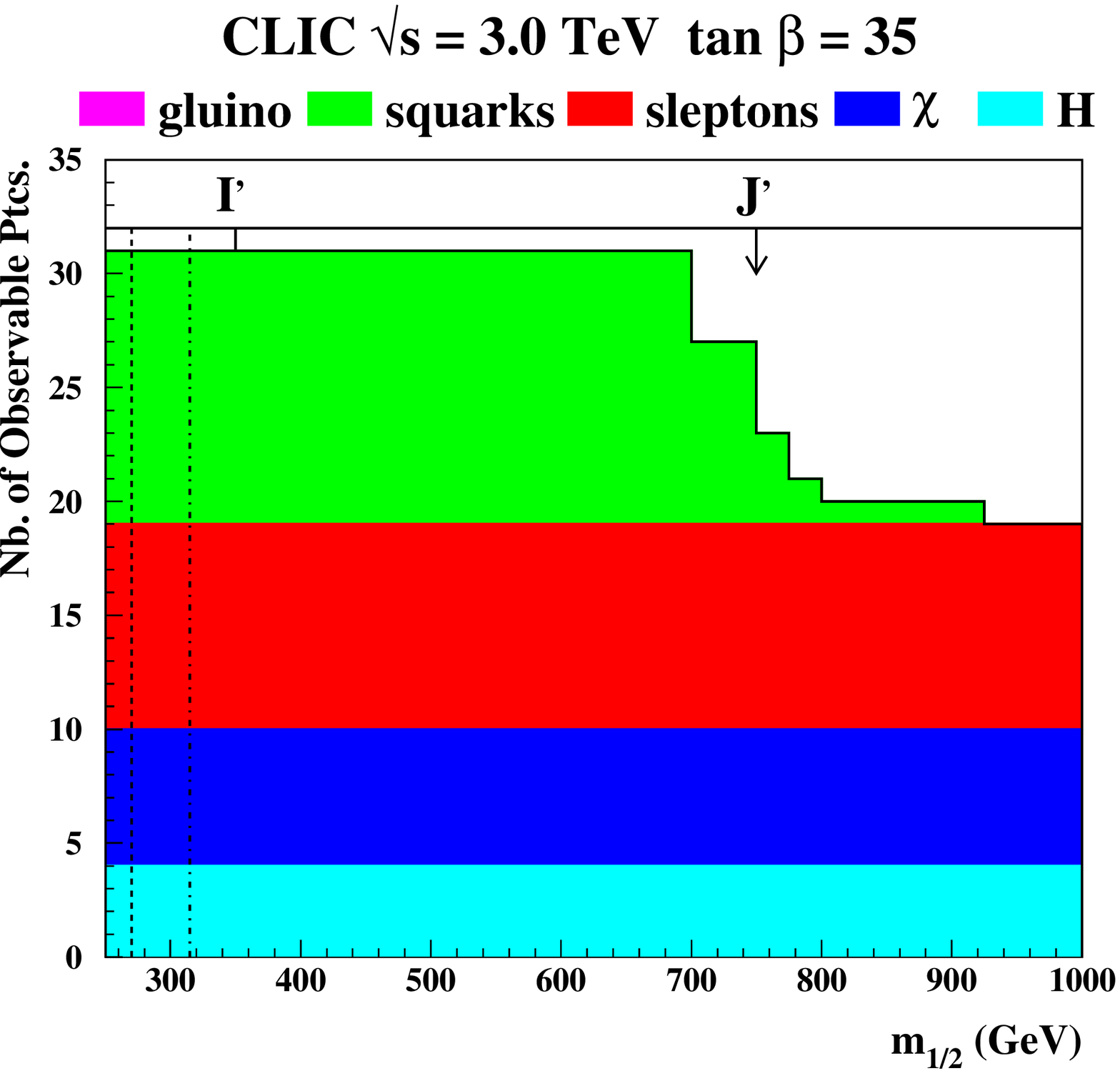}}
\centerline{\epsfxsize = 0.4\textwidth 
\epsffile{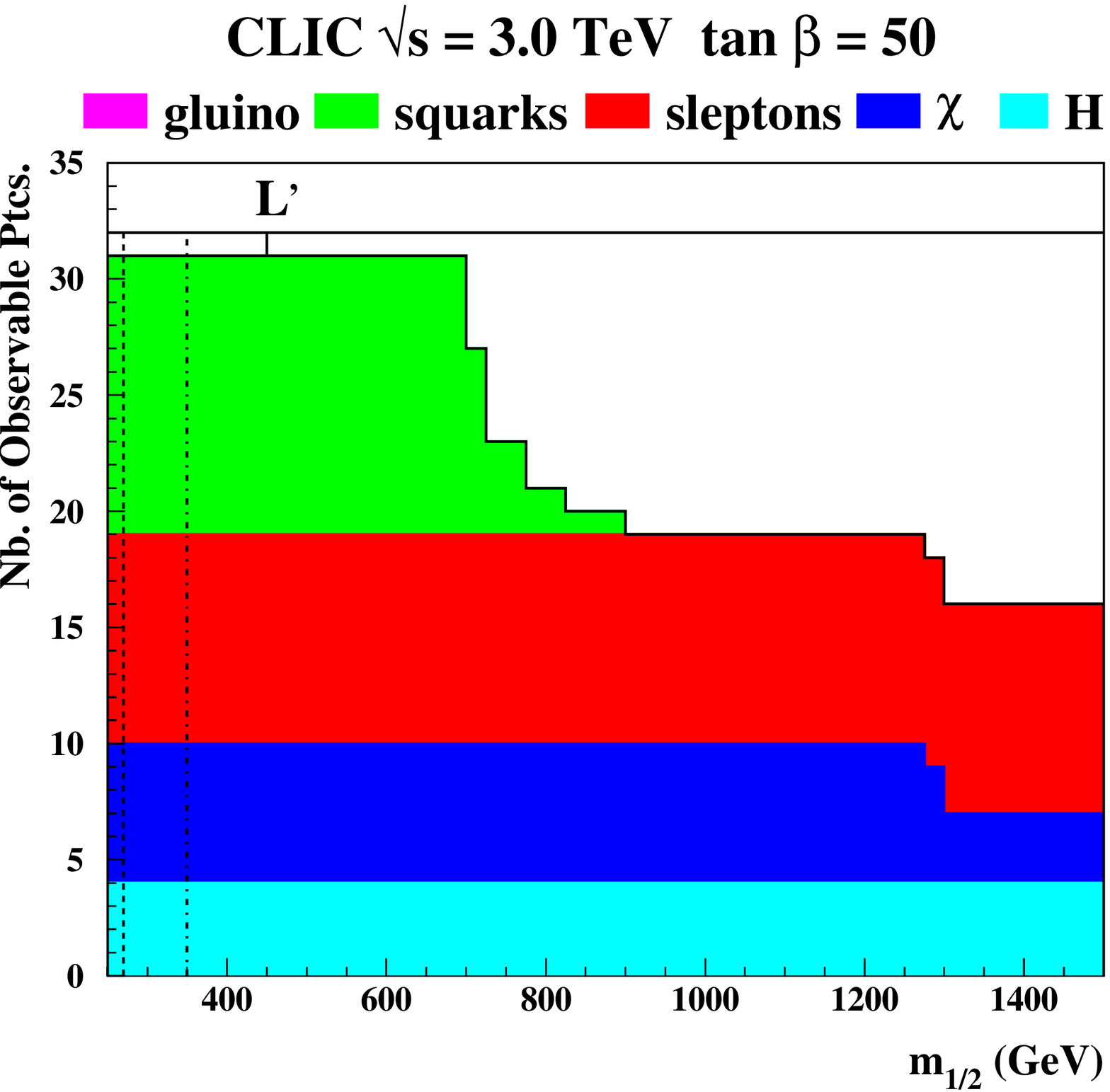}
\hfill \epsfxsize = 0.4\textwidth \epsffile{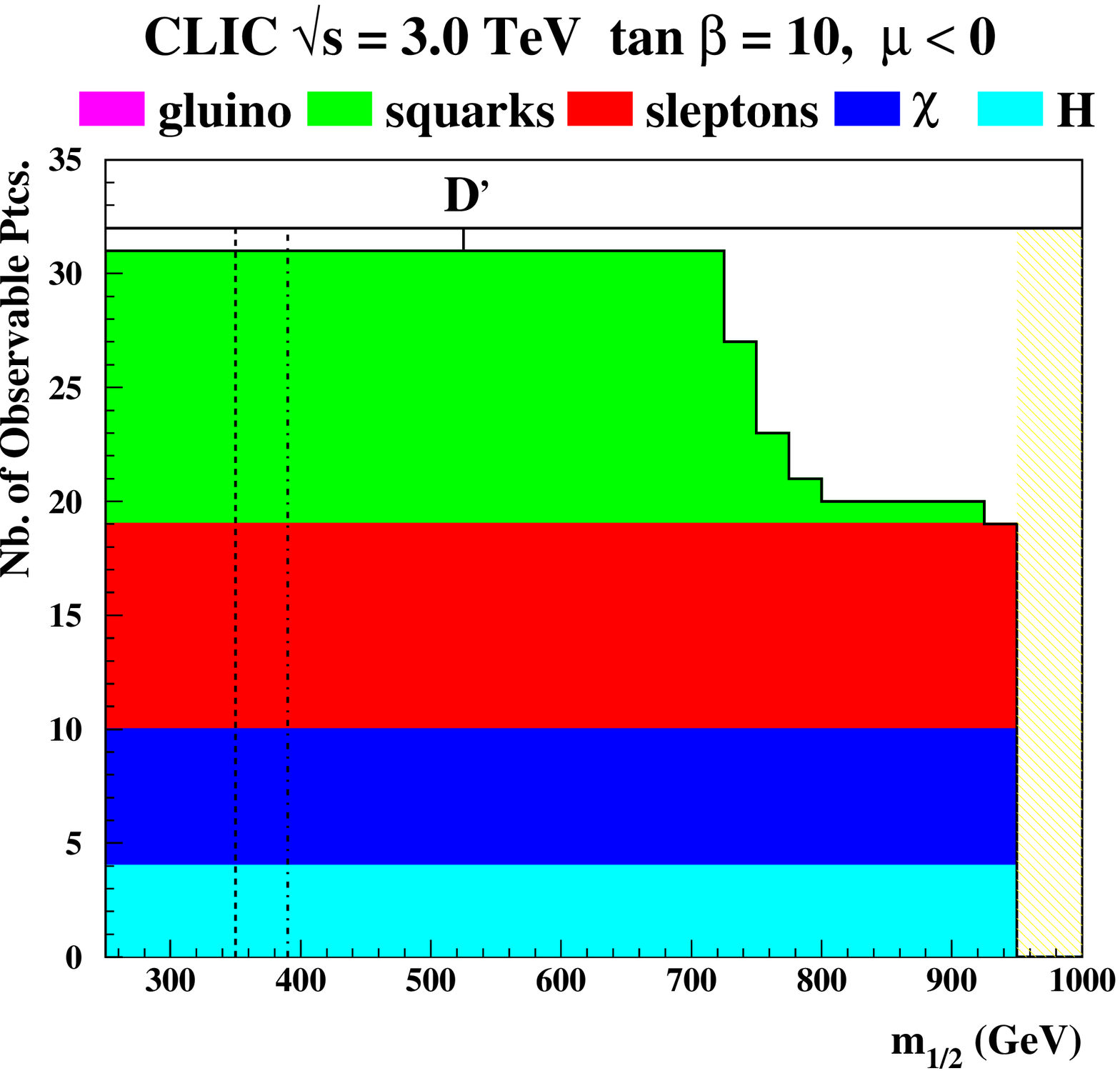}}
\caption{\it
Estimates of the numbers of MSSM particles that may be detectable at the   
3-TeV version of the CLIC linear $e^+ e^-$ collider
as functions of $m_{1/2}$ along the WMAP lines for $\mu > 0$ and $\tan   
\beta = 5, 10, 20, 35$ and $50$, and for $\mu < 0$ and $\tan \beta = 10$.
The locations of updated benchmark points along these WMAP lines are
indicated, as are the nominal lower bounds on $m_{1/2}$ imposed by $m_h$
(dashed lines) and $b \to s \gamma$ (dot-dashed lines).
\vspace*{0.5cm}}  
\label{fig:lc3000}   
\end{figure}

A 3-TeV lepton collider, such as CLIC, is expected to access almost all the
sparticle spectrum for $m_{1/2} < 700$~GeV, as seen in
Fig.~\ref{fig:lc3000}. This would enable it, for example, to distinguish
and provide detailed measurements of the different flavours of squarks
that will not be possible at the LHC. Moreover, at larger $m_{1/2}$ CLIC
would be able to observe (almost) all the spectrum of Higgs bosons,
sleptons, charginos and neutralinos~\cite{CLICPhys}. CLIC will also be 
able to observe gluinos via squark decays at the focus-point benchmark 
E~\footnote{As already noted, we do not discuss here the possibility of 
observing gluinos in $\gamma \gamma$ collisions.}. It would therefore
provide full complementarity with the LHC along the full extension of the
lines.

\begin{figure}[t]
\centerline{\epsfxsize = 0.4\textwidth 
\epsffile{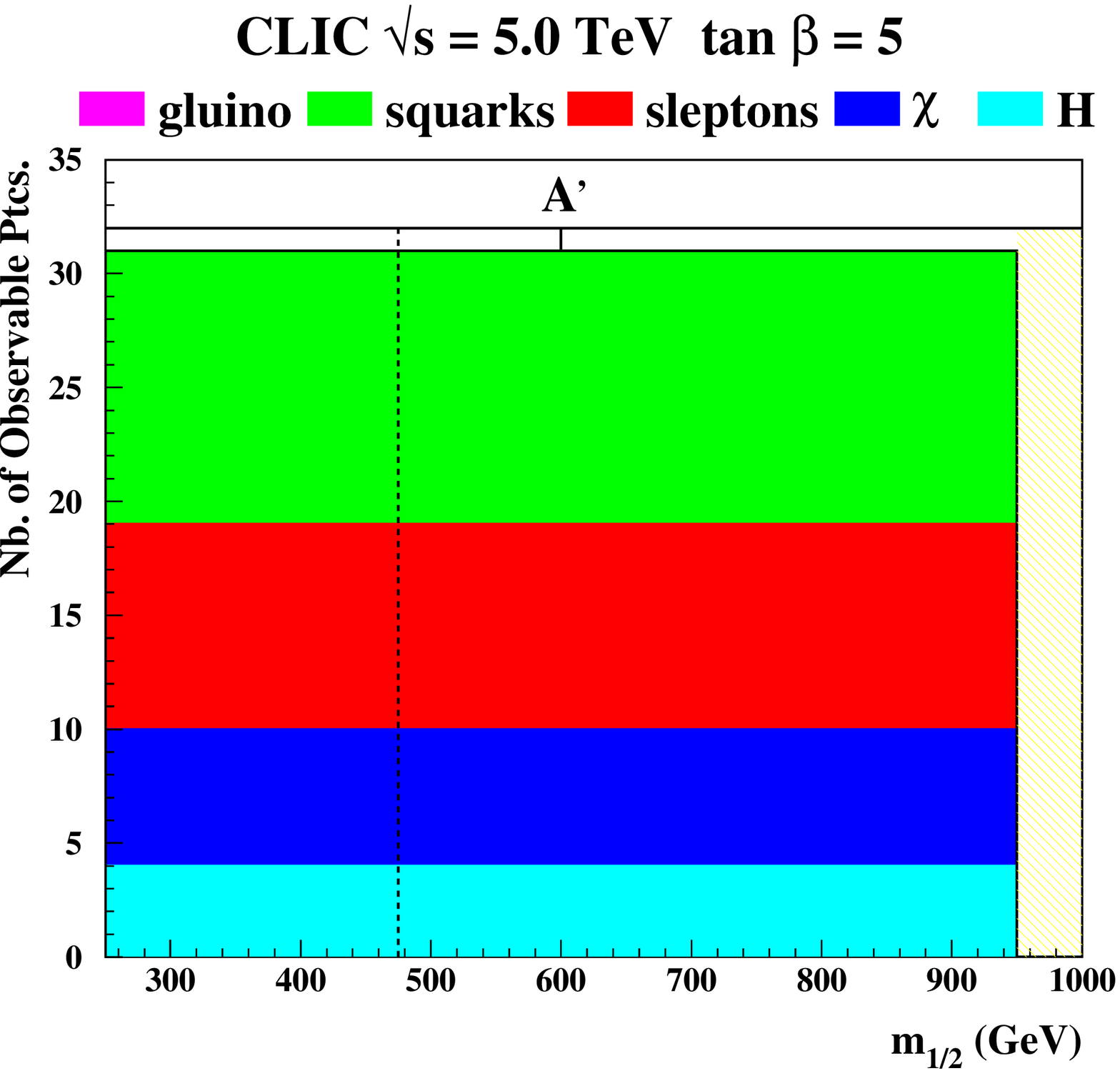}
\hfill \epsfxsize = 0.4\textwidth \epsffile{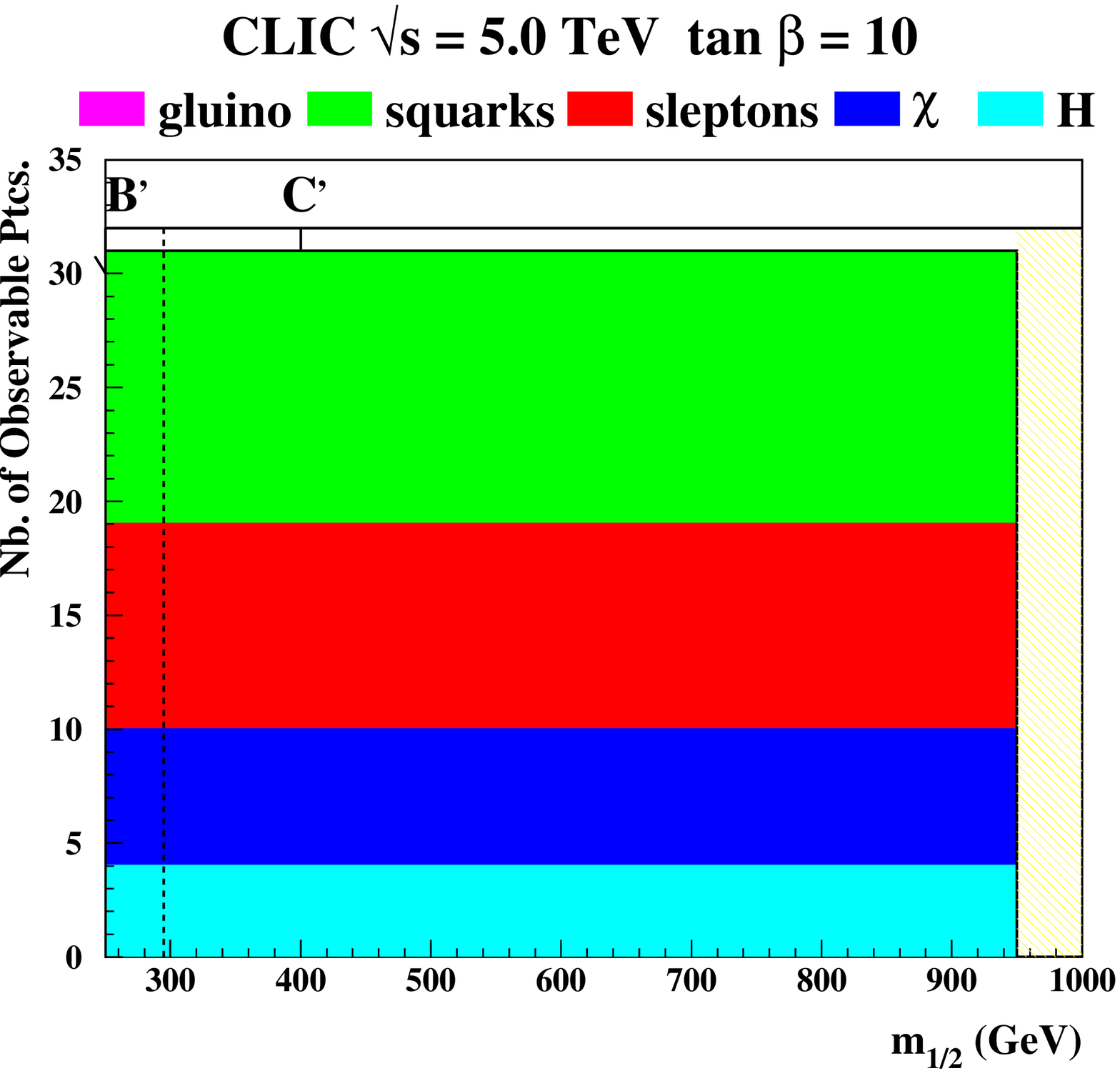}}
\centerline{\epsfxsize = 0.4\textwidth 
\epsffile{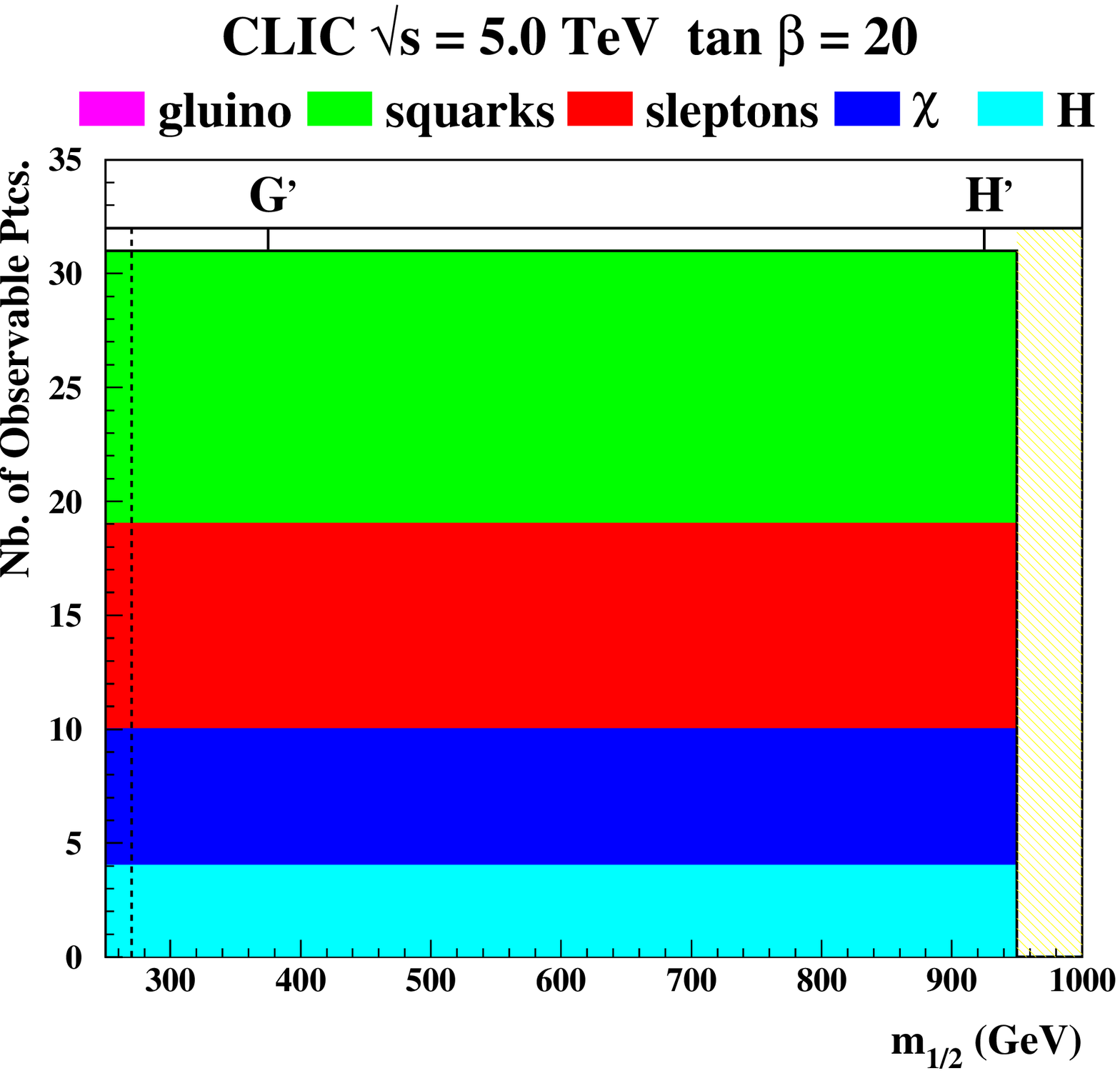}
\hfill \epsfxsize = 0.4\textwidth \epsffile{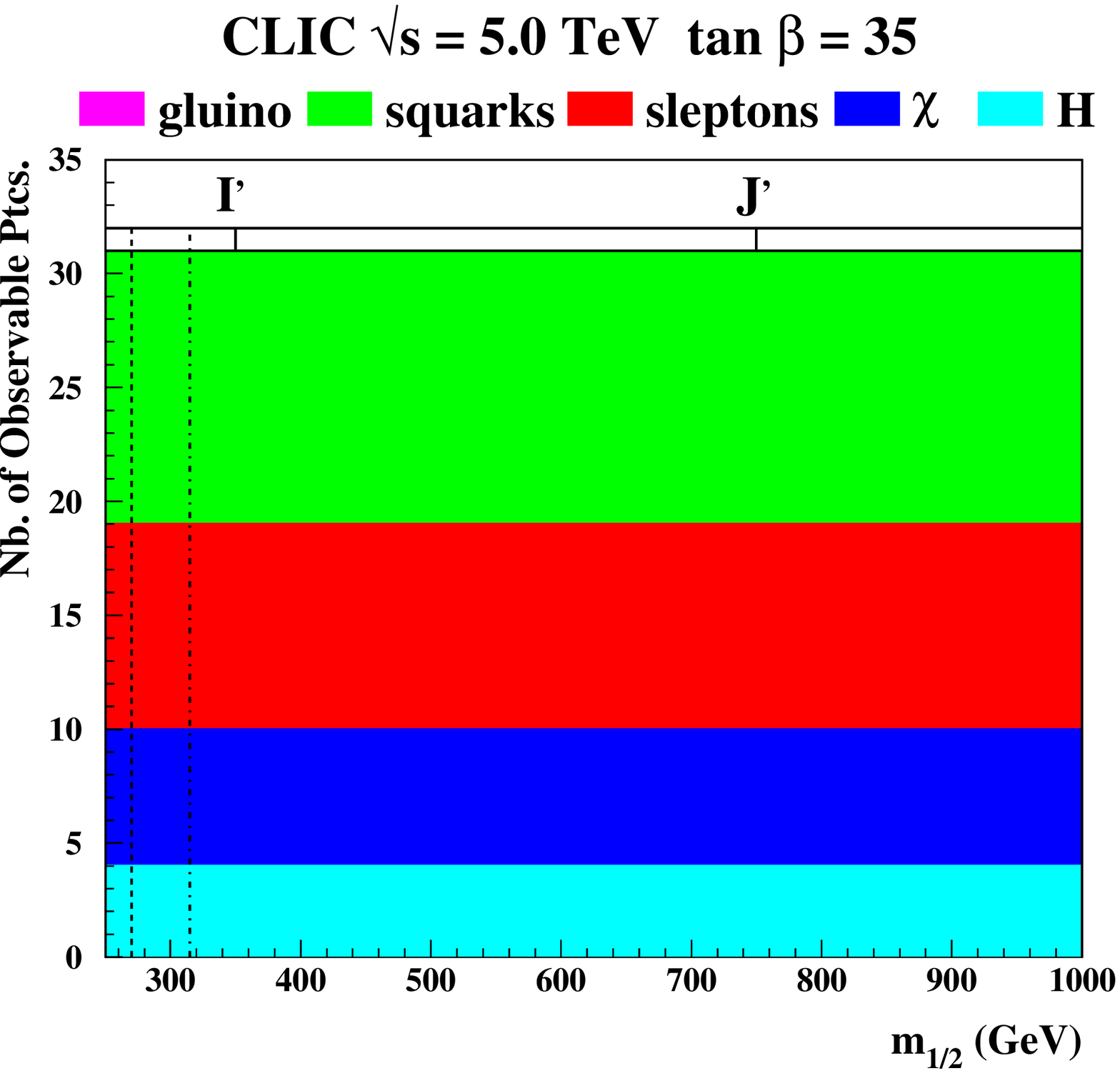}}
\centerline{\epsfxsize = 0.4\textwidth 
\epsffile{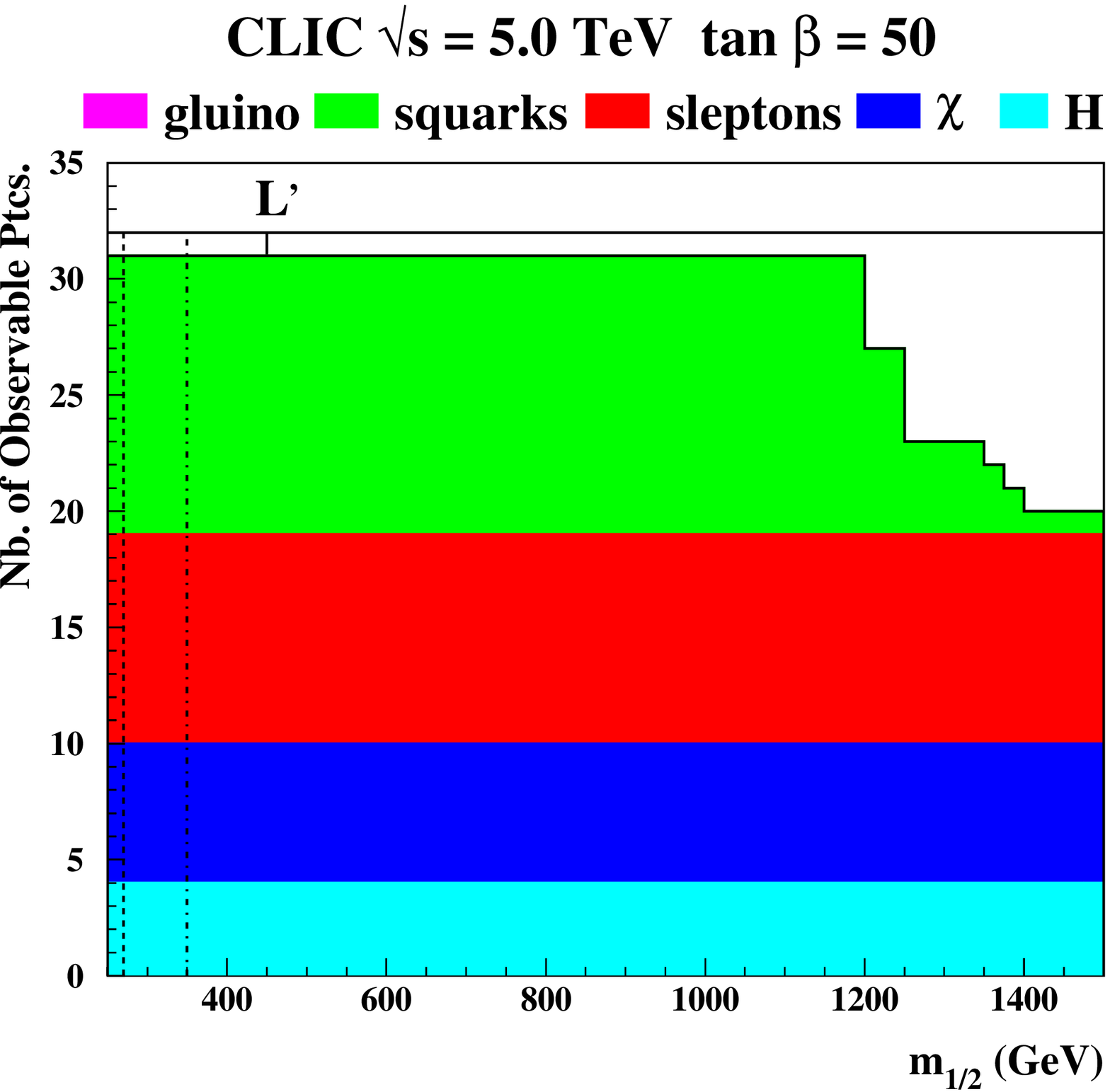}
\hfill \epsfxsize = 0.4\textwidth \epsffile{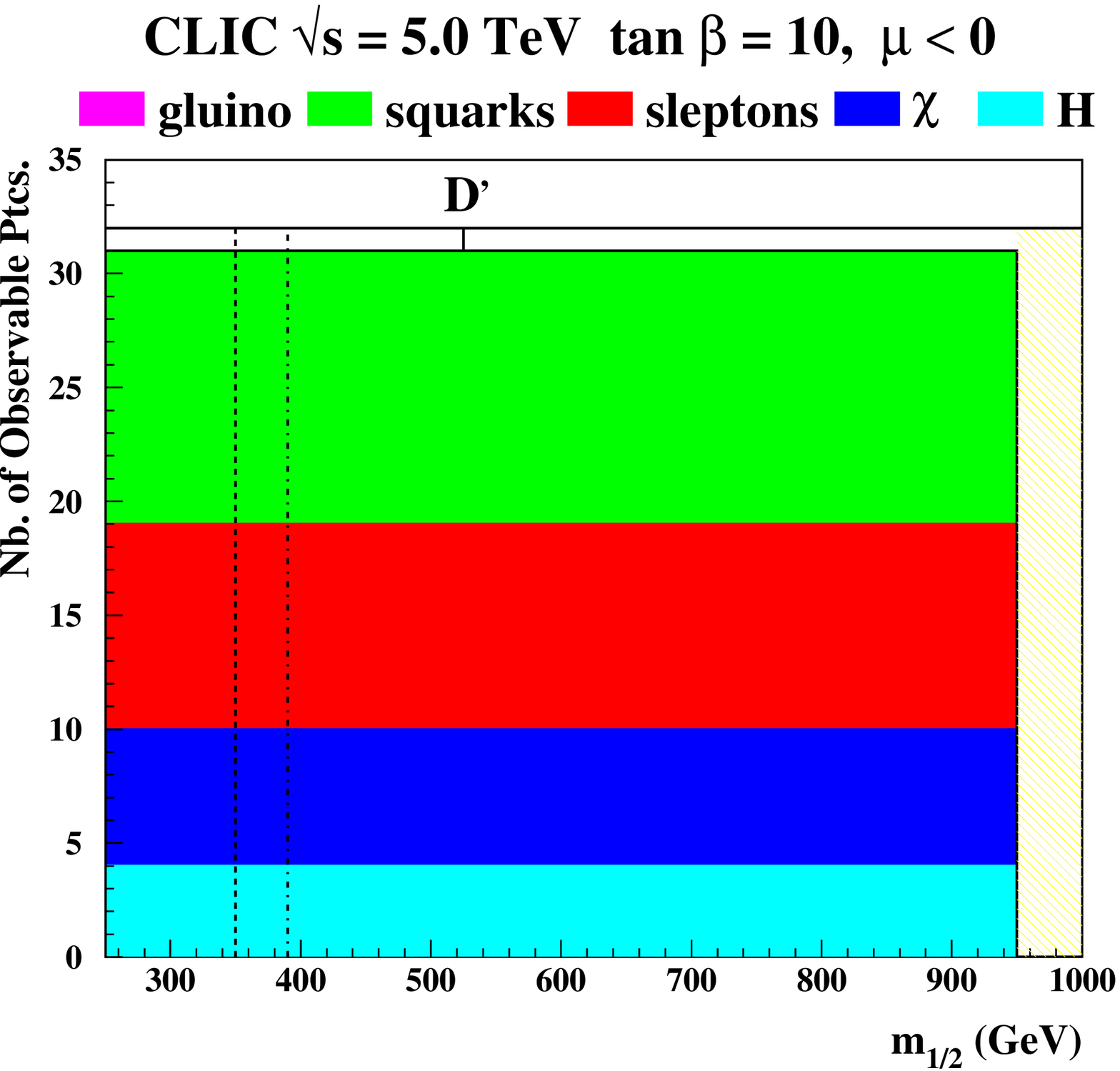}}
\caption{\it
Estimates of the numbers of MSSM particles that may be detectable at the   
5-TeV version of the CLIC linear $e^+ e^-$ collider
as functions of $m_{1/2}$ along the WMAP lines for $\mu > 0$ and $\tan   
\beta = 5, 10, 20, 35$ and $50$, and for $\mu < 0$ and $\tan \beta = 10$.
The locations of updated benchmark points along these WMAP lines are
indicated, as are the nominal lower bounds on $m_{1/2}$ imposed by $m_h$
(dashed lines) and $b \to s \gamma$ (dot-dashed lines).
\vspace*{0.5cm}}  
\label{fig:lc5000}   
\end{figure}

Finally, we display in Fig.~\ref{fig:lc5000} the capabilities of a 5-TeV 
lepton collider such as CLIC. We see that all the MSSM particles are 
detected along all the WMAP lines, with the exception of the gluino that 
would generally have been seen at the LHC, and, for $\mu > 0$ and $\tan 
\beta = 50$, squarks towards the tip of the corresponding WMAP line, 
including the point M.

Despite the larger machine-induced backgrounds and beam energy spread,
CLIC is expected to perform measurements of the properties of accessible
suspersymmetric particles with good accuracy~\cite{CLICPhys}. Slepton and
heavy Higgs boson masses can be determined to ${\cal{O}}$(1\%) accuracy,
also when accounting for realistic experimental conditions and
resolutions.  A similar accuracy can typically be obtained also for
sparticles reconstructed through cascade decays, such as the the $\chi_2$
discussed below. The availability of polarized beams is not only
beneficial to increase the signal cross sections (such as in the case of
slepton-pair production)  but also as an analyzing tool. A combination of
measurements of the stop-pair production with different polarisation
states can be used to determine the stop mixing angle~\cite{CLICPhys}.

\subsection{A Neutralino Case Study}

\begin{figure}
\begin{center}
\begin{tabular}{c c}
\epsfig{file=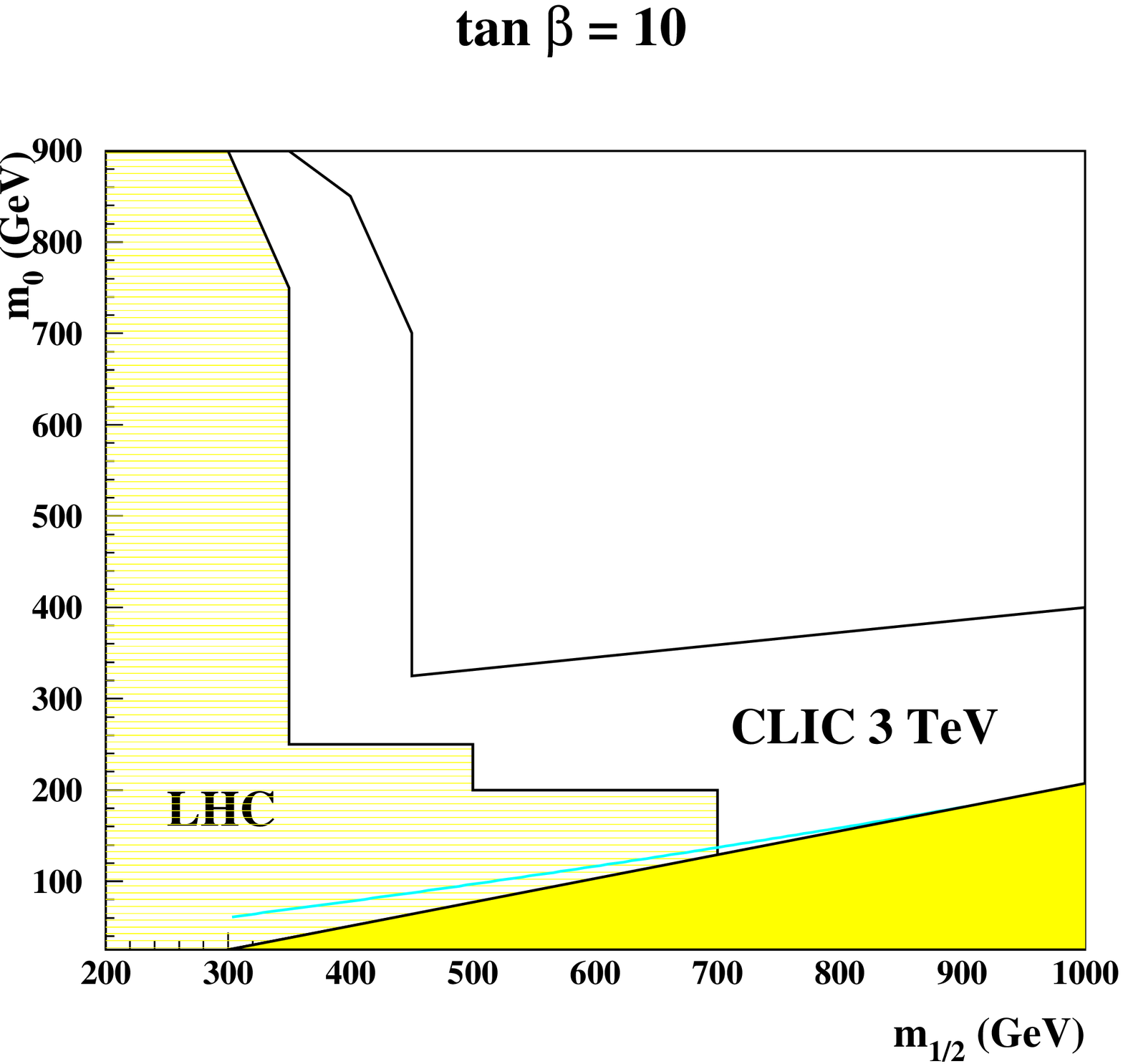,width=7cm} &
\epsfig{file=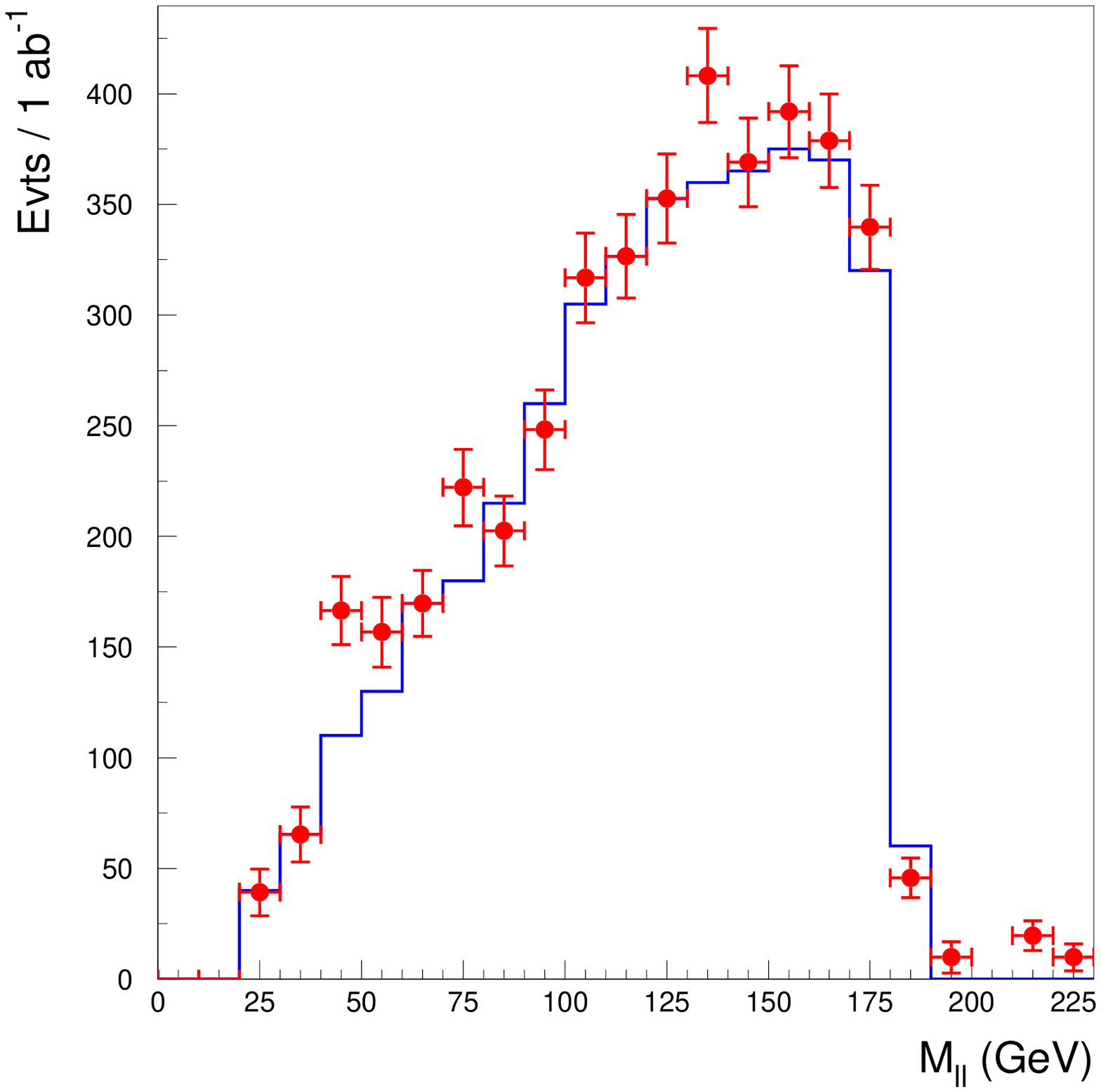,width=6.5cm} \\
\end{tabular}
\end{center}
\caption[]{\it (a) Comparison of the sensitivities in the $(m_{1/2}, 
m_0)$ plane of the LHC and CLIC at 3~TeV
to $\chi_2 \to \chi \ell^+ \ell^-$ decay and (b) the dilepton mass 
spectrum from this decay observable at 3-TeV CLIC for the CMSSM point
$m_{1/2} = 700$~GeV, $m_0 = 150$~GeV, $\mu > 0$ and $\tan \beta = 10$
discussed in the text.} 
\label{fig:neutralino}
\end{figure}

Many of the most interesting differences in the capabilities of the
different colliders discussed above arise in the spectrum of charginos and
neutralinos. In particular, while the LHC and a TeV-scale linear $e^+ e^-$
collider are largely complementary, they may not be able to discover all
the charginos, neutralinos and sleptons when $m_{1/2} \gappeq 400$~GeV,
whereas CLIC can in principle observe all of them along all of the WMAP
lines. To confirm these statements, we now compare the physics reaches of
different accelerators for neutralinos in the $(m_{1/2}, m_0)$ plane, with
the conclusions shown in Fig.~\ref{fig:neutralino}(a).

For the LHC, we show Fig.~\ref{fig:neutralino}(a) the region of the
$(m_{1/2}, m_0)$ plane for $\tan \beta = 10$ in which dilepton structures
due to the decay chains $\chi_2 \to \ell \tilde{\ell}, \tilde{\ell} \to
\ell \chi$ and $\chi_2 \to X Z^0, Z^0 \to \ell^+ \ell^-$ are expected to
be observable in the cascade decays of heavier sparticles.  
We consider $\ell = e, \mu$ only, though the case $\ell = \tau$ would also
be interesting at large $\tan \beta$, and require $\sigma \times$ dilepton
branching ratio to exceed 0.01~pb, as discussed above. For our purposes,
the relevant parts of the $(m_{1/2}, m_0)$ plane are those allowed by the
WMAP dark matter constraint. We see that here the LHC coverage extends up
to $m_{1/2} \simeq 500$~GeV, as also reflected in Fig.~\ref{fig:lhc}.


In order to see how accurately the 3-TeV version of CLIC could measure
heavier neutralinos, we have made a new simulation of the process $e^+ e^-
\to \chi \chi_2$ followed by the decay chains $\chi_2 \to \ell
\tilde{\ell}, \tilde{\ell} \to \ell \chi$ and $\chi_2 \to X Z^0, Z^0 \to
\ell^+ \ell^-$, using {\tt SIMDET}~\cite{Simdet} to simulate the detector
response and {\tt PYTHIA~6.215}~\cite{PYTHIA} interfaced to {\tt
ISASUGRA}~\cite{ISASUGRA} to simulate the signal. Events with at least two
leptons and significant missing energy were selected. Both slepton and
Standard Model gauge-boson pair-production backgrounds were considered.
Combinatorial backgrounds were subtracted by taking the difference of the
pair $e^+e^- + \mu^+\mu^-$ events and the mixed $e^{\pm} \mu^{\mp}$
events, and a sliding window has been used to search for a $>$5-$\sigma$
excess in the $M_{\ell\ell}$ mass distribution. The results shown in
Figure~\ref{fig:neutralino}(a) for $\tan \beta = 10$ make manifest the
extended reach provided by CLIC, covering all the range of $m_{1/2}$ 
allowed by WMAP. At larger values of $\tan \beta$, decay
chains involving the $\tilde{\tau}$ become more significant, requiring a
more detailed study that should include $\tau$ reconstruction.

To benchmark the 3-TeV CLIC capabilities for measuring the masses of heavy
neutralinos, a representative point has been chosen at $m_{1/2} =
700$~GeV, $m_0 = 150$~GeV, $\mu > 0$ and $\tan \beta = 10$, along the
corresponding WMAP line. Here, $m_{\chi_2} = 540$~GeV, $m_{\chi} =
290$~GeV and $m_{\ell_L} = 490$~GeV. As seen in
Fig.~\ref{fig:neutralino}(b), at CLIC the dilepton invariant mass
distribution shows a clear upper edge at 120~GeV due to $\chi_2 \to \ell^+
\tilde{\ell^-}_L$ followed by $\tilde{\ell^-}_L \to \ell^- \chi$, which
can be very accurately measured with 1~ab$^{-1}$ of data. However, in
order to extract the mass of the $\chi_2$ state, the masses of both the
$\tilde{\ell}_L$ and $\chi$ need to be known. In fact, making a
two-parameter fit to the muon energy distribution, the masses of the
$\tilde{\mu}_L$ and $\chi$ can be extracted with an accuracy of 3~\% and
2.5~\%, respectively, with 1~ab$^{-1}$ of data~\cite{smuon}. Improved
accuracy can be obtained with higher luminosity and using also a threshold
energy scan. We therefore assume that these masses can be known to 1.7~\%
and 1.5~\% respectively, which gives an uncertainty of 8~GeV, or 1.6~\%,
on the $\chi_2$ mass, accounting for correlations. This uncertainty is
dominated by that in the $\tilde{\ell}$ and $\chi$ masses. The accuracy on
the determination of the end point alone would correspond to a $\chi_2$
mass determination to 0.3~GeV, fixing all other masses, which is not
sensitive to the details of the beamstrahlung and accelerator-induced
backgrounds.


\section{Conclusions}

We have seen in this paper how different colliders could provide
complementary information about the MSSM spectrum in updated CMSSM
benchmark scenarios and along lines in the $(m_{1/2}, m_0)$ planes for
different values of $\tan \beta$ and both signs of $\mu$. In addition to
direct and indirect laboratory constraints, we have implemented the
constraints on cold dark matter imposed by WMAP and other astrophysical
and cosmological data.

We emphasize that the LSP is stable in any supersymmetric model in which
$R$ parity is conserved, such as the MSSM, in which case astrophysical and
cosmological constraints on dark matter must be taken into account. In
principle, the LSP might be some different sparticle, such as the
gravitino or axino.  Specific scenarios of this type include models with
gauge-mediated supersymmetry breaking, and benchmarks for such models have
been proposed elsewhere~\cite{SPS}. We recall, however, that the WMAP data
on reionization of the early Universe disfavour models with warm dark
matter~\cite{WMAP}, so that all such models must grapple with the
constraint on cold dark matter provided by WMAP and other data.

Alternatively, one might seek to avoid the cold dark matter constraints by
postulating some amount of $R$ violation. The collider signatures of any
such model would differ from those of the MSSM if $R$-violating couplings
are not sufficiently small, and an additional set of astrophysical and
cosmological constraints come into play if the $R$-violating
couplings are not very large. Any $R$-violating scenario must address
these issues.

We have updated a set of thirteen CMSSM benchmark points proposed
previously~\cite{Bench} in light of the current WMAP constraint on
supersymmetric dark matter. Since this constraint has reduced the
dimensionality of the CMSSM parameter space, we have also introduced `WMAP
lines' in the $(m_{1/2}, m_0)$ plane for different values of $\tan \beta$,
and have studied the physics reach of different accelerators, by charting
the decay signatures and numbers of observable particles. It is
interesting to observe that some important signatures, such as the pattern
of $\chi_2$ decays, remain rather uniform along these lines. However, we
note that different signatures may be recovered by relaxing some of the
assumptions taken here. An interesting example is the appearence of large
$\chi_2$ decay branching fractions into $h^0$ or $Z^0$ in models with
non-universal Higgs masses. As these are of importance for defining the
phenomenology of early runs at the LHC, we plan to consider them
separately in conjunction with the corresponding dark matter density
constraints.

The observation at the LHC of an excess of events containing large missing
transverse energy could immediately indicate that supersymmetry is
present, and that $R$ parity is conserved, so that the Universe should
contain relic LSPs. However, it is possible that the relic LSP density
might not saturate the bound on cold dark matter from WMAP, since there
could be other significant contributions to the cold dark matter, {\it
e.g.}, from axions or superheavy metastable particles. It is therefore
interesting and important to study, within the CMSSM framework considered
here, to what extent colliders can verify that the LSP `does its job' of
providing the cold dark matter~\cite{EHNOS}, yielding a relic density that
is neither too small nor too large.

The answer to this question requires a detailed study of the accuracy with
which a given collider (or combination of colliders) can measure sparticle
masses. General measurements of supersymmetric final states, based on the
methods already established by ATLAS and CMS several years ago, involving,
{\it e.g.}, the end-points of kinematic distributions, should allow the
masses of the visible sparticles to be measured with precisions of 10~\%
or better in many cases. This should in turn provide constraints on the
fundamental parameters of upersymmetry at the level of a few percent in
minimal models such as the CMSSM. At this point one should be able to test
the compatibility of the region of parameter space favoured by the LHC
with that preferred by cold dark matter experiments and hypotheses, {\it
e.g.}, with the WMAP lines.

To go further, one needs to understand the sensitivity of calculations of
$\ohsq$ to variations in the supersymmetric model parameters~\cite{EO}.
These sensitivities depend on their values, are different for different
benchmark scenarios, and vary along the WMAP lines. Determining in general
the accuracy with which $\ohsq$ could be calculated would require detailed
simulations going beyond the scope of this paper. However, existing
studies of the previous version of benchmark point B yield a provisional
answer in this case.

\begin{figure}
\begin{center}
\epsfig{file=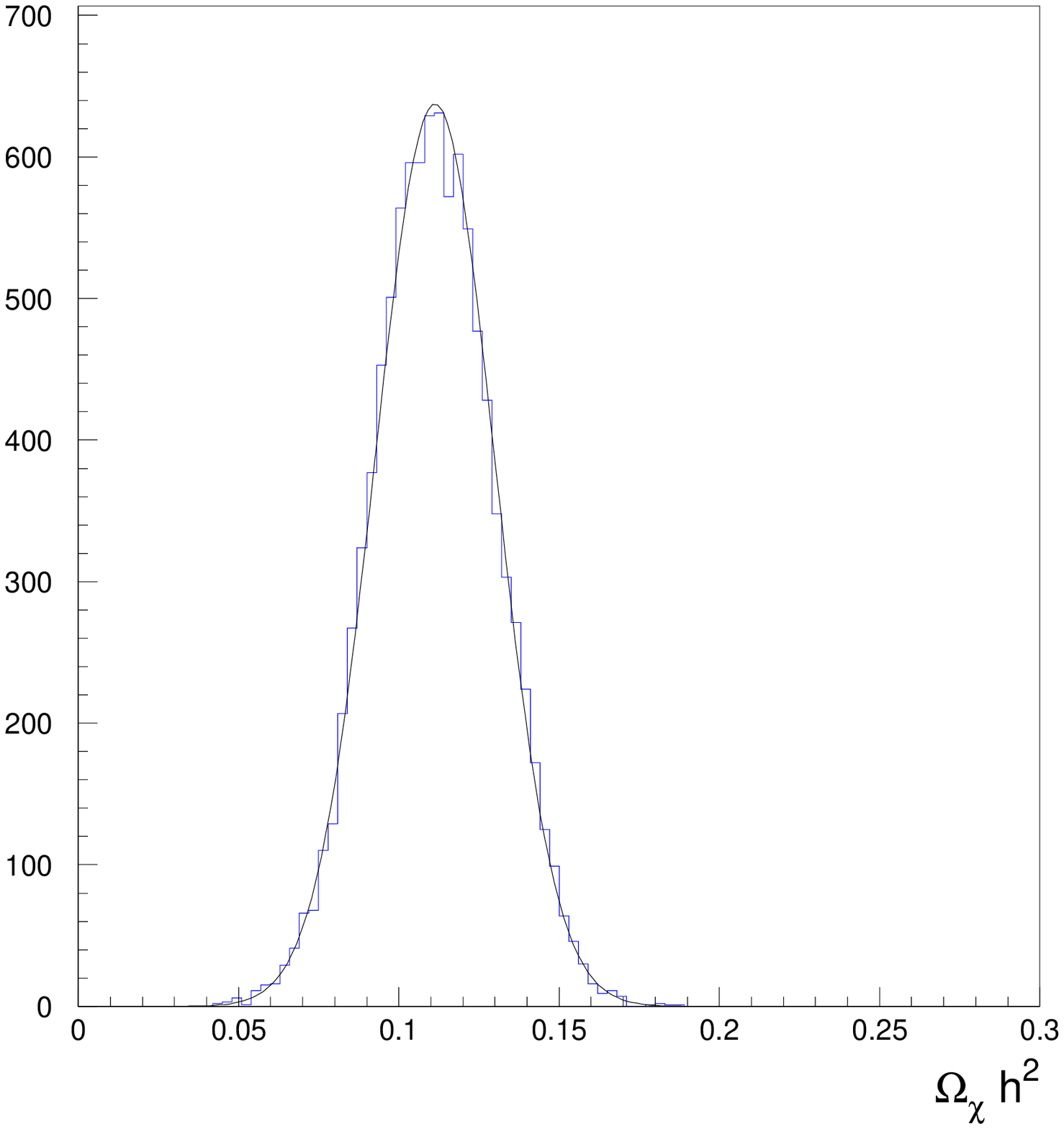,height=5in}
\end{center}
\caption{\label{fig:omega}
{\it
The estimated accuracy with which $\ohsq$ could be predicted on the basis 
of LHC data for the updated benchmark point B, using the projected 
experimental errors reported in ~\cite{Polesello}, a fit to {\tt ISASUGRA 
7.67} parameters and the {\tt Micromegas} code. We find similar results 
using {\tt SSARD}. 
}}
\end{figure}

The following sparticle masses are thought to be measurable with the 
indicated accuracies at the LHC~\cite{Polesello} at SPS point 
1a~\cite{SPS}, which is equivalent for our purposes to benchmark B:
\begin{eqnarray}
m_{\tilde g} & = & 595.1 (8.0), \nonumber \\
m_{\tilde q_L} = 540.3 (8.8), \; m_{\tilde 
q_L} = 520.4 (11.8) & , & m_{\tilde b_1} = 491.9 (7.5), \; m_{\tilde 
b_2} = 524.5 (7.9) \nonumber \\
m_{\tilde \ell_L} = 202.3 (5.0), \; m_{\tilde \ell_R} & = & 143.1 (4.8), 
\; m_{\tilde \tau_1} = 132.5 (6.3), \nonumber \\
m_\chi = 96.2 (4.8), \; m_{\chi_2} & = & 176.9 (4.7), \; m_{\chi_4} = 
377.9 (5.1)
\label{Polesello}
\end{eqnarray}
Since these simulations were made using {\tt ISASUGRA}, we fit the 
determinations (\ref{Polesello}) to the CMSSM parameters $m_{1/2}, m_0$ 
and $\tan \beta$. We assume that the sign of $\mu$ is known, and set
$A_0 = 0$, recalling $A_0$ does not in any case have a large impact on 
$\ohsq$, and find:
\begin{equation}
m_0 = (103 \pm 8)~{\rm GeV}, ~~m_{1/2} = (240 \pm 3)~{\rm GeV}, 
~~\tan \beta = 10.8 \pm 2.
\label{fit}
\end{equation}
We have adjusted the central value of $m_0$ to correspond to the 
updated benchmark value for point B, and then 
propagated the above uncertainties into the calculation of $\ohsq$ 
using {\tt Micromegas}, with the result shown in Fig.~\ref{fig:omega}. 

As seen in Fig.~\ref{fig:omega}, the calculated relic density has a 
well-behaved distribution that is well fit by a Gaussian with 
\begin{equation} 
\ohsq = 0.111 \pm 0.018
\label{ohsq}
\end{equation}  
We have made a similar analysis using {\tt SSARD}, finding an identical
central value of $\ohsq$ and errors $^{+0.014}_{-0.020}$, 
$^{+0.001}_{-0.000}$ and $^{+0.010}_{-0.017}$ associated with $m_0$, 
$m_{1/2}$ and $\tan \beta$, respectively.
We conclude that collider measurements should, in principle enable the 
expected value of $\ohsq$ to be calculated with interesting accuracy. 
Measurements at linear $e^+ e^-$ colliders would increase the accuracy in 
(\ref{ohsq}) and also enable $\ohsq$ to be calculated for other benchmarks 
where data from the LHC alone might be insufficient.

The LHC and linear $e^+ e^-$ collider measurements would also provide 
crucial input for predicting the cross sections expected in direct
dark matter searches or the LSP annihilation rate in the Sun. For this, it
would in general be important to measure the higgsino and gaugino
components of the LSP, which may be possible at the LHC in some
regions of the parameter space by measuring the decay modes of the
heavier gauginos or by comparing the rates of $\chi_2 \to \ell \ell \chi$
via a (virtual) slepton and $\chi_2 \to  \chi h$.

Dark matter physics is another example of the complementarity of LHC and
linear $e^+ e^-$ colliders for exploring new physics in the TeV energy
range. Very possibly, Nature does not choose the CMSSM, and it is
desirable to formulate benchmarks for other possibilities~\cite{SPS}.
However, the CMSSM remains one of the prime options for physics beyond the
Standard Model that might be accessible to the next generation of
colliders. We hope that this paper contributes to a better understanding
of what they might achieve, given the present constraints on supersymmetry
from laboratory experiments and cosmology, as updated here using
information from WMAP.

\vskip 0.5in
\vbox{
\noindent{ {\bf Acknowledgments} } \\
\noindent
The work of K.A.O. was supported partly by DOE grant
DE--FG02--94ER--40823. We thank G.~Belanger, F.~Boudjema and A.~Pukhov 
for useful discussions about {\tt Micromegas}.}

\end{document}